%% file: Main.tex
\documentclass[journal,10pt,a4paper,twocolumn,twoside]{IEEEtran}
\IEEEoverridecommandlockouts

\usepackage{cite}
\usepackage{amsmath,amssymb,amsfonts}
\usepackage{algorithmic}
\usepackage{graphicx}
\usepackage{graphbox} % for vertical alignment
\usepackage{textcomp}
\usepackage{xcolor}
\usepackage{tikz}
\usetikzlibrary{quantikz}
\usepackage{float}
\usepackage{subcaption}
\usepackage[T1]{fontenc}
\usepackage{algorithm}
\usepackage{verbatim}
\usepackage{hyperref}
\usepackage{makecell}
\usepackage{adjustbox}
\usepackage{balance}
\usepackage{anyfontsize}

\usepackage{lipsum}
\usepackage{xcolor}
\definecolor{RED}{rgb}{1,0,0}

\usepackage[utf8]{inputenc}

\newcommand{\eqdef}{\stackrel{\triangle}{=}}

\usepackage{amsthm}
\def\BibTeX{{\rm B\kern-.05em{\sc i\kern-.025em b}\kern-.08em
    T\kern-.1667em\lower.7ex\hbox{E}\kern-.125emX}}

\newtheorem{prop}{Proposition}

\newtheorem{lem}{Lemma}
\newtheorem{cor*}{Corollary}
\newtheorem*{mes}{Pauli Measurements}
\newtheorem*{prob}{Problem}
\newtheorem*{res}{QLAN Entanglement Resource}
\newtheorem*{res2}{Inter-QLAN Entanglement Resource}

\theoremstyle{remark}

\theoremstyle{definition}
\newtheorem{defin}{Definition}
\newtheorem*{remark}{Remark}

\begin{document}

\title{Entanglement-Based Artificial Topology:\\ Neighboring Remote Network Nodes

\author{Si-Yi~Chen, Jessica Illiano, Angela~Sara~Cacciapuoti,~\IEEEmembership{Senior~Member,~IEEE}, \\ and Marcello~Caleffi,~\IEEEmembership{Senior~Member,~IEEE}
}
\thanks{The authors are with the Quantum Internet Research Group, University of Naples Federico II, Naples, NA 80125, Italy}
\thanks{CORRESPONDING AUTHOR: Angela Sara Cacciapuoti (e-mail: angelasara.cacciapuoti@unina.it).}
\thanks{This work has been funded by the European Union under Horizon Europe ERC-CoG grant QNattyNet, n.101169850. Views and opinions expressed are however those of the author(s) only and do not necessarily reflect those of the European Union or the European Research Council Executive Agency. Neither the European Union nor the granting authority can be held responsible for them.  
The work has been also partially supported by PNRR MUR RESTART-PE00000001 and J. Illiano acknowledges PNRR MUR NQSTI-PE00000023. A preliminary conference version of this work has been accepted in the Proceedings of IEEE QCE’24 \cite{CheIllCac-24-1}.}}

\maketitle

\begin{abstract}
    Entanglement is unanimously recognized as the key communication resource of the Quantum Internet. Thus, the possibility of implementing unparalleled network functionalities by exploiting entanglement is gaining huge attention. However, the research efforts in this context are mainly focused on \textit{bipartite entanglement}, often discarding the wide unexplored classes of  entanglement shared among more than two parties, known as \textit{multipartite entanglement}. In this paper, we aim at exploiting multipartite entanglement as \textit{inter-network resource}. Specifically, we consider the interconnection of different Quantum Local Area Networks (QLANs), and we show that multipartite entanglement allows to dynamically generate an inter-QLAN \textit{artificial} topology, by means of local operations only, that overcomes the limitations of the physical QLAN topologies. To this aim, we first design the multipartite entangled state to be distributed within each QLAN. Then, we show how such a state can be engineered to: i) interconnect nodes belonging to different QLANs, and ii) dynamically adapt to different inter-QLAN traffic demands. Our contribution aims at providing the network engineering community with a hands-on guideline towards the concept of artificial topology and artificial neighborhood. 
\end{abstract}


%--------------------------------------------------------------
% Sec I
%--------------------------------------------------------------

\section{INTRODUCTION}
\label{sec:1}
Entanglement is the foundational resource for the Quantum Internet \cite{KozWehVan-22, BriBroDur-09,CacCalTaf-20,CacCalVan-20,DurLamHeu-17,WalPirDur-19,RamDur-20}. However, the research efforts have mainly focused towards a specific network functionality, namely, the distribution of bipartite entanglement. The objective has been sharing Einstein–Podolsky–Rosen (EPR) pairs \cite{FrePirDur-24} between remote nodes, by leveraging quantum repeaters \cite{BriDurCir-98} and well-known strategies such as entanglement swapping and entanglement purification for extending the entanglement range over end-to-end paths, as discussed in \cite{IllCalMan-22}.

Yet, entanglement does not limit to EPR pairs. In fact, multipartite entanglement\footnote{By oversimplifying, bipartite entanglement denotes quantum correlation shared between two subsystems or parties, such as two qubits. In such composite systems, one can distinguish entangled states from unentangled (or separable) states. Quantum correlation shared among more than two subsystems -- such as three qubit systems -- is referred to as multipartite entanglement. Notably, in multipartite systems, the state classification becomes broader and different classes of entangled states can be individuated, such as W states or graph states. The definition of multipartite entangled states requires a dedicated treatise and represents an ongoing exploration which goes beyond the scope of this contribution. However, the interested reader can refer to \cite{IllCalMan-22}, for an introduction of multipartite entanglement from a communication perspective and to \cite{RiePol-11} for a formal introduction. Furthermore, we refer to \cite{HeiDurEis-06} for an in depth treatise on graph states.} -- i.e., entanglement shared between more than two parties -- represents a powerful resource for quantum communication networks \cite{PirDur-18,PirDur-19,RamPirDur-21,IllCalMan-22,CacIllCal-23,KruAndDur-04}, enabling the design of unparalleled   functionalities \cite{CacIllCal-23,CheCacChe-23}. In this paper, we exploit such a resource for one of the fundamental task of the network layer, namely, the interconnection of remote nodes.

Specifically, entanglement enables a new form of connectivity, referred to as \textit{entanglement-enabled connectivity} \cite{IllCalMan-22,CacIllCal-23}, which enables half-duplex unicast links between pairs of nodes sharing entanglement, regardless of their relative positions within the underlying physical network topology. Thus, \textit{the entanglement-enabled proximity builds an overlay topology} -- referred to in the following as \textit{artificial topology} -- where pair of nodes are connected via artificial links despite their physical proximity, as long as they share some form of entanglement. In this context, by exploiting only bipartite entanglement, the identities of the nodes, connected via artificial links, have to be fixed a-priory. On the contrary, through multipartite entanglement, it is possible to decide on-demand -- i.e., at run-time -- the identities of the nodes interconnected within the artificial topology. In other words, by exploiting multipartite entanglement, it is possible to secure two promising features in a quantum network: i) the ability to enable an overlaying artificial topology that differs from the fixed physical topology, coupled ii) with the ability to dynamically change the node neighborhood according to the communication demands. However,  the aforementioned capability activated by multipartite entanglement has yet to be fully explored.

For this, by departing from traditional literature on multipartite entangled states (mainly concerning the analysis of their properties such as the amount of entanglement \cite{HeiDurEis-06,KruAndDur-04,ClaPer-24} and the \textit{pairability} \cite{BraShaSze-22,HahPapEis-19,MorDur-23,CauClaMha-24}) we are motivated to provide the community with an operational and easy-to-use guide to fully leverage and manipulate the artificial topology enabled by multipartite entangled states. Indeed, as recently highlighted in \cite{CacIllCal-23}, providing the aforementioned hands-on guide is preliminary and pivotal for facilitating the design of communication protocols, able to effectively exploit entanglement and its unprecedented peculiarities. As a consequence, our overall objective is to engineer the connectivity in large-scale quantum networks towards envisioned applications characterized by variable traffic demands \cite{WanRahRui-24}.

\begin{figure}
    \centering
        \resizebox{0.49\textwidth}{!}{
            \input{Figures/Fig-01}
    }
    \caption{Schematic representation of the considered physical quantum network architecture. The network comprises several QLANs. Within each QLAN, a super-node generates and distributes resources -- namely, multipartite entangled states -- to a set of quantum nodes -- referred to as \textit{clients} -- with a star-like topology. Inter-QLAN connectivity is enabled by point-to-point quantum channels interconnecting different super-nodes. For the sake of illustration consistency, we maintain the super-node and client node icons used in this figure also in the following figures.}
    \label{fig:01}
    \hrulefill
\end{figure}
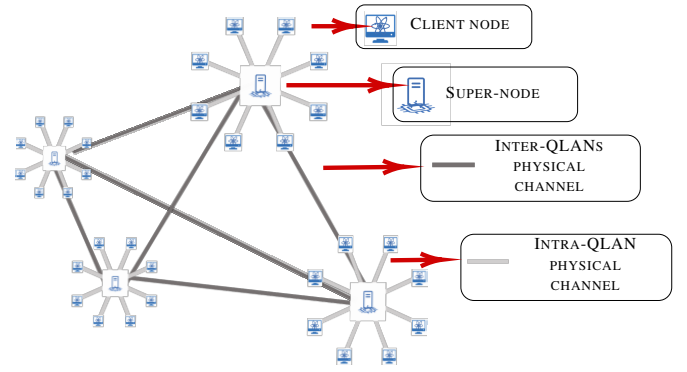

In light of the above motivations, in this paper we consider the interconnection of different Quantum Local Area Networks (QLANs) as building-block of the Quantum Internet, as depicted in Fig.~\ref{fig:01}, and we engineer multipartite entangled states distributed across the QLANs. The aim is to obtain an inter-QLAN artificial topology, where multiple artificial links among remote nodes are dynamically generated according to the traffic demands, overcoming so the constraints imposed by the physical topology. Remarkably, we show that multiple artificial links can be dynamically obtained among remote nodes belonging to different QLANs, by means of local operations only, i.e., via free operations from a quantum communication perspective. This enriches the quantum network functioning with adaptability and flexibility, while maintaining ``low-cost'' requirements, by avoiding to deploy additional physical connections. Specifically, our main contributions can be summarized as follows.

\begin{itemize}
    \item[-] We discuss how to build an artificial topology that matches the underlying physical topology, and that can be used as resource for connectivity between different Quantum Local Area Networks.
    \item[-] We provide an easy-to-use guide for artificial topology manipulations with the tools of graph theory.
    \item[-] We engineer the artificial topology to enable several inter-QLAN artificial topologies, where multiple artificial links among remote nodes are \textit{dynamically generated according to the traffic demands}. These topologies include: \textit{peer-to-peer} (hierarchical and pure), \textit{role delegation} (type I and type II), \textit{clients hand-over}, and \textit{extranet}.
    \item[-] We achieve the above inter-QLAN artificial topologies while overcoming the constraints imposed by the physical topology \textit{using only local operations}, which represent free operations from a quantum communication perspective.
\end{itemize}

We kindly refer the reader to App.~\ref{app:0} for an illustrated guideline on our main contributions.

%--------------------------------------------------------------
\subsection{RELATED WORK}
\label{sec:1.2}

\renewcommand{\arraystretch}{1.2}

\begin{table*}[t!]
\centering
\caption{Summary of related works on multipartite entangled states in quantum networks.}
\label{tab:related_works}

\fontsize{8pt}{8pt}\selectfont
\begin{tabular}{|c|p{0.35\linewidth}|p{0.52\linewidth}|}
\hline
\hline
\multicolumn{1}{|c|}{\textbf{Ref.}} & \multicolumn{1}{c|}{\textbf{Topic}} & \multicolumn{1}{|c|}{\textbf{Key Contribution}} \\ 
\hline

\cite{HeiDurEis-06} & Discuss the basic notions and properties of graph states &  A tutorial introduction into the theory of graph states  \\

\cite{ClaPer-24} & Structure and properties of local sets in graphs & The graphical counterpart
of quantum qudit graph states. \\

\cite{BenHajVan-23} & All-photonic quantum repeaters & Graph states exploited for quantum repeaters. \\  

\cite{HahPapEis-19} & EPR extraction from cluster states & X-protocol extracts EPR pairs while preserving part of the 2D cluster state. \\  

\cite{FrePirDur-24} & Multi-EPR extraction & Multiple EPR extraction while restoring the original 2D cluster paths. \\ 

\cite{BraShaSze-22} & The bounds of EPR extraction & Low bounds of EPR extraction between any $k$ disjoint pairs from $n$-party states.\\

\cite{MorDur-23} & EPR extraction from shared multipartite states & Protocols for extracting EPRs from different pre-shared multipartite states and fidelity analisys. \\  

\cite{MorWalDur-25} & EPR extraction from linear cluster state & Optimization of a linear cluster state for EPR extraction and analysis of impact of noise on the extracted EPR.\\

\cite{CauClaMha-24} & Stabilizer state extraction from multipartite states & Demonstration of an $n$-qubit quantum state  to induce any stabilizer state on any $k$ qubits.\\

\cite{DurBri-04} & Study of persistence of entanglement under noise & Analysis of noise effects on multipartite entanglement. \\  
\cite{DurBri-07} & Entanglement purification & Review on entanglement purification for bipartite and multipartite entangled states\\

\cite{JiLiuZha-24} & Optimization of EPR distribution & LC-orbits of graph states used to minimize required EPRs. \\ 

\cite{CacIllVis-23} & Optimization of multipartite entanglement distribution & Knowing when to stop entanglement distribution in noisy scenarios. \\

\cite{IllCalVis-23} & Entanglement access control &  EAC strategy for coordinating quantum nodes in accessing entangled resources. \\

\cite{KruAndDur-04} & Preparation and distribution of high-fidelity multipartite states & Identify the optimal purification strategies for entangled states. \\
\hline
\hline
\end{tabular}
\end{table*}

Within the context of multipartite entangled states for quantum networks, graph states have been recently exploited for an all-photonic implementation of quantum repeaters \cite{BenHajVan-23}. Furthermore, promising results have been discussed with reference to a particular class of multipartite entangled states, namely, the 2D cluster states, which have been proposed as network resource to be distributed for ``routing'' EPRs through network nodes. As instance, in \cite{HahPapEis-19} the so-called \textit{X-protocol} extracts one EPR between two remote nodes, by leaving part of the remaining graph state intact. This result has been extended in \cite{FrePirDur-24} where multiple EPRs among disjoint paths are extracted. Differently, some works focused on the impact of noise on the distribution of multipartite entangled states \cite{DurBri-04,DurBri-07} as well as on the extraction of EPRs from a multipartite state previously shared between network nodes  \cite{MorDur-23,MorDur-23-1,MorWalDur-25}. Notably, the ``additional material'' in \cite{MorWalDur-25} includes an explicit example of the effects of decoherence on operations different Pauli measurements on graph states and the extraction of EPR pairs. Additionally, in \cite{JiLiuZha-24} the LC-orbits of graph states are exploited to minimize the number of EPR required for their distribution. For multipartite entanglement distribution, in \cite{CacIllVis-23}, a framework based on a Markov decision process is developed for determining when it is convenient to stop early the distribution process. To solve the contention problem arising in accessing a multipartite entangled resource, an entanglement access control (EAC) \cite{IllCalVis-23} strategy is proposed to coordinate the access of quantum nodes. The above contributions represent promising relevant results on multipartite entanglement as they highlight how this heterogeneous resource can be exploited for extracting one or more EPRs.

Yet, our main goal is not to extract EPR pairs or explore applications on a given network architecture. Conversely, we rather aim at engineering a multipartite entangled state able to easily adapt to different traffic patterns.

Indeed, while the physical topology is associated with a fixed pre-determined connectivity, the artificial topology is associated with a connectivity that can be changed dynamically. Such a promising possibility represents the main motivation of our contribution, which aims at exploiting the features of multipartite entanglement for interconnecting nodes belonging to different networks according to traffic patterns. In this paper, we show that this can be achieved by engineering the multipartite entangled state shared among quantum network nodes. In this context, a set of remote nodes is directly connected via artificial links regardless their relative physical position. In other words, \textit{remote} nodes can be in each other's \textit{proximity} within the artificial topology. From this concept, it can be observed that entanglement redefines the concept of neighborhood and, as our title suggests, it allows to ``neighbor'' remote network nodes. 

From the above, it follows that, rather than focusing on the number of EPR pairs consumed or extracted, we focus on the features of the topologies arising from multipartite entanglement. Specifically, we aim at enabling an inter-QLANs artificial topology that can be easily tailored to different traffic patterns by local operations only.  

\vspace{9pt}

The rest of this paper is organized as follows. In Sec.~\ref{sec:2} we provide the reader with the preliminary concepts about graph theory, graph states and the description of operations on graph states. In Sec.~\ref{sec:3} we present the system model by detailing the considered network architecture as well as the distribution of the considered graph states. In Sec.\ref{sec:4} we discuss our main results, by proving a discussion of the choice of the initial multipartite entangled state distributed in each QLAN. Then, we show that multiple artificial links can be dynamically obtained among remote nodes belonging to different QLANs, by means of local operations only. Finally, in Sec.~\ref{sec:5} we conclude the paper.

%--------------------------------------------------------------
% Sec II
%--------------------------------------------------------------
\section{PRELIMINARIES}
\label{sec:2}

%--------------------------------------------------------------
\subsection{GRAPH THEORY BASICS}
\label{sec:2.1}

Here, we collect some definitions that will be used in the paper and we introduce the adopted notation, as summarized in Table.~\ref{tab:00}. We refer the readers to \cite{MazCalCac-25,HeiDurEis-06} for a wider treatise of graph states, and the correspondence between operations on graph states and their associated graph.

A graph is a collection of vertices\footnote{We resort to graph theory for modeling a communication network. Hence, in the following, we equivalently refer to vertices as nodes, by assuming that each vertex models a network node and an edge models an artificial link, i.e., shared entanglement constituting a quantum communication resource between nodes. The mapping between vertices and nodes is defined in Sec.~\ref{sec:4.1}.} interconnected by edges, and it is formally defined as the pair $G=(V,E)$ of the two (finite) sets of vertices $V$ and edges $E\subseteq V^2$, respectively.

\begin{remark}
    In the following, we will restrict our attention on \textit{finite} graphs, namely, graphs with finite $V$ and $E$. Furthermore, we will consider \textit{undirected} and \textit{simple} graphs,
     since these two properties are required for the mapping between graphs and graph states \cite{MazCalCac-24-QCNC,HeiDurEis-06}, as analyzed in the following subsection.
     We further highlight that \textit{undirected} and \textit{simple} graphs correspond to graphs with un-directed edges\footnote{In the following, with a small notation abuse, we denote un-directed edges as $(i,j)$ -- rather than with angle brackets as $\{i,j\}$ -- for notation simplicity.}, i.e., $(x,y) \equiv (y,x)$, and with edges that cannot connect the same vertex, i.e., $E \subseteq V^2 \eqdef \big\{ (x,y): x,y \in V \wedge x \neq y \big\}$.
\end{remark}

\renewcommand{\arraystretch}{1.2}
\begin{table*}[t]
\centering
\caption{{Adopted Notation}}
\label{tab:00}

\fontsize{8pt}{8pt}\selectfont
\begin{tabular}{|c|p{0.35\linewidth}|c|c|p{0.35\linewidth}|}
\hline
\hline
\multicolumn{1}{|c|}{\textbf{Symbols}} & \multicolumn{1}{c|}{\textbf{Definitions}} & &
\multicolumn{1}{c|}{\textbf{Symbols}} & \multicolumn{1}{c|}{\textbf{Definitions}}   
\\ 
\hline

    $G$ & graph, collection of edges and vertices &  & $\tau(G)$ & complement graph of a graph $G$\\
            
    $E$& set of edges &  & $\tau_i(G)$ &  local complementation of the graph $G$ at vertex $i$ \\

    $E_A$ & set of edges with endpoints belonging to $A$ &  & $\overline{E}$ & complement edge set of $E$ \\
            
    $V$& set of vertices &  & $G-i$ & deletion of a vertex $i$ from a graph $G$\\

    $|V|$ & number of elements in the set $V$ &  & $K_{n_1,n_2}$ & complete bipartite graph with $|P_1| = n_1$, $|P_2| = n_2$\\

    $V^2$ & set of edges connecting every pair of vertices in $V$ &  & $\ket{G}$ & graph state associated with the graph $G$\\
            
    $N_i$ & neighborhood of a vertex $i$ &  & $S_n$ & a $n$-vertices star graph\\

    $P_i$ & vertex set partition &  & $S_{n_1,n_2}$ & binary star graph with $|P_1| = n_1$ and $|P_2| = n_2$\\
        
    $G[A]$ & subgraph of graph $G=(V,E)$ induced by $A\subset V$ &  & $\dot{v}^1_1$ & super-node in QLAN 1\\
            
    $K_n$& complete graph with $n$ vertices &  & $\ddot{v}^1_1$ & super-node in QLAN 2\\
            
    $A \times B$ &  set of all the possible edges having one &  & $\dot{v}^2_i$ & $i-th$ client node in QLAN 1\\

    & endpoint in vertex set $A$ and the other in vertex set $B$ &  & $\ddot{v}^2_i$ & $i-th$ client node in QLAN 2\\

\hline
\hline
\end{tabular}
\end{table*}

Whenever two vertices $i$ and $j$ are the endpoints of an edge $(i,j)$, the two vertices are defined as \textit{adjacent}. We equivalently refer to adjacent vertices as neighbour nodes, as in the following. 

\begin{defin}[\textbf{Neighborhood}]
    \label{def:01}
    Given a graph $G = (V,E)$ and a vertex $i \in V$, the neighborhood of $i$ is the set $N_i$ of vertices adjacent to $i$:
    \begin{align}
        \label{eq:01}
        N_i= \{j \in V : (i,j) \in E \}.
    \end{align}
\end{defin}

\begin{defin}[\textbf{Induced Subgraph}]
    \label{def:02}
    Given a graph $G=(V,E)$ and a subset of vertices $A \subseteq V$, the \textit{subgraph induced by subset} $A$, denoted with $G[A]$, is the graph having: i) as vertices, the set $A$, and ii) as edges, the edges in $E$ that have both endpoints in $A$. Formally: 
    \begin{equation}
        \label{eq:02}
        G[A] = (A, E_{A}),
    \end{equation}
with  $E_A\eqdef\big\{(i,j) \in E : i \in A \wedge j \in A \big\}$.
\end{defin}
Whether the subset of vertices $A$ is the neighborhood $N_i$ of some vertex $i$, then $G[N_i]$ is referred to as the \textit{subgraph induced by the neighborhood of $i$}.

\begin{defin}[\textbf{Complete Graph}]
    \label{def:03}
    A complete graph of order $n=|V|$ is a graph $K_n$ such that every pair of vertices is adjacent:
    \begin{equation}
    \label{eq:03}
        K_n=(V, V^2).
    \end{equation}
\end{defin}

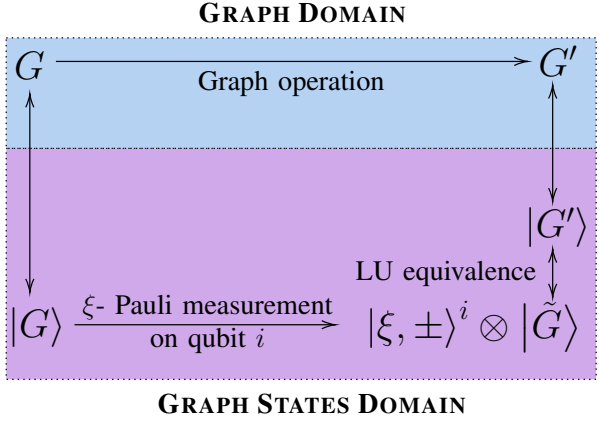
\begin{figure}
    \centering
    \begin{adjustbox}{width=0.9\columnwidth}
    
        \input{Figures/Fig-G}
    
    \end{adjustbox}
    \caption{Schematic diagram of the correspondence between \textit{graph} domain and \textit{graph state} domain, i.e, of the mapping between projective measurements through Pauli operators on graph states and transformations of the associated graphs.}
    \label{fig:02}
    \hrulefill
\end{figure}

In the following, given two vertex sets $A,B\subseteq V$, we widely use the symbol $A \times B \subseteq V^2$ to denote the set of all the possible edges having one endpoint in $A$ and the other in $B$:
\begin{equation}
    \label{eq:04}
    A \times B \eqdef \big\{ (i,j) \in V^2: i\in A \wedge j \in B \big\} \subseteq V^2.
\end{equation}

\begin{defin}[\textbf{Graph Complementation}]
    \label{def:04}
    The complement (or inverse) of a graph $G=(V,E)$ is the graph $\tau(G) = (V, \overline{E})$ defined on the same vertex set $V$, such that two vertices of $\tau(G)$ are adjacent \textit{iff} they are not adjacent in $G$. Formally:
    \begin{align}
        \label{eq:05}
        \tau(G) &= (V, \overline{E}),
    \end{align}
    with $\overline{E}$ given by:
    \begin{equation}
        \label{eq:06}
        \overline{E} \eqdef V^2 \setminus E = \big\{ (i,j) \in V^2 : (i,j) \not\in E \big\}.
    \end{equation}  
\end{defin}

Graph complementation can be done also with respect to the subgraph $G[N_i]$ induced by the neighbourhood $N_i$ of vertex $i$. In this case, it is usually referred to as \textit{local complementation of $G$ at vertex $i$} and denoted as $\tau_i(G)$.

\begin{defin}[\textbf{Local Complementation}]
    \label{def:05}
    Given a graph $G=(V,E)$, the \textit{local complement of $G$ at vertex $i$} is the graph $\tau_i(G)$ obtained by complementing the subgraph $G[N_i]$ induced by neighbourhood $N_i$ of vertex $i$, while leaving the rest of the graph unchanged:
    \begin{align}
        \label{eq:08}
        \tau_i(G) = \big( V, ( E \cup N_i^2 ) \setminus E_{N_i}\big)
     \end{align}
    with $E_{N_i}$ defined in Def.~\ref{def:02}.
\end{defin}

\begin{defin}[\textbf{Vertex Deletion}]
    \label{def:06}
    The deletion of a vertex $i$ from a graph $G=(V,E)$ generates a new graph, denoted with a small notation abuse as $G - i$, which is obtained by removing both vertex $i$ and all the edges connecting $i$ with its adjacent vertexes:
    \begin{align}
        \label{eq:07}
        G - i = \big(V \setminus \{i\}, E \setminus ( \{i\} \times N_i ) \big)
    \end{align}
    Hence, the edge-set of $G-i$ is the set of edges in $G$ without the edges with vertex $i$ as endpoint. 
\end{defin}

%--------------------------------------------------------------
\subsection{GRAPH STATES}
\label{sec:2.2}

\textit{Graph states} constitute a notable class of multipartite entangled states from a network engineering perspective \cite{HeiEisBri-04,HeiDurEis-06}. This class of multipartite entangled states can be described by leveraging the graph theory tools presented in Sec.\ref{sec:2.1}. More into details and according to a constructive definition, associated with a \textit{graph state} $\ket{G}$ there exists a graph $G$. The mapping between $G$ and $\ket{G}$ can be described as follows: each vertex in $G$ corresponds to a qubit belonging to the state $\ket{G}$ and prepared in the state $\ket{+}$; furthermore, each edge in $G$ corresponds to a Controlled-Z (\texttt{CZ}) logical gate acting on the qubits pair corresponding to the endpoints of the given edge \cite{HeiEisBri-04}.
\begin{remark}
    The motivation of such a mapping arises from the correspondence between graph edges and interactions -- i.e., Ising interactions -- between subsystems of the composite  entangled system. Accordingly, vertices represent physical quantum subsystems and edges represent their interactions. 
\end{remark}

We recall that the \texttt{CZ} operation is an entangling operation. As a consequence, the distribution of a graph state among remote nodes of a quantum network establishes an entanglement-based connectivity among remote nodes.

Formally, the graph state $\ket{G}$ associated to graph $G=(V,E)$ can be expressed as\footnote{With a (widely-used) notation abuse in \eqref{eq:09}, since the application of the $\texttt{CZ}_{(i,j)}$ gate on the state $\ket{+}^{\otimes n}$ requires a reference to $n-2$ identity operations $I$ acting on all the qubits different from $i$ or $j$.}:
\begin{equation}
    \label{eq:09}
    \ket{G}=\prod_{(i,j)\in E}\texttt{CZ}_{(i,j)}\ket{+}^{\otimes n}
\end{equation}
with $\ket{+}=\frac{\ket{0}+\ket{1}}{\sqrt{2}}$, $n = |V|$ and $\texttt{CZ}_{(i,j)}$ denoting the $\texttt{CZ}$ gate applied to the qubits associated to the neighbours $i,j \in V$.

A graph state $\ket{G}$ uniquely corresponds to a graph $G$. This means that two different graphs $G$ and $G'$ do not describe the same graph state. However, graph states of two different graphs might be equivalent accordingly to some criteria, determining some equivalence classes among such states.
From a network engineering perspective, an equivalency class of interest is represented by the so-called \textit{local unitary} (LU) equivalence, since local unitaries do not change the amount of entanglement. Hence they do not change the communication resources available at the network nodes.

\begin{defin}[\textbf{LU equivalence}]
    \label{def:07}    
    Given two $n$-qubit quantum states, say $\ket{G}$ and $\ket{G'}$, then $\ket{G}$ and $\ket{G'}$ are LU-equivalent \textit{iff} there exists $n$ local-unitary operators $\{ U_i \}$ so that \cite{Kra-10}:
    \begin{equation}
        \label{eq:10}
        \ket{G'}=\bigotimes_i U_i \ket{G}
    \end{equation}
\end{defin}

The mapping between graph states and graphs is of a paramount importance, beyond the expression given in \eqref{eq:09} and, hence beyond a merely representation purpose. 

In fact, a projective measurement through one of the three Pauli operators -- namely, $\sigma_x, \sigma_y,$ or $\sigma_z$ -- on a qubit of the graph state $\ket{G}$ yields to a new graph state $\ket{\tilde{G}}$ among the remaining unmeasured qubits. Notably, as discussed in \cite{HeiDurEis-06,HeiEisBri-04}, this new graph state $\ket{\tilde{G}}$ can be obtained -- up to local unitaries -- by means of simple transformations on the graph $G$ associated to the original graph state $\ket{G}$, such as vertex deletion and the local complementation introduced in Defs.~\ref{def:06} and \ref{def:05}, respectively. 

As illustrated in Fig.~\ref{fig:02}, $\ket{\tilde{G}}$ denotes the new graph state that is LU-equivalent to $\ket{G'}$, i.e.,  the graph state obtained by performing manipulation on the graph $G$ associated to the original graph state $\ket{G}$.

Since projective measurements through Pauli operators are exploited in Sec.~\ref{sec:3} for engineering the artificial connectivity enabled by entanglement, it is convenient to summarize their effects on an arbitrary graph state $\ket{G}$ \cite{HeiDurEis-06,HeiEisBri-04}.

\begin{mes}
    \label{mes:01}
    The projective measurement via a Pauli operator $\sigma_\xi^i$ on the $i$-th qubit of the graph state $\ket{G}$ -- namely, on the qubit associated to vertex $i$ in graph $G$ -- yields to a new graph state $\ket{\tilde{G}}$\footnote{With a mild notation abuse, the dependence on qubit $i$ is omitted for the sake of notation simplicity.} among the unmeasured qubits, which is LU-equivalent to the graph state $\ket{G'}$ associated to the graph $G'$ obtained with  vertex deletion and local complementations:
    \begin{align}
        \label{eq:11}
         G' &\equiv \begin{cases}
                 G  - i & \text{if } \sigma_{\xi}^i = \sigma_z \\
                 \tau_i(G) - i & \text{if } \sigma_{\xi}^i = \sigma_y \\
                \tau_{k_0} \left( \tau_i\big(\tau_{k_0}(G)\big)-i \right)  & \text{if } \sigma_{\xi}^i= \sigma_x.
            \end{cases}
    \end{align}
In \eqref{eq:11}, $k_0 \in N_i$ denotes an arbitrary neighbor of vertex $i$. For more details please refer to \cite{MazCalCac-24-QCNC}.
\end{mes}

%--------------------------------------------------------------
% Sec III
%--------------------------------------------------------------
\section{SYSTEM MODEL}
\label{sec:3}

%--------------------------------------------------------------
\subsection{NETWORK TOPOLOGY}
\label{sec:3.1}

\begin{figure*}
    \centering
        \resizebox{0.9\textwidth}{!}{
    \input{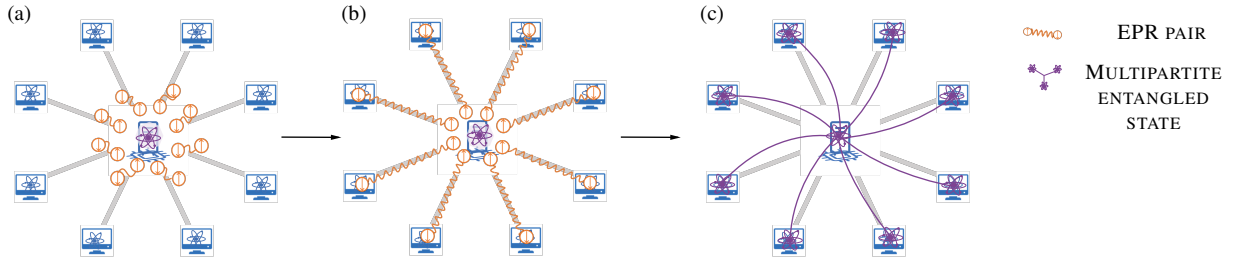}
    }
    \caption{Pictorial representation of the multipartite entanglement distribution process within a single QLAN of Fig.~\ref{fig:01}. (a) The super-node is responsible for entanglement generation and distribution within each QLAN. Accordingly, it locally generates the multipartite entanglement state and distribute it via teleportation. For this, one EPR pair per client must be generated at the super-node. (b) Once an EPR pair is shared between super-node and each client, one e-bit of the multipartite entangled state can be teleported to the client by consuming such an EPR. (c) Eventually, the multipartite entangled state is distributed to the clients so that all the QLAN nodes, including the super-node, are entangled.}
    \label{fig:03}
    \hrulefill
\end{figure*}
We consider, as the archetype of the future Quantum Internet, the network resulting from the interconnection of different Quantum Local Area Networks (QLANs) \cite{MazCalCac-24-QCNC}. 

Entanglement generation is a complex hardware-demanding task, that becomes even more challenging when it comes to multipartite entanglement. For this, as commonly adopted in literature \cite{EppKamBru-17,AviRozWeh-22,VarGuhNai-21-1,BugCouOma-23,IllCalVis-23}, it is pragmatic to assume each QLAN organized in a star-like physical topology, as represented in Fig.~\ref{fig:01}, with a set of \textit{clients} connected to a specialized \textit{super-node}, which is responsible for entanglement generation and distribution.
Accordingly, multipartite entangled states are generated locally at each super-node, and then distributed to the corresponding clients via teleportation process, as represented in Fig.~\ref{fig:03}. 

The rationale for this strategy -- i.e, for distributing multipartite entanglement via teleportation rather than via direct transmission -- lies in the higher robustness against losses and tolerance to different persistence levels exhibited by different classes of multipartite states \cite{RiePol-11,BruVer-12,IllCalVis-23}.

While intra-QLAN topologies are pragmatically assumed as star-like physical topologies for the reasons above, no constraints are enforced to inter-QLAN physical connectivity, which is enabled by quantum links interconnecting different super-nodes as shown in Fig.~\ref{fig:01}.

%--------------------------------------------------------------
\subsection{PROBLEM STATEMENT}
\label{sec:3.2}

Stemming from the network architecture introduced so far, we can now formally define our problem, by focusing on the toy-model constituted by two QLANs interconnected by a single physical link between the corresponding super-nodes. We highlight that we develop our analysis in a worst-case scenario, namely the scenario where for each use of this channel, only an EPR pair can be distributed.

\begin{prob}
    Given two QLANs, interconnected by a single physical link between the corresponding super-nodes, the goal is to design and engineer a multipartite entanglement state distributed in each QLAN so that artificial links among nodes belonging to different QLANs can be dynamically obtained on-demand, by overcoming so the constraints induced by the physical topology.
\end{prob}

In essence, an \textbf{artificial link} represents a virtual communication link established between two remote nodes, since they share some entanglement. 
This concept can be exemplified by considering an EPR pair shared between two nodes. Specifically, as long as the two node share an EPR, they can fulfill a communication task. As instance, they can transmit a qubit through the teleportation process, even in absence of a physical quantum link. In this sense, we say that the maximally entangled state acts as a virtual communication link. 

In EPR-based networks, artificial links between distant nodes can be obtained by relying only on bipartite entanglement and thus by performing swapping operations at intermediate nodes \cite{CacCalVan-20} so that an EPR pair between remote nodes is eventually obtained. Yet, this strategy presents a drawback: the identities of the nodes to be artificially linked must be decided a-priori. In other words, for each EPR pair\footnote{Hereafter, we obviously refer to maximally-entangled EPR pairs, neglecting any noise affecting the EPR generation and distributing for the sake of exposition simplicity.} distributed through the inter-QLAN link only one artificial link among distant nodes can be obtained. This implies that artificial links via entanglement swapping is reminiscent of \textit{reactive} classical routing strategies, where the source-destination path is discovered when a packet is ready to be transmitted.

Conversely, in multipartite-based networks, multiple artificial links between distant nodes can be obtained by properly choosing the initial multipartite state and by wisely manipulating it via local operations, i.e., via \textit{free operations}\footnote{Pauli measurements and local complementation operations are considered free operations from a quantum communication perspective as they are \textit{local operations}, namely, operations that do not require entanglement distribution nor quantum communications for being performed. Indeed, the assumption of being freely able to measure one qubit and to perform single-qubit gates represents a fair and commonly-adopted assumption, ubiquitous to different research areas ranging from computing \cite{CalAmoFer-22}, through cryptography \cite{PirAndBan-20} to sensing \cite{DegReiCap-17}. From a hardware perspective, Pauli measurements can be implemented through superconducting nanowires, photon detectors, polarizing beam splitters (PBS) and homodyne detection, which enable projection via Pauli operators \cite{AltJefKwi-05}. Furthermore, local complementation is achieved through single qubit operations \cite{HeiDurEis-06}, which can be implemented through lasers \cite{VitScaWil-15}, specific pulses \cite{AleKanEgg-20} or linear optical components \cite{KokMunNem-07}, depending on the underlying quantum technology.} from a quantum communication perspective. Hence, it is possible to build an \textit{artificial network topology} interconnecting multiple remote nodes via multiple artificial links upon the physical topologies. In such a way, we can pro-actively generate and distribute entanglement among subset of nodes of different QLANs so that the identities of the nodes eventually communicating can be chosen dynamically at run time.  Clearly, this strategy is reminiscent of \textit{proactive} classical routing strategies, where source-destination paths are discovered in advance, and they remain ready to be used eventually, when the necessity of transmitting a packet arises.

\begin{remark}
    The artificial topology will be eventually manipulated -- without the need of further quantum communications -- when a communication request will be ready to be served by extracting the ultimate artificial link interconnecting the effective source-destination pair. In this regard, we further clarify an important aspect underlying the concept of artificial link. Each artificial link represents the ``possibility'' to extract one EPR pair between the two nodes -- i.e., the two endpoints of the artificial link -- from a larger multipartite entangled state, that is a multipartite entangled state shared between a larger set of nodes. Nevertheless, the number of EPR pair that can be extracted from a given multipartite entangled state at the same time strictly depends on the structure of the considered multipartite entangled state. Furthermore, some of the artificial link of the multipartite entangled state are consumed during the extraction process.
\end{remark}

%--------------------------------------------------------------
% Sec IV
%--------------------------------------------------------------
\section{NEIGHBORING REMOTE NODES}
\label{sec:4}

%--------------------------------------------------------------
\subsection{ENGINEERING MULTIPARTITE ENTANGLEMENT}
\label{sec:4.1}

\begin{figure*}[t]
    \centering
    \begin{minipage}[t]{\textwidth}
        \begin{minipage}[t]{0.14\textwidth}
            \centering\includegraphics[width=0.8\textwidth]{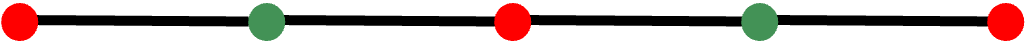}
            \subcaption{Path graph $P_n$.}
            \label{fig:04-a}
        \end{minipage}
        \begin{minipage}[t]{0.14\textwidth}
    	   \centering
    	   \includegraphics[width=0.8\textwidth]{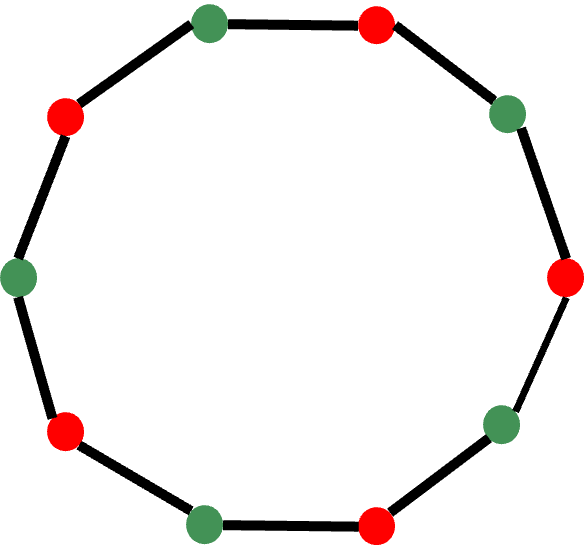}
    	   \subcaption{Even Cycle graph $C_{2k}$.}
    	   \label{fig:04-b}
        \end{minipage}
        \begin{minipage}[t]{0.14\textwidth}
    	   \centering
    	   \includegraphics[width=0.8\textwidth]{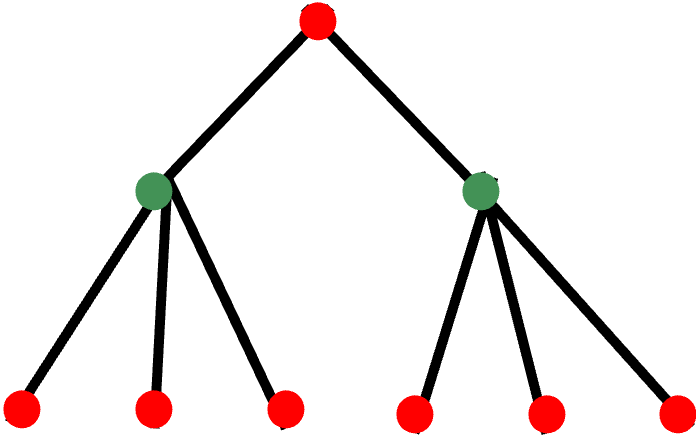}
    	   \subcaption{Tree graph $T$.}
    	   \label{fig:04-c}
        \end{minipage}
        \begin{minipage}[t]{0.14\textwidth}
    	   \centering
    	   \includegraphics[width=0.8\textwidth]{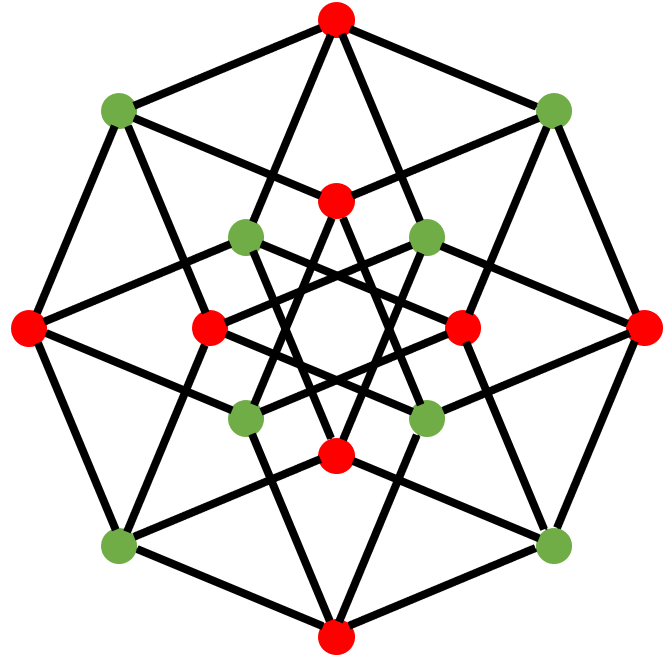}
    	   \subcaption{Hypercube graph $Q_n$.}
    	   \label{fig:04-d}
        \end{minipage}
        \begin{minipage}[t]{0.13\textwidth}
    	   \centering
    	   \includegraphics[width=0.7\textwidth]{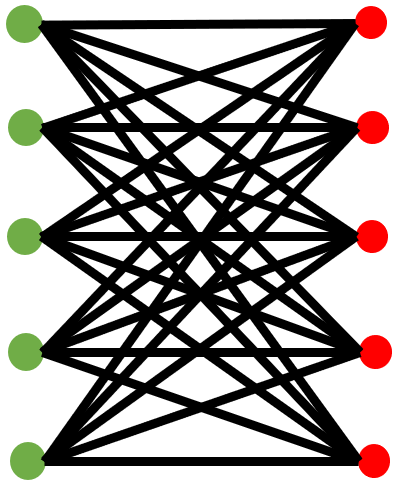}
    	   \subcaption{Complete bipartite graph $K_{n_1,n_2}$. }
    	   \label{fig:04-e}
        \end{minipage}
        \begin{minipage}[t]{0.13\textwidth}
    	   \centering
    	   \includegraphics[width=0.7\textwidth]{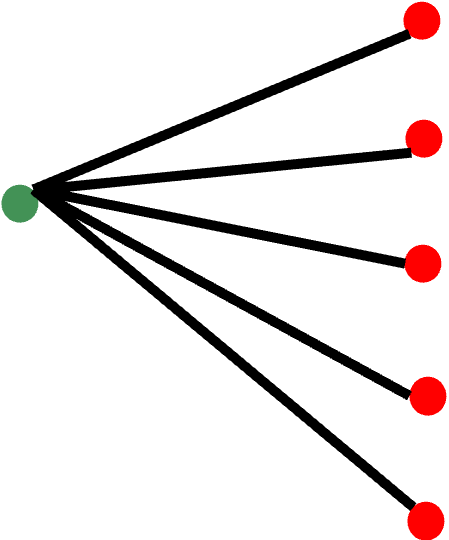}
    	   \subcaption{Star graph $S_n$.}
    	   \label{fig:04-f}
        \end{minipage}
        \begin{minipage}[t]{0.13\textwidth}
    	   \centering
    	   \includegraphics[width=0.7\textwidth]{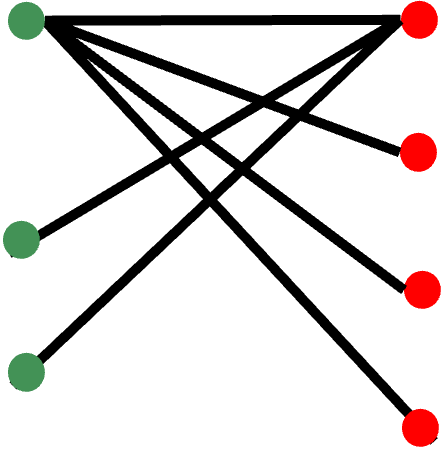}
    	   \subcaption{Binary star graph $S_{n_1,n_2}$.}
    	   \label{fig:04-g}
        \end{minipage}
    \end{minipage}
    \caption{Notable examples of two-colorable graphs.} 
    \label{fig:04}
    \hrulefill
\end{figure*}

As anticipated, we aim at engineering the entanglement-based artificial topology for \textit{neighboring distant nodes}, namely, for creating artificial links among network nodes belonging to different QLANS. To this aim, the choice of the initial multipartite entangled state distributed in each QLAN is of paramount importance.

As discussed in Sec.~\ref{sec:2.2}, we focus our attention on graph states due to the useful mapping between operations on a graph state $\ket{G}$ and transformations of the associated graph $G$. Yet, graph states represents a wide class of multipartite entangled states. In the next sections, we design and manipulate a specific class of graph states that allows us to enable multiple artificial links among distant nodes belonging to different QLANs. Before formally introducing this specific graph state in Def.~\ref{def:11}, the following preliminaries are needed. 

More into details, the so-called \textit{chromatic number} of a graph denotes the lowest number of colors needed for coloring\footnote{In general, coloring assigns labels -- namely, colors -- to elements of a graph according to arbitrary partition constraints. In the following, we adopt the most widely-used partition constraint based on vertex adjacency, since other coloring problems can be transformed into vertex coloring instances.} the vertices of a graph, so that no adjacent vertices are colored with the same color. Graphs with chromatic number equal to $k$ are often defined as $k$\textit{-colorable}, and in the following we formally define two-colorable graphs, also referred to as \textit{bipartite graphs}.

\begin{defin}[\textbf{Two-colorable Graph} or \textbf{Bipartite Graph}]
    \label{def:08}
    A graph $G=(V,E)$ is two-colorable if the set of vertices $V$ can be partitioned\footnote{A partition of a set is a grouping of its elements into non-empty subsets, so that every element is included in exactly one subset. Formally, the sets $\{ P_i \}$ are a partition of $V$ if: i) $P_i \neq \emptyset \; \forall \, i$, ii) $\bigcup_{i} P_i = V$, iii) $P_i \cap P_j = \emptyset \; \forall \, i \neq j$.} into two subsets $\{ P_1, P_2 \}$ so that there exist no edge in $E$ between two vertices belonging to the same subset. Two-colorable graph $G=(V,E)$ can be also denotes as $G=(P_1, P_2, E)$.
\end{defin}

\begin{remark}
    We focus our attention on two-colorable graph states without loss of generality, and the rationale is that any graph state can be converted  under relaxed conditions in a two-colorable one \cite{HeiDurEis-06}. Indeed, any graph is two-colorable iif it does not contain cycles of odd length. Furthermore, two-colorable graphs model a wide range of different network topologies. Notable examples of two-colorable graphs are represented in Fig.~\ref{fig:04}, These include the path-graph, the even cycle graph and the star graph, which represent relevant network topologies -- i.e., bus, ring and star \cite{MazCalCac-24-QCNC} -- commonly investigated within the classical networking framework. Furthermore, more complex topologies such as tree and hypercube are two-colorable graphs as well.
\end{remark}

In the following, we will label the vertices in $P_1$ and $P_2$ as follows, for the sake of notation simplicity:
\begin{align}
    \label{eq:17}
    P_1 &= \{v^1_1,\cdots,v^1_{n_1}\} \\
    \label{eq:18}
    P_2 &= \{v^2_1,\cdots,v^2_{n_2}\}
 \end{align}
with $n_1 + n_2 = n$.

\begin{defin}[\textbf{Complete Bipartite Graph}]
    \label{def:09}
    Let $G = (P_1,P_2,E)$ be a bipartite graph with $|P_1| = n_1$ and $|P_2| = n_2$. If $E = P_1 \times P_2$, i.e., if
    \begin{equation}
        \label{eq:19}
        \forall \, v_i^1 \in P_1 \wedge v_j^2 \in P_2, \exists \, (v_i^1,v_j^2) \in E,
    \end{equation}
    $G$ is defined as \textit{complete bipartite graph} and denoted as $K_{n_1,n_2}$.
\end{defin}

Hence, in a complete bipartite graph, any vertex belonging to one part is connected to every vertex belonging to the complementary part by one edge, as shown in Fig.~\ref{fig:04-e}.

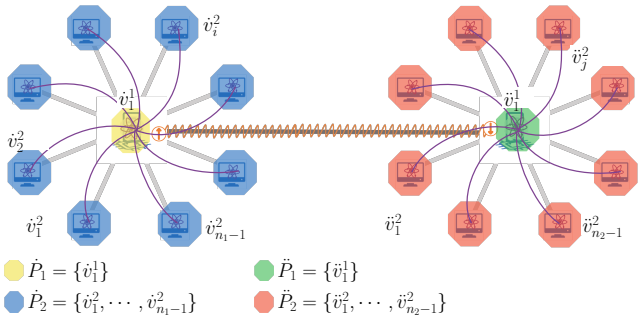
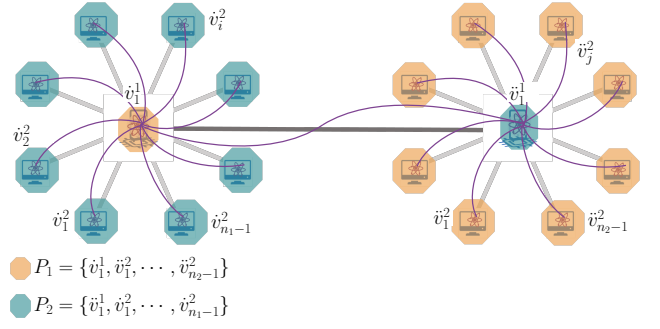
\begin{figure*}
    \begin{minipage}[t]{0.48\textwidth}
        \centering
        \resizebox{\textwidth}{!}{
            \input{Figures/Fig-02a}
        }
        \subcaption{Each super-node generates and distributes, as described in Fig.~\ref{fig:03}, a star graph state in each QLAN, denoted as $\ket{\dot{S}_{n_1}} = (\dot{P}_1,\dot{P}_2,\dot{E})$ and $\ket{\ddot{S}_{n_2}} = (\ddot{P}_1,\ddot{P}_2,\ddot{E})$, respectively. Furthermore, an EPR is shared between the two super-nodes.}
        \label{fig:05-a}
    \end{minipage}
    \hfill
    \begin{minipage}[t]{0.48\textwidth}
        \centering
        \resizebox{\textwidth}{!}{
            \input{Figures/Fig-02b}
        }
        \subcaption{By exploiting the EPR shared between the two QLANS for performing a remote \texttt{CZ}, a new graph state shared among all the nodes of the two QLANs is obtained. Remarkably, also this new graph state is a two-colorable graph state, and specifically it is a binary star graph $\ket{S_{n_1,n_2}}$.}
        \label{fig:05-b}
    \end{minipage}
    \caption{Interconnecting two QLANs through a binary graph state $\ket{S_{n_1,n_2}}$ distributed among all the nodes, obtained from two star graphs distributed in each QLAN and an EPR pair shared between the two super-nodes. A preeminent feature of the distributed state $\ket{S_{n_1,n_2}}$ is that the qubits associated with the vertices of one part -- say $P_1$ -- are distributed so that: i) one qubit is at super-node of rightmost QLAN, and ii) the remaining qubits are at clients of the leftmost QLAN.}
    \label{fig:05}
    \hrulefill
\end{figure*}

\begin{defin}[\textbf{Star Graph}]
    \label{def:10}
    Let $K_{n_1,n_2}$ be a complete bipartite graph. If $n_1$ (or equivalently $n_2$) is equal to $1$  -- namely, if one part is composed by a single vertex -- then the graph is called \textit{star graph} and denoted equivalently as either $K_{1,n-1}$ or $S_{n}$, where $n-1$ is the cardinality of the other part. From \eqref{eq:19}, it results:

    \begin{equation}
        \label{eq:20}
        E = {v^1_1} \times P_2
    \end{equation}
    and $v^1_1$ is referred to as the \textit{center} of the star graph.
\end{defin}

\begin{res}
    The \textit{star graph state} $\ket{S_n}$ associated to a star graph $S_n$ shown in Fig.~\ref{fig:04-f} represents the multipartite entangled state generated and distributed in each QLAN, with each qubit of state $\ket{S_n}$ distributed to a different node. Specifically, by following the labelling of \eqref{eq:17} and \eqref{eq:18}, qubit corresponding to vertex $v^1_1$ is stored at the super-node, whereas qubits corresponding to vertices $\{ v^2_1, \ldots, v^2_{n-1} \}$ are distributed to the clients.
\end{res}

Such a state corresponds to a graph that perfectly matches with the QLAN physical topology and it is easy to generate \cite{WalPirZwe-19,AviRozWeh-22,BugCouOma-23}. It is worthwhile to mention that it represents the worst-case scenario, since from a star graph it is possible to extract only one EPR pair, thus limiting the communication dynamics within the single QLAN. Despite this, in the next sections we will prove that by properly manipulating the multipartite states in the different QLANs, the limitations of the physical topologies can be overcome.

Stemming from the concept of star graph $S_{n}$, we are ready now to introduce a two-colorable graph that will be extensively used in the following, and referred to as \textit{binary star graph} $S_{n_1,n_2}$.

\begin{defin}[\textbf{Binary Star Graph}]
    \label{def:11}
    A binary star graph $S_{n_1,n_2}$ is a bipartite graph $G = (P_1,P_2,E)$ with the edge set $E$ defined as:
    \begin{align}
        \label{eq:21}
        E = \{ v^1_1 \} \times P_2 \cup P_1 \times \{ v^2_1 \}
    \end{align}
    with $P_1$ and $P_2$ given in \eqref{eq:17} and \eqref{eq:18}, respectively.
\end{defin}
From \eqref{eq:21}, it results that only one vertex in each part of a binary star graph -- namely, $v^1_1 \in P_1$ and $v^2_1 \in P_2$ -- is fully connected, as shown in Fig.~\ref{fig:04-g}.

In Sec~\ref{sec:4.2}, we will show that -- by locally manipulating at some specific network nodes a binary star state $\ket{S_{n_1,n_2}}$ shared between the two QLANs -- artificial links among remote nodes are dynamically generated. Yet, before operating on such a state, the state must be distributed among all the nodes. Hence, one preliminary question naturally arises: \textit{how expensive is it -- from a quantum communication perspective -- to distribute such a state within the two QLANs}? Or, in other words, how many EPR pairs must be consumed for distributing such a state? One might believe that the required number of EPR pairs should somehow depend on the number of artificial links that must be generated among remote nodes belonging to different QLANs.

We answer to this question with the following proposition.

\begin{prop}
    \label{prop:01}
    Let's assume that a star state $\ket{\dot{S}_{n_1}}$ has been distributed in the first QLAN and that another star state $\ket{\ddot{S}_{n_2}}$ has been distributed in the second QLAN. Then, a binary star state $\ket{S_{n_1,n_2}}$ can be distributed among all the nodes by consuming \textit{only one} EPR pair at the two super-nodes.
    \begin{IEEEproof}
        Please refer to App.~\ref{app:01}.
    \end{IEEEproof}
\end{prop}

Fig.~\ref{fig:05} depicts the building process of the binary star state, by also labeling each network node with the vertex-label corresponding to the associated qubit. We can now define our global entanglement resource.

\begin{res2}
The binary star state $\ket{S_{n_1,n_2}}$ represents the inter-QLAN multipartite entangled resource, which is locally manipulated at network nodes for dynamically enabling multiple artificial links among remote nodes.
\end{res2}
\begin{remark}
    In this paper, the binary star state $\ket{S_{n_1,n_2}}$ represents the entangled resource resulting after the generation and distribution over intra- and inter-QLAN quantum channels. Being entanglement a communication resource (and not information), it is possible to overcome the constraints imposed by the no-cloning theorem by adopting entanglement purification protocols, either at the generation stage or at the distribution stage. As an instance, multiple rounds of entanglement purification can be employed to counteract the impact of noise and obtain an inter-QLAN resource above some fidelity threshold \cite{DurBri-07}. Furthermore, even in the worst-case of absorbing quantum channels -- which corresponds to the case of irremediably failed distribution, thus no farther purification can be attempted -- it is possible to model the entanglement distribution on noisy quantum channels as Markov chain process, as proved in \cite{IllCalVis-23}. Accordingly, noisy absorbing quantum channels reduce the probability of successful distribution of the global entanglement resource, by reducing the number of clients successfully connected to the super-node. Nevertheless, the multipartite entangled state distribution can be engineered such that some requirements, such as for example the minimum number of connected clients or the maximum time to be devoted to distribution, are met according to the application needs, as detailed in \cite{CacIllVis-23}. Stemming from the above, in the following we refer the reader to the wide existing literature analyzing the noisy effects on entangled resources (such as \cite{BugCouOma-23,MorDur-23-1,MorDur-23,NegDirMun-24}), as we focus on the overlooked topic of entanglement-connectivity manipulation. Indeed, our aim is to pave the way for engineering the connectivity in large-scale quantum networks beyond the classical physical connectivity concept. More into details, we aim at shedding the light on the artificial neighborhood concept activated by entanglement to dynamically adapt to the traffic demands, regardless of the physical distance among the nodes.
\end{remark}
%--------------------------------------------------------------
\subsection{DYNAMIC ARTIFICIAL LINKS}
\label{sec:4.2}

Here, we show that multiple artificial links can be dynamically obtained among remote nodes belonging to different QLANs, by means of local operations only, by overcoming the limitations induced by the physical inter-QLAN physical connectivity. Specifically, the set of employed operations limits to single qubit gates, single qubit Pauli measurements and classical communications, which all represent \textit{free operations} from a quantum communication perspective.

To this aim, we consider four different archetypes summarizing different traffic patterns, namely:
\begin{itemize}
    \item[i)] \textit{peer-to-peer},
    \item[ii)] \textit{role delegation},
    \item[iii)] \textit{clients hand-over},
    \item[iv)] \textit{extranet,} 
\end{itemize}
and for each archetype we discuss different artificial topologies that equivalently satisfy the communication demand. 

\begin{figure*}
    \centering
    \begin{minipage}[t]{0.32\textwidth}
        \resizebox{\textwidth}{!}{
            \input{Figures/Fig-05-b}
        }
        \subcaption{\textit{Hierarchical peer-to-peer} artificial topology discussed in Prop.~\ref{prop:02}: an artificial fully-connected topology among all the clients of the same QLAN and one super-node of a different QLAN.}
        \label{fig:06-a}
    \end{minipage}
    \hfill
    \begin{minipage}[t]{0.32\textwidth}
        \resizebox{\textwidth}{!}{
            \input{Figures/Fig-05-c}
        }
        \subcaption{\textit{Role delegation type I} artificial topology discussed in Cor.~\ref{cor:01}: an artificial star topology among the same set of nodes of Fig.~\ref{fig:06-a}, but centered at a client of the same QLAN.}
        \label{fig:06-b}
    \end{minipage}
    \hfill
    \begin{minipage}[t]{0.32\textwidth}
        \resizebox{\textwidth}{!}{
            \input{Figures/Fig-05.a}
        }
        \subcaption{\textit{Clients hand-over} artificial topology discussed in Cor.~\ref{cor:02}: an artificial star topology among the same set of nodes of Fig.~\ref{fig:06-a}, but centered at a super-node of a different QLAN.}
        \label{fig:06-c}
    \end{minipage}
    \caption[Caption for LOF]{Different artificial inter-QLAN topologies matching with a traffic pattern involving the super-node of one QLAN and clients of the other QLAN. The three topologies represent three LU-equivalent graph states that can be obtained by local operations\footnotemark. Remarkably, all the artificial inter-QLAN topologies are obtained by manipulating a binary star graph state with local  operations and measurements only.}
    \label{fig:06}
    \hrulefill
\end{figure*}
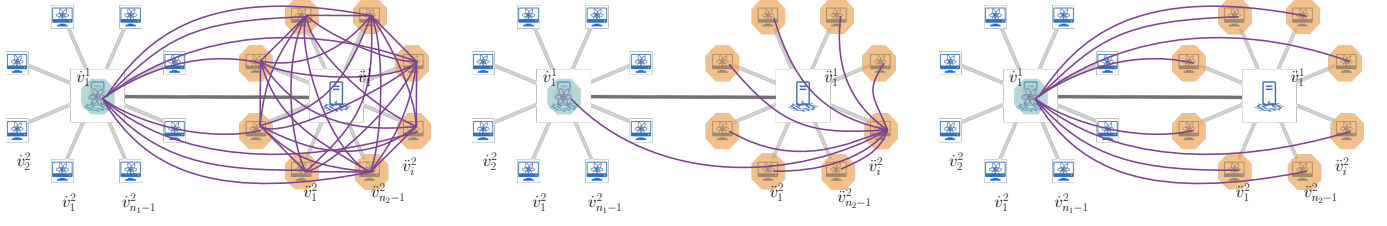
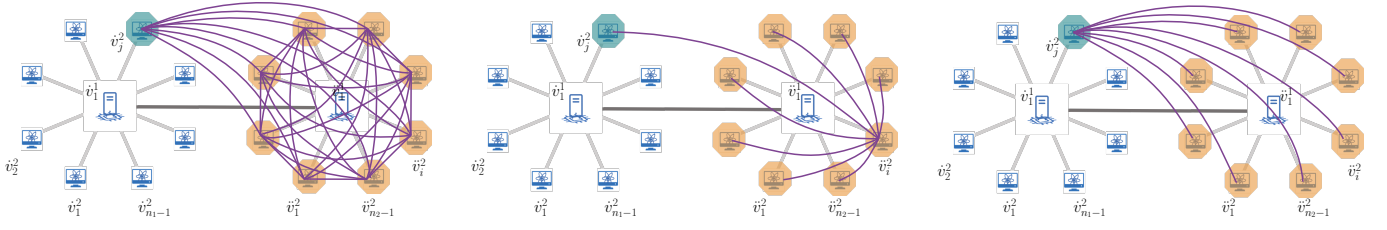
\begin{figure*}
    \centering
    \begin{minipage}[t]{0.32\textwidth}
        \resizebox{\textwidth}{!}{
            \input{Figures/Fig-07.a}
        }
        \subcaption{\textit{Pure peer-to-peer} artificial topology discussed in Prop~\ref{prop:03} : an artificial fully-connected topology among all the clients of the same QLAN and one client node of a different QLAN.}
        \label{fig:07-a}
    \end{minipage}
    \hfill
    \begin{minipage}[t]{0.32\textwidth}
        \resizebox{\textwidth}{!}{
            \input{Figures/Fig-07.b}
        }
        \subcaption{\textit{Role delegation type II} artificial topology, case $1$, discussed in Cor.~\ref{cor:03}: an artificial star topology among the same set of nodes of Fig.~\ref{fig:07-a}, but centered at a client of the rightmost QLAN.} 
        \label{fig:07-b}
    \end{minipage}
    \hfill
    \begin{minipage}[t]{0.32\textwidth}
        \resizebox{\textwidth}{!}{
            \input{Figures/Fig-07.c}
        }
        \subcaption{\textit{Role delegation type II} artificial topology, case $2$, discussed in Cor.~\ref{cor:04}: an artificial star topology among the same set of nodes of Fig.~\ref{fig:07-a}, but centered at a cleint of a different QLAN.}
        \label{fig:07-c}
    \end{minipage}
    \caption[Caption for LOF]{Different artificial inter-QLAN topologies matching wih a traffic pattern involving one client of 
    one QLAN and clients of the other QLAN. These three topologies correspond to three LU-equivalent graph states that can be obtained by a sequence of local operations\footref{fn:LC}. Remarkably, the action of ``sacrificing'' the artificial link between the two QLAN super-nodes enables inter-QLAN peer-to-peer artificial topologies particular useful for designing distributed network functionalities relying on client communication capabilities.}
    \label{fig:07}
    \hrulefill
\end{figure*}

\footnotetext{\label{fn:LC} 
Let us consider the graph $G$ associated with the \textit{hierarchical peer-to-peer} artificial topology represented in Fig.~\ref{fig:06-a}. Then, the \textit{role delegation type I} artificial topology represented in Fig.~\ref{fig:06-b}, and associated with the graph $G'$, can be obtained from the \textit{hierarchical peer-to-peer} by performing local complement on $G$ with respect to the leftmost QLAN client $\ddot{v}_i^2$, namely, $G'= \tau_{\ddot{v}_i^2}(G)$. Similarly, by performing local complement on $G$ with respect to the super-node $\dot{v_1^1}$, we obtain the graph $G''$ corresponding to the \textit{clients hand-over} artificial topology represented in Fig.~\ref{fig:06-c}, namely, $G''=\tau_{\dot{v_1^1}}(G)$. Clearly, one can obtain $G'$ from $G''$ and vice versa with local complement operations as well. Specifically, $G'$ can be converted back to $G$ by applying an additional local complement operation to the client node $\ddot{v}_i^2$. Then, $G$ can be transformed into $G''$. Hence, we achieve the mutual transitions between $G'$ and $G''$. The same considerations hold for the artificial topologies represented in Fig.~\ref{fig:07}.}

\begin{prop}[\textbf{Hierarchical Peer-to-Peer}]
    \label{prop:02}
    Starting from a binary star graph state $\ket{S_{n_1,n_2}}$ shared between $n=n_1+n_2$ nodes belonging to two different QLANs, a $n_i$-complete graph state $\ket{K_{n_i}}$ (with $i=1,2$) shared between the super-node of a QLAN and all the $(n_i-1)$-clients of the other QLAN can be obtained.
    \begin{IEEEproof}
        Please refer to App.~\ref{app:02} for more details.
    \end{IEEEproof}
\end{prop} 

The effect of Prop.~\ref{prop:02}, depicted in Fig.~\ref{fig:06-a}, is particularly relevant from a communication perspective. Specifically, the obtained complete (i.e., fully connected) graph is LU-equivalent to a GHZ state, which allows the deterministic extraction of one EPR pair between \textit{any couple} of nodes sharing it. As a consequence, the complete graph obtained exhibits a remarkable flexibility on the choice of the identities of the nodes exploiting the ultimate artificial link, aka the extracted EPR. Notably, the completely connected graph includes the other QLAN super-node. Hence, from a topological perspective, the involved clients and super-node act as peer-to-peer entities, which can exploit the shared entangled state either to accommodate intra-QLAN traffic requests or inter-QLAN traffic requests. This consideration induced us to label this proposition as ``hierarchical peer-to-peer'' artificial topology, by including the differentiation -- aka hierarchy -- in terms of hardware requirements between clients and super-node.

It is worthwhile to mention that the ``Hierarchical Peer-to-Peer'' artificial topology is particularly advantageous whenever no information is available on the actual network traffic features of the QLAN clients. Specifically, if a client may equally need to communicate with clients belonging to the same QLANs or with clients belonging to a different QLAN, then the communication request will be ready to be served by proactively manipulating the ``hierarchical peer-to-peer'' artificial topology. And, remarkably, the communication request is easily served by simply performing local unitaries and measurements, without any additional quantum communication. Notably, ``Hierarchical Peer-to-Peer'' implements a pure mesh topology between different network nodes, such as the super-node of one QLAN and the clients nodes of a different QLAN. This is a particularly useful topology for applications such as multiparty communications and quantum conference key agreement \cite{HahJonPap-20}.

Furthermore, from Prop.~\ref{prop:01} two further results follow.

\begin{cor*}[\textbf{Role Delegation Type I}]
    \label{cor:01}
    Starting from a binary star graph state $\ket{S_{n_1,n_2}}$ shared between $n=n_1+n_2$ nodes belonging to two different QLANs, a $n_i$-star graph state $\ket{S_{n_i}}$ (with $i=1,2$) centered at one QLAN client node and connecting all the remaining $(n_i-2)$-clients of the same QLAN and the super-node of the other QLAN can be obtained.
    \begin{IEEEproof}
        Please refer to App.~\ref{app:03} for more details.
    \end{IEEEproof}
\end{cor*}

The result of Cor.~\ref{cor:01} is represented in Fig.~\ref{fig:06-b} with the client node $\ddot{v}^2_i$ being the center of the star graph. By looking at both Fig.~\ref{fig:06-a} and Fig.~\ref{fig:06-b} it is evident that -- differently from the ``hierarchical peer-to-peer'' artificial topology exhibiting $\frac{n_i(n_i-1)}{2}$ artificial links -- the ``role delegation'' artificial topology provides less freedom for selecting the identities of the nodes that can proactively extract the ultimate artificial link.  However, this is not less advantageous as the star graph comes in handy whenever the traffic pattern likely involves a specific client node $\ddot{v}^2_i$, which may equally need to communicate with clients belonging to its QLANs or need a link outside its QLAN. Due to the particular structure of this artificial topology -- having one client as center of the star graph instead of the super-node -- we were induced to label this topology as ``role delegation'' topology of type I. Indeed, a different type of role delegation topology, named type II, is introduced in Cors.~\ref{cor:03} and \ref{cor:04}.

Another remarkable corollary of Prop.~\ref{prop:02} is the possibility of building the so-called ``\textit{clients hand-over}'' artificial topology between two QLANs, as detailed in the following corollary.

\begin{cor*}[\textbf{Clients Hand-Over}]
    \label{cor:02}
     Starting from a binary star graph state $\ket{S_{n_1,n_2}}$ shared between $n=n_1+n_2$ nodes belonging to two different QLANs, a $n_i$-star graph state $\ket{S_{n_i}}$, (with $i=1,2$) centered at one QLAN super-node and connecting all the $(n_i-1)$-clients of the other QLAN can be obtained.
    \begin{IEEEproof}
    The proof follows by adopting similar reasoning as in the proof of Cor.~\ref{cor:01}, and by setting as arbitrary neighbouring node $k_0$ the super-node $\dot{v}^1_1$.
    \end{IEEEproof}
\end{cor*}

We note that, accordingly to Cor.~\ref{cor:02}, artificial links are built between a super-node of one QLAN and the clients of a different QLAN, as shown in Fig~\ref{fig:06-c}. This, from a topological perspective, is equivalent to virtually \textit{move} the clients of a QLAN into a different QLAN, resembling thus a sort of \textit{clients hand-over} from one QLAN to the other.
Furthermore, it is worthwhile to anticipate that the star-like artificial topologies depicted in Fig.~\ref{fig:06-b} and Fig.~\ref{fig:06-b} are widely used in the satellite-to-ground networks \cite{CheZhaChe-21}, where the satellite nodes play the role of the super-nodes interconnecting multiple ground stations, namely, clients.

\begin{remark}
    As proved in the Appendices, the artificial topologies represented in Fig.~\ref{fig:06} are obtained by leveraging suitable sequences of Pauli measurements, which, thus, represent a tool for engineering the artificial connectivity. To elaborate more, the graph represented in Fig.~\ref{fig:06-a} leverages a sequence of Pauli-$z$ and Pauli-$y$ measurements. Remarkably, by replacing the Pauli-$y$ measurement at the super-node with a Pauli-$x$ measurement, we obtain the LU-equivalent graph state corresponding to a star graph among all the nodes, as shown in Fig.~\ref{fig:06-b}.
    A Pauli-$x$ measurement is equivalent to the sequence of graph operations given in \eqref{eq:11} that involves the arbitrary neighbor $k_0$. And indeed $k_0$ represents an engineering parameter. In fact, by choosing a client as $k_0$ -- as instance, say $k_0 = \ddot{v}^2_i$ -- we obtain a graph state corresponding to a star graph centered at $\ddot{v}^2_i$ as shown in Fig.~\ref{fig:06-b}. Differently, if we set $k_0 = \dot{v}^1_1$, the star graph is centered at $\dot{v}^1_1$, as shown in Fig.~\ref{fig:06-c}.
\end{remark}

From the above, it may be concluded that only artificial topologies involving super-node and clients can be built upon the binary star graph state $\ket{S_{n_1,n_2}}$. The following Prop.~\ref{prop:03} and the subsequent Cors.~\ref{cor:03} and \ref{cor:04} prove, instead, that artificial topologies involving only clients belonging to different QLANs can be built by properly manipulating the binary star graph state.

\begin{prop}[\textbf{Pure Peer-to-Peer}]
    \label{prop:03}
    Starting from a binary star graph state $\ket{S_{n_1,n_2}}$ shared between $n=n_1+n_2$ nodes belonging to two different QLANs, a $n_i$-complete connected graph state $\ket{K_{n_i}}$ , with $i=1,2$, shared between one client node of a QLAN and all the $(n_i-1)$-clients of the other QLAN can be obtained.
    \begin{IEEEproof}
        Please refer to App.~\ref{app:05} for more details.
    \end{IEEEproof}
\end{prop}

As shown in Fig.~\ref{fig:07-a}, Prop.~\ref{prop:03} allows an arbitrary client belonging to a QLAN to share an artificial link with any client belonging to a different QLAN. Hence, it generates an artificial QLAN topology among \textit{peer} client entities -- thus, the naming ``pure'' -- by neighboring remote nodes, despite the original constraints imposed by the physical topologies.
 
Indeed, a pure peer-to-peer artificial topology extends the flexibility on the choice of the identities of the nodes exploiting the ultimate artificial link, by involving only clients at different QLANs.  This could be particularly advantageous for designing distributed network functionalities relying on clients communication capabilities. Indeed, if a client needs to communicate with a client belonging to a different QLAN, then -- by
proactively manipulating the artificial topology -- the communication request is ready to be served, without further orchestration at the super-node. Indeed, the communication request is fulfilled by performing local unitaries without any additional quantum communications. And, actually, the client of the other QLAN can be selected by properly manipulating the initial binary star graph. Hence their identities can be engineered on-demand. 

From a communication perspective, as the hierarchical peer-to-peer, the pure peer-to-peer forms a fully-connected artificial topology, where each node shares an artificial link with all others. Hence, this property is particularly beneficial in multiparty scenarios which require consensus and coordination such as quantum secret sharing \cite{HilBuVla-99, GaeKurBou-07} and quantum voting systems \cite{MisThaPar-21, HilZimBie-06}. 

\begin{cor*}[\textbf{Role Delegation Type II - Case 1}]
    \label{cor:03}
    Starting from a binary star graph state $\ket{S_{n_1,n_2}}$ shared between $n=n_1+n_2$ nodes belonging to two different QLANs, a $n_i$-star graph state $\ket{S_{n_i}}$ (with $i=1,2$) centered at one QLAN client node and connecting all the remaining $(n_i-2)$-clients of the same QLAN and a client node of the other QLAN can be obtained.
    \begin{IEEEproof}
    Please refer to App.~\ref{app:06} for more details.
    \end{IEEEproof}
\end{cor*}

Similarly to Corollary~\ref{cor:01}, due to the particular structure of this artificial topology shown in Fig.~\ref{fig:07-b} -- which has one client as center of the star graph instead of the super-node -- we are induced to label this topology as ``role delegation topology type II'', since (differently from type I) no super-node is connected within the star graph. A different realization of this topology -- labeled as ``case 2'' for distinguish from the previous one that is labeled as ``case 1'' -- is derived with the following corollary.

\begin{cor*}[\textbf{Role Delegation Type II - Case 2}]
    \label{cor:04}
     Starting from a binary star graph state $\ket{S_{n_1,n_2}}$ shared between $n=n_1+n_2$ nodes belonging to two different QLANs, a $n_i$-star graph state $\ket{S_{n_i}}$ (with $i=1,2$) centered at one QLAN client node and connecting all the $(n_i-1)$ clients of the other QLAN can be obtained.
    \begin{IEEEproof}
    The proof follows similarly to Cor.~\ref{cor:03}, by setting $k_0 = \dot{v}^2_j$.
    \end{IEEEproof}
\end{cor*}

The result of this corollary establishes that artificial links are built between a client of one QLAN and all the clients of the other QLAN, as shown in Fig.~\ref{fig:07-c}. This topology  comes in handy whenever a client node $\dot{v}_j^2$ needs to communicate with clients belonging to a different QLAN.

\begin{remark}
The theoretical results established via Props.~\ref{prop:02} and \ref{prop:03} and their corollaries make evident that it is possible to overcome the constraints imposed by the physically topologies by locally and properly manipulating engineered multipartite entangled states, without any further use of quantum links. This is equivalent to neighbor remote nodes, where the concept of neighboring is not meant in terms of physical proximity via physical links, but it rather should be intended -- as discussed in \cite{CacIllCal-23} -- as ``\textit{entangled proximity}''. 
\end{remark}

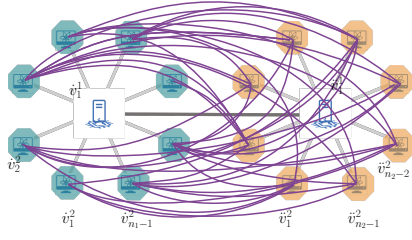
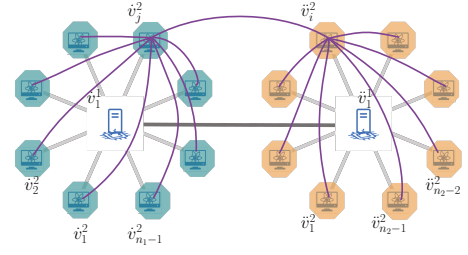
\begin{figure*}
    \centering
    \begin{minipage}[t]{0.48\textwidth}
        \centering
        \resizebox{!}{100pt}{
            \input{Figures/Fig-10}
        }
        \subcaption{\textit{Extranet} artificial topology discussed in Prop.~\ref{prop:04}: an artificial complete bipartite topology interconnecting each client of one QLANs with each client of the other QLAN.}
        \label{fig:08-a}
    \end{minipage}
    %\hspace{0.16\textwidth}
    \hfill
    \begin{minipage}[t]{0.48\textwidth}
    \centering
        \resizebox{!}{100pt}{
            \input{Figures/Fig-10.b}
        }
        \subcaption{\textit{Double Role Delegation} artificial topology of Cor.~\ref{cor:05}: an artificial binary star topology among the same set of clients of Fig.~\ref{fig:08-a}, centered at two client nodes belonging to two different QLANs.}
        \label{fig:08-b}
    \end{minipage}
    \caption[Caption for LOF]{Different artificial inter-QLAN topologies matching with a traffic pattern potentially involving all the clients of the two QLAN. The two topologies represent two LU-equivalent graph states that can be obtained by leveraging different local operations\footnotemark. Although LU-equivalent, the two artificial topologies are able to accommodate different client traffic patterns.}
    \label{fig:08}
    \hrulefill
\end{figure*}
\footnotetext{Let us consider the graph $G$ corresponding to the \textit{extranet} artificial topology represented in Fig.~\ref{fig:08-a}. The graph $G$ can be successfully transformed into $G'$ corresponding to the \textit{double role delegation} artificial topology represented in Fig.~\ref{fig:08-b} by performing the sequence of local complementation operations $\tau_{\dot{v}_j^2}(\tau_{\ddot{v}_i^2}(\tau_{\dot{v}_j^2}(G)))$. Clearly, $G'$ can be converted back to $G$ by performing local operations on some specific nodes of the double role delegation artificial topology.}

The degrees of freedom offered by the binary star graph to select the identities of the nodes sharing the ultimate artificial link are further highlighted by the following proposition and its subsequent corollary. 

\begin{prop}[\textbf{Extranet}]
    \label{prop:04}
    Starting from a binary star graph state $\ket{S_{n_1,n_2}}$ shared between $n=n_1+n_2$ nodes belonging to two different QLANs, a $(n_1+n_2-2)$-complete bipartite graph state $\ket{K_{n_1-1,n_2-1}}$, shared between all the $(n_1-1)$-clients of one QLAN and all the $(n_2-1)$-clients of the other QLAN can be obtained. 
    \begin{IEEEproof}
        Please refer to App.~\ref{app:08} for more details.
    \end{IEEEproof}
\end{prop}

\begin{remark}
    The effects of the result in Prop.~\ref{prop:04}, depicted in Fig.~\ref{fig:08-a}, are particularly relevant form a communication perspective. Notably, allowing connections only between nodes in different sets enables the “extranet” to effectively reflect cross-set interactions.  Specifically, artificial links are created among clients belonging to different QLANs. Thus, inter-QLANs communication needs can be promptly fulfilled, by selecting on-demand -- i.e. accordingly to the traffic patterns -- the identities of the clients sharing the ultimate artificial link. With respect to the topology in Fig.~\ref{fig:07-c}, it is evident that the degrees of freedom in selecting these identities are higher. This comes without paying the price of additional quantum communications, but only by engineering the proper local operations to be performed -- in such a case -- at the super-nodes. This, from a topological perspective, can be seen as obtaining an ``extranet'' interconnecting all the clients of the two original ``intranets'', thus the name.  Finally, as a practical example, we note that in quantum software architecture blueprints for the cloud \cite{CorMarGio-23}, dynamic supply-demand relationship between sellers and buyers reflects an interaction pattern similar to an extranet topology. 
\end{remark}

\begin{cor*}[\textbf{Double Role Delegation}]
    \label{cor:05}
     Starting from a binary star graph state $\ket{S_{n_1,n_2}}$ shared between $n=n_1+n_2$ nodes belonging to two different QLANs, a $(n_1+n_2-2)$-binary star graph state $\ket{S_{n_1-1,n_2-1}}$ shared between the $(n_1-1)$-clients of one QLAN and the $(n_2-1)$-clients of the other QLAN, centered at one client from each QLAN, can be obtained.
    \begin{IEEEproof}
        Please refer to App.~\ref{app:09} for more details.
    \end{IEEEproof}
\end{cor*}

The result of Cor.~\ref{cor:05}, depicted in Fig.~\ref{fig:08-b}, extends the dynamics offered by the artificial topology in Fig.~\ref{fig:07-b}. Indeed, differently from the ``extranet'' topology, with the ``\textit{double role delegation}'' artificial topology both intra-QLAN and inter-QLAN traffic demands can be accommodated. The price is having less degree of freedom in selecting the identities of the clients sharing the ultimate artificial link. 
The same considerations made for role delegation type I and clients hand-over topologies hold for the remaining role delegation (type II and double role delegation) topologies, as they all exhibit a star-like artificial topology. Remarkably, this topology is widely used in the satellite-to-ground quantum network\cite{CheZhaChe-21} and is envisioned to play a key role in wide-area quantum networks.

\begin{remark}
    As highlighted in  Sec.~\ref{sec:4.1}, our contribution can be applied also in presence of noisy distribution process, by adopting flexible noise network tools and frameworks as the ones in \cite{IllCalVis-23,CacIllVis-23}. Indeed, considerations on noisy local operations can be easily included too in our framework. Specifically, the action of local Pauli noise is equivalent to undesired Pauli-$z$ and Pauli-$y$ measurements \cite{MorDur-23,MorDur-23-1}. Clearly, such noisy operations reduce the number of network nodes joining the artificial topology and hence impact the inter-QLAN connectivity. Thus, similar considerations developed in Sec.~\ref{sec:4.1} continue to hold. 
\end{remark}

%--------------------------------------------------------------
% Sec V
%--------------------------------------------------------------
\section{CONCLUSIONS AND FURTHER WORKS}
\label{sec:5}

The goal of this paper is to provide an operational and easy-to-use guide for understanding and manipulating the artificial topology enabled by multipartite entangled states, with the aim to facilitate the design and engineering of dynamic communication protocols. As a consequence, we departed from the traditional combinatorial study of graph states, to focus on the theoretical analysis of different artificial inter-QLAN topologies. Such topologies are enabled by the proper manipulation of two-colorable graph states via local operations, and they are able to accommodate different inter-QLAN traffic patterns.

More in details, the design of two-colorable graph states used as inter-QLAN entangled resource follows the simplest scenario of classical LAN interconnection, such as two switched LANs interconnected by a single physical channel. Remarkably, the designed state is the simplest multipartite entangled state allowing inter-QLAN connectivity, since it requires for its preparation only a single EPR to be distributed between the different QLANs. Furthermore, even with this minimal requirement, it enables entanglement-based connectivity between the client nodes of the two considered QLANs, which are not connected in the physical topology. Indeed, this resource enables a pure mesh topology between different QLANs clients, such as the peer-to-peer topologies.

As mentioned in Sec.~\ref{sec:1}, our proposal represents a preliminary step toward the design of traffic-dependent communication scenarios. Besides the use-cases analyzed in Sec.~\ref{sec:4}, further applications of the proposed model can be envisioned in scenarios involving a variable entanglement demand, where the variability may concern the identities of the involved nodes as well as the artificial links interconnecting them. One possible use-case in this direction is the interconnection of multiple quantum sensors \cite{DegReiCap-17,LenStrMue-16}, where the set of involved sensors can be dynamically determined at runtime and according to local feedback. For instance, suppose the super-nodes retrieve information about geographical information on target quantity to be measured. In that case, by properly manipulating the multipartite entangled state, the identities of sensors actively involved in the protocol can be adjusted accordingly. Similarly, our model can support entanglement-based quantum key agreement, where the set of participants in the agreement can be modified on-demand \cite{MurGraKam-20}.

We aim at extending the proposed framework by investigating several future research directions, as:
\begin{itemize}
    \item[-] impact of decoherence on the artificial topology;
    \item[-] analysis of the scalability properties exhibited by our proposal, as the number of QLANs and clients scale;
    \item[-] simulation of the proposal with traffic patterns tailored for specific applications, through selected quantum network simulators.
\end{itemize}

More in details, noise can affect multiple stages of the proposal. 
An analysis of noisy entanglement distribution can be included in the presented framework by leveraging Markov chains as done in \cite{IllCalVis-23,CacIllVis-23}. Furthermore, noisy controlled-operations impact the average waiting time for fulfilling an inter-network communication request. In this context, we envision that the waiting time exhibited by an EPR-based approach is somehow fixed, being related to the time required to perform entanglement swapping. Differently, we envision that the multipartite-based approach could exhibit null waiting times for inter-network request arrival rates lower than the time required to prepare the inter-QLAN entangled resource. However, this requires a quantitative and formal analysis, which we aim at developing in a future work.

Another important future investigation includes the analysis of the impact of Pauli noise on the proposed framework, by leveraging the equivalence between this type of noise and undesired Pauli-$z$ and Pauli-$y$ measurements on graph states, as shown in \cite{MorDur-23,MorDur-23-1,MorWalDur-25}. We expect that such noisy measurements reduce the number of network nodes connected via the artificial topology. Nevertheless, we observe that undesired measurements would also destroy artificial links in EPR-based approaches. 

Regarding, instead, the scalability properties exhibited by our proposal as the number of QLANs and clients scale, we envision that  our approach can maintain topological consistency with the archetypes analyzed for the artificial topologies in Sec.~\ref{sec:4}, by changing the designed entangled resource, i.e., by moving from a bi-star state to a $n$-star state. However, this has to be confirmed by a formal analysis. Indeed, further investigation includes the analysis of multiple-QLANs interconnection with minimal requirements on the inter-QLAN physical channels. Additionally, it would be beneficial to design inter-QLAN topologies that enable disjoint node subsets to be interconnected separately, concurrently, and on-demand.

Finally, another promising direction is simulating the proposed framework in realistic use-cases, using network simulators.

We underline that our contribution should be intended as hands-on guideline for the network engineering community to recognize the marvels -- with no-classical counterpart -- enabled by entanglement for networking and inter-networking tasks.

%--------------------------------------------------------------
% Appendices
%--------------------------------------------------------------
\begin{appendices}

%--------------------------------------------------------------
\section{A GUIDED EXAMPLE OF PROPOSED FRAMEWORK}
\label{app:0}

\begin{figure*}[!t]
    \centering    \includegraphics[width=\textwidth]{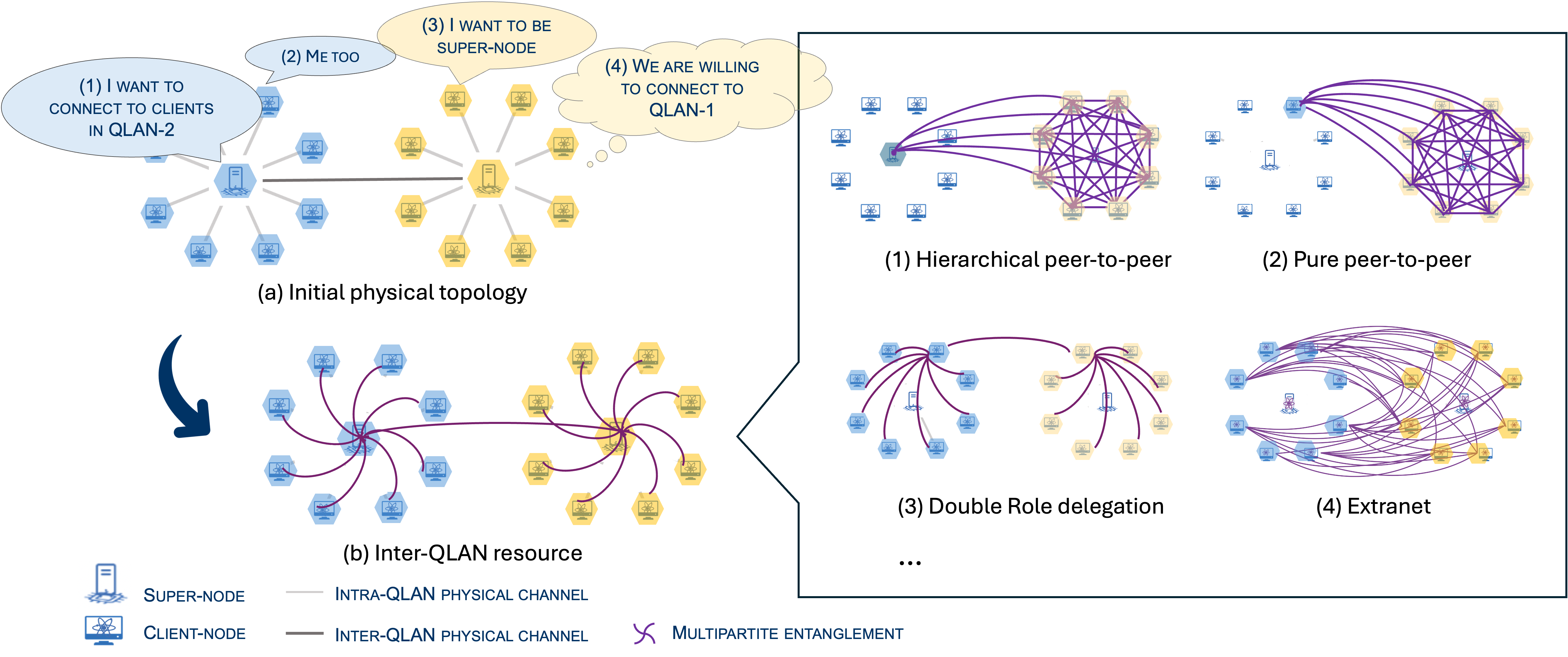}
    \caption{Summary of the paper contribution and methodology. \textit{(a) Initial physical topology.} The initial scenario consists of two QLANs: QLAN-1 (blue) and QLAN-2 (yellow). Each QLAN is equipped with a super-node that serves as a central hub and is connected to a set of local client nodes through quantum physical channels (\textbf{\textit{intra}}-QLAN physical channels). The two QLAN are connected through one point-to-point quantum physical channel (\textbf{\textit{inter}}-QLAN physical channel) which has the two QLAN super-nodes as endpoints. \textit{(b) Inter-QLAN resource}. We generate and distribute a particular graph state, referred to as \textit{inter-QLAN resource} which is associated with an artificial topology that matches the physical topology. The cost of the inter-QLAN resource corresponds to the cost of distributing one EPR over each physical channel and perform controlled-operations at the super-nodes. Stemming from the inter-QLAN resource, the topologies represented in subfigures (1),(2),(3),(4) can be extracted ``for free'', namely, by manipulations of the inter-QLAN resource though local operations and Pauli measurements. 
    Notably, we present four of the most representative artificial topologies that correspond to different traffic patterns: \textit{hierarchical peer-to-peer}, \textit{pure peer-to-peer}, \textit{double role delegation}, and \textit{extranet}. Remarkably, all these artificial topologies are derived from the inter-QLAN resource using only local operations and measurements. }
    \label{fig:09}
    \hrulefill
\end{figure*}

In the following paragraph we provide a guided example for our framework, summarizing the main steps of our contribution. Specifically, we refer to a pictorial illustration in Fig.~\ref{fig:09} as a summary of the contribution. We start from the physical topology shown in the upper left corner of Fig.~\ref{fig:09}. This represents the considered quantum network architecture which comprises two QLANs: QLAN-1 (blue) and QLAN-2 (yellow). Each QLAN is equipped with a super-node that serves as a central hub and is connected to a set of local client nodes through quantum physical channels (\textbf{\textit{intra}}-QLAN physical channels). The two QLAN are connected through one point-to-point quantum physical channel (\textbf{\textit{inter}}-QLAN physical channel) which has the two QLAN super-nodes as endpoints. Stemming from this physical topology, we establish an artificial topology that matches the physical topology. This is achieved by distributing the multipartite entangled state, referred to as \textit{inter-QLAN resource} and represented in Fig.~\ref{fig:09}.b. Such a distribution process requires, as a communication cost, the distribution of one EPR pair for each physical channel and controlled-operations to be performed at the super-node.\\
Our aim is to be able to obtain artificial links (purple lines in Fig.~\ref{fig:09}.b) between nodes that are not connected in the physical topology, without additional quantum communication overhead. Furthermore, we aim at achieving different artificial topologies for being able to accommodate different traffic patterns.

Notably, we show that, once the inter-QLAN resource is distributed, different artificial topologies -- that significantly differ from the physical one -- can be achieved without further cost from a quantum communication perspective, as no additional entanglement distribution or controlled-operations are required. More in details, the inter-QLAN resource is manipulated and transformed by the means of only local operations, that can be considered as free operations from a quantum communication perspective. 
Here, we provide four examples of the most representative inter-QLAN traffic patterns for illustration:

\begin{itemize} 
    \item [1)] \textit{The super-node in QLAN-1 requests to connect with the clients belonging to QLAN-2};
    \item [2)] \textit{One client node -- instead of the QLAN-1 super-node-- requests to connect with the clients belonging to QLAN-2};
    \item [3)] \textit{One client node in QLAN-2 requests to be delegated as a super-node};
    \item [4)] \textit{The entire set of clients belonging to QLAN-2 collectively request to connect with the entire set of clients belonging to QLAN-1}.
\end{itemize}

By engineering multipartite entangled states distributed across the QLANs, we first construct a dynamic inter-QLAN artificial topology, as shown in lower left corner of Fig.~\ref{fig:09}. Here "dynamic" implies that such an artificial topology can be manipulated on-demand so that the identities of the nodes interconnected within the artificial topology change at run-time. In this work, we flexibly adjust the artificial inter-QLAN topologies to different traffic patterns. For example, in response to the traffic patterns mentioned above, we can exploit the following artificial topologies:

\begin{itemize} 
    \item [1)] \textit{Hierarchical peer-to-peer} artificial topology: an artificial fully-connected topology among all the clients of QLAN-2 and the super-node of QLAN-1. This configuration appears particularly advantageous whenever no information is available on the actual network traffic features of the QLAN-2 clients;
    \item [2)] \textit{Pure peer-to-peer} artificial topology: an artificial fully-connected topology among all the clients of QLAN-2 and the client node of QLAN-1. This could be particularly advantageous for designing distributed network functionalities relying on clients communication capabilities;
    \item [3)] \textit{Double Role Delegation} artificial topology: an artificial binary star topology among each client of QLAN-1 and QLAN-2, centered at those two client nodes belonging to two different QLANs. This results advantageous to release the super-nodes resources while being able to preserve the artificial topology structure;
    \item [4)] \textit{Extranet} artificial topology: an artificial complete bipartite topology interconnecting each client of QLAN-1 with each client of QLAN-2. This configuration comes handy whenever no detailed information on the clients inter-QLAN traffic features is available.
\end{itemize}

%--------------------------------------------------------------
\section{PROOF OF PROPOSITION~\ref{prop:01}}
\label{app:01}

\begin{figure*}
    \centering
    \resizebox{\textwidth}{!}{
        \input{Figures/FigAv1}
    }
    \caption{Generation of a \textit{binary star state} starting from two \textit{star states} distributed in each QLAN, with physical topology (quantum links) omitted for the sake of simplicity. (a) Initial scenario where three entangled resources are shared: i) star state $\ket{\dot{S}_{n_1}}$ among QLAN1 nodes, ii) star state $\ket{\ddot{S}_{n_2}}$ among QLAN2 nodes, and iii) an EPR pair between the two super-nodes $\dot{v}^1_1$ and $\ddot{v}^1_1)$. (b) By consuming the EPR shared between the two super-nodes, a remote \texttt{CZ} operation is performed. This results in the generation of the additional edge $(\dot{v}^1_1,\ddot{v}^1_1)$, represented by the purple wavy line. (c) By re-coloring the overall graph, it becomes evident that it is a two-colorable graph corresponding to the binary star graph state $\ket{S_{n_1,n_2}}$.}
    \hrulefill
    \label{fig:11}
\end{figure*}
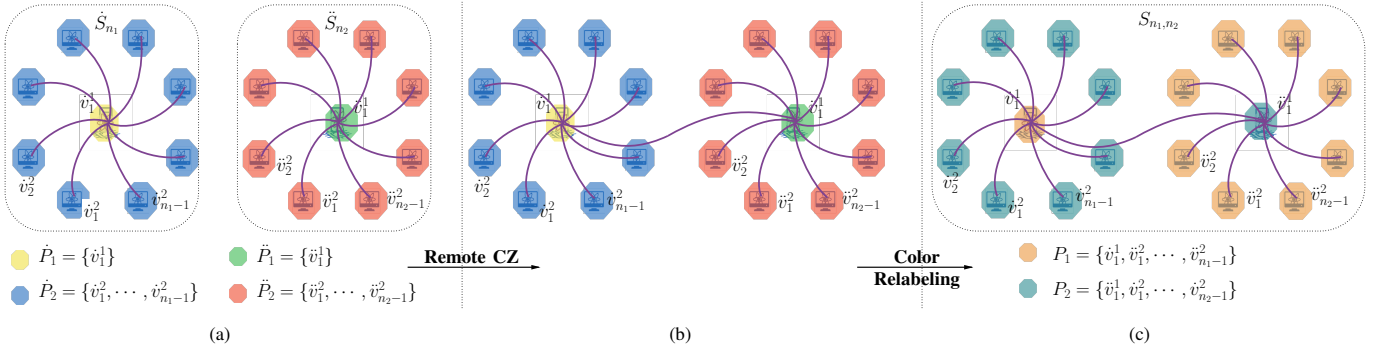

Let us label the vertices of the graphs $\dot{S}_{n_1}= (\dot{P}_1,\dot{P}_2,\dot{E})$ and $\ddot{S}_{n_2} = (\ddot{P}_1,\ddot{P}_2,\ddot{E})$, associated to the graph states $\ket{\dot{S}_{n_1}}$ and $\ket{\ddot{S}_{n_2}}$ distributed in the first and second QLANs respectively, according to \eqref{eq:17} and \eqref{eq:18}:
\begin{align}
    \label{app:1.1}
    \dot{P}_1 &= \{\dot{v}^1_1 \} \wedge  \dot{P}_2 = \{\dot{v}^2_1,\cdots,\dot{v}^2_{n_1-1}\},\\
    \label{app:1.1.bis}
    \ddot{P}_1 &= \{\ddot{v}^1_1 \} \wedge  \ddot{P}_2 = \{\ddot{v}^2_1,\cdots,\ddot{v}^2_{n_2-1}\},
\end{align}
with $ \dot{E} = \dot{P}_1 \times \dot{P}_2 $ and $ \ddot{E} = \ddot{P}_1 \times \ddot{P}_2$. Accordingly and as shown in Fig.~\ref{fig:11}-a, the qubits of graph state $\ket{\dot{S}_{n_1}}$ are distributed among the nodes of the first QLAN, with the super-node storing the qubit associated to vertex $\dot{v}^1_1$ and the clients storing the qubits associated to $\dot{v}^2_1, \ldots, \dot{v}^2_{n_1-1}$. Similarly, the qubits of the graph state $\ket{\ddot{S}_{n_2}}$ are distributed among the nodes of the second QLAN, with the super-node storing the qubit associated to the vertex $\ddot{v}^1_1$ and the clients storing the qubits associated to $\ddot{v}^2_1, \ldots, \ddot{v}^2_{n_1-1}$.
By consuming an EPR pair, the two super-nodes can perform a \texttt{CZ} operation between the two qubits at their sides, which corresponds to adding edge $(\dot{v}^1_1,\ddot{v}^1_1)$ between the associated vertices in the corresponding graph, as shown in Fig.~\ref{fig:11}-b. Hence, this additional edge connects the (only) vertex in $\dot{P}_1$ with the (only) vertex in $\ddot{P_2}$, and it results:
\begin{align}
    \label{app:1.2}
    E &= \dot{E} \cup \ddot E \cup \{ (\dot{v}^1_1,\ddot{v}^1_1) \}. 
\end{align}
        
If we want to color the overall graph, then these two vertex sets must be colored with two different colors, say orange and cyan. However, no edges connect the vertex in $\dot{P}_1$ with vertices in $\ddot{P_2}$, and hence all these vertices can be colored with the same color, orange. Similarly, no edges connect vertex in $\ddot{P}_1$ with the vertices in $\dot{P}_2$, hence all these vertices can be colored with the same color.        
It follows that the overall graph $G$ is a two-colorable graph with parts $P_1$ and $P_2$ given by:
\begin{align}
    \label{app:1.3}
    {P}_1 &= \dot{P}_1 \cup \ddot{P}_2 = \{ \dot{v}^1_1, \ddot{v}^2_1,\cdots,\ddot{v}^2_{n_2-1}\} \\
    \label{app:1.4}
    {P}_2 &= \ddot{P}_1 \cup \dot{P}_2 = \{ \ddot{v}^1_1, \dot{v}^2_1,\cdots,\dot{v}^2_{n_1-1} \}
\end{align}
and with edge-set $E$ in \eqref{app:1.2}. The proof follows by acknowledging that \eqref{app:1.2} coincides with \eqref{eq:21}.

%--------------------------------------------------------------
\section{PROOF OF PROPOSITION~\ref{prop:02}}
\label{app:02}

Let us adopt Fig.~\ref{fig:05} labeling and let us suppose that the state $\ket{K_{n_i}}$ to be obtained must interconnect the clients of the right-most QLAN, which implies $n_i=n_2$\footnote{Clearly, the same final state shared among a set of $n_i'<n_i$ nodes can be obtained straightforwardly, by simply removing the undesired nodes with Pauli-$z$ measurements at their qubits.}. Accordingly, the final state must be the complete graph state $\ket{K_{n_2}}$, which corresponds to complete graph $K_{n_2} = (P_1, P_1^2)$, with $P_1$ defined in \eqref{app:1.3}. 

The proof follows by performing: i) $(n_1-1)$-Pauli-$z$ measurements on the qubits stored at the clients of QLAN-1, and ii) a Pauli-$y$ measurement on the qubit stored at the super-node of the second QLAN. From \eqref{eq:11}, it results that the action of $(n_1-1)$-Pauli-$z$ measurements is equivalent to remove all the client vertices in $P_2$ in \eqref{app:1.4}, which yields to the graph:
\begin{equation}
    \label{app:2.1}
    S_{n_1,n_2} - P_2 \setminus \{ \ddot{v}^1_1 \} = \big( P_1 \cup \{ \ddot{v}^1_1 \}, P_1 \times \{ \ddot{v}^1_1 \} \big) = S'_{n_2+1}
\end{equation}
We observe that the graph $S'_{n_2+1}$ in \eqref{app:2.1} corresponds to a star graph connecting the vertices in the set $P_1 \cup \{\ddot{v}^1_1\}$, with the super-node $\ddot{v}^1_1$ being the center of the star.

Then, a Pauli-$y$ measurement is performed on the qubit stored at super-node of second QLAN and associated to the vertex $\ddot{v}^1_1$. From \eqref{eq:11}, the action of this measurement is equivalent to the local complementation of the graph $S'_{n_2+1}$ at vertex $\ddot{v}^1_1$, followed by the deletion of $\ddot{v}^1_1$ from the graph, i.e., $\tau_{\ddot{v}^1_1} \left( S'_{n_2+1} \right)-\ddot{v}^1_1$. Step-by-step, we first perform the local complementation $\tau_{\ddot{v}^1_1} \left( S'_{n_2+1} \right)$, which yields to the graph:
\begin{align}
    \label{app:2.2}
    \tau_{\ddot{v}^1_1} \left( S'_{n_2+1} \right) 
    &=\big( P_1 \cup \{ \ddot{v}^1_1 \}, P_1 \times \{ \ddot{v}^1_1 \} \cup N_{\ddot{v}^1_1}^2 \setminus E_{N_{\ddot{v}^1_1}} \big) \nonumber \\
    &= \big( P_1 \cup \{ \ddot{v}^1_1 \}, (P_1 \cup \{ \ddot{v}^1_1 \})^2 \big)
\end{align}
This equals to adding to the edge-set of the star graph $S'_{n_2+1}$ all the possible edges having both endpoints in the subset $P_1$. As a result, we obtain the complete graph connecting the set $P_1 \cup \{ \ddot{v}^1_1 \}$. We then proceed by removing the super-node $\ddot{v}^1_1$, and the proof follows: $\tau_{\ddot{v}^1_1} \left( S'_{n_2+1} \right) - \ddot{v}^1_1= \big( P_1 , P_1 ^2 \big)=K_{n_2}$

%--------------------------------------------------------------
\section{INTERMEDIATE RESULTS: PAULI-$x$ MEASUREMENT AT STAR GRAPH CENTER}
\label{app:00}
In the following, we prove an intermediate result widely used thorough the remaining appendices, namely, the Pauli-$x$ measurement on the qubit stored at the center of a star graph state. Hence, it is convenient to collect it in a separate appendix for the sake of clarity.

\begin{lem}
    Performing an $x$-measurement of the qubit held at the center $\dot{v}_1^1$ of a star graph $\dot{S}_{n_1}= (\dot{P}_1,\dot{P}_2,\dot{E})$ with $k_0$ set equal to $\dot{v}_j^2$ yields to a new star graph $\dot{S}_{n_1-1}$ among the set of nodes $\dot{P}_2$ with center node $\dot{v}_j^2$.
    \begin{IEEEproof}
        Let us adopt Fig.~\ref{fig:05} labeling and let us consider the initial graph $\dot{S}_{n_1}$ as defined in App.~\ref{app:01}. According to \eqref{eq:11}, the $x$-measurement on the qubit corresponding to vertex $\dot{v}_1^1$ is equivalent to perform the following sequence of graph operations, where we set $k_0= \dot{v}^2_j$ as from the thesis:
        \begin{equation}
            \label{eq:app:x-star1}
            \tau_{\dot{v}^2_j} \left( \tau_{\dot{v}^1_1}\big(\tau_{\dot{v}^2_j}(\dot{S}_{n_1})\big)-\dot{v}^1_1 \right).
        \end{equation}
        We proceed step-by-step. First, the local complementation at node $\dot{v}^2_j$ does not alter the graph, i.e., $\tau_{\dot{v}^2_j}(\dot{S}_{n_1})= \dot{S}_{n_1}$, since the neighborhood $N_{\dot{v}^2_j}$ of $\dot{v}^2_j$ has cardinality equal to $1$. Then, the local complementation at $\dot{v}^1_1$ adds to the edge-set of $\dot{S}_{n_1}$ all the possible edges among the vertexes in $\dot{P}_2$. As a result, the complete graph $\dot{K}_{n_1}$ connecting the set $\dot{P}_2 \cup \{ \dot{v}^1_1 \}$ is obtained:
        \begin{align}
            \label{eq:app:x-star2}
            \tau_{\dot{v}^1_1}(\dot{S}_{n_1}) &= \big( \dot{P}_2 \cup \{ \dot{v}^1_1 \}, ( \dot{P}_2 \times \{ \dot{v}^1_1 \} ) \cup N_{{\dot{v}^1_1}}^2 \setminus E_{N_{\dot{v}^1_1}} \big) \nonumber \\
            &= \big( \dot{P}_2 \cup \{ \dot{v}^1_1 \}, ( \dot{P}_2 \times \{ \dot{v}^1_1 \} ) \cup {\dot{P}_2}^2 \big) = \dot{K}_{n_1}.
        \end{align}
        Then, by removing $\dot{v}^1_1$, the resulting graph is:
        \begin{align}
            \label{eq:app:x-star4}
            \dot{K}_{n_1} - \dot{v}^1_1=\big( \dot{P}_2,{\dot{P}_2}^2 \big)=\dot{K}_{n_1-1},
        \end{align}
        which corresponds to the complete graph involving the set of nodes $\dot{P}_2$. Finally, the local complementation at node $\dot{v}^2_j$ yields to the graph:
        \begin{align}
            \label{eq:app:x-star3.bis}
            \tau_{\dot{v}^2_j}(\dot{K}_{n_1-1})&= \big( \dot{P}_2,  (\dot{P}_2^2 \cup N_{\dot{v}^2_j}^2)\setminus E_{N_{\dot{v}^2_j}} \big) \\
            &\nonumber= \big( \dot{P}_2, \{\dot{v}_j^2\} \times (\dot{P}_2 \setminus \{\dot{v}_j^2\}) \big)= \dot{S}_{n_1-1}
        \end{align}
    where $N_{\dot{v}^2_j}^2=E_{N_{\dot{v}^2_j}}=\dot{P}_2\setminus \{\dot{v}^2_j\}$. Accordingly, the local complementation at $\dot{v}^2_j$ removes from the edge-set of $\dot{K}_{n_1-1}$ all the possible edges among the vertexes in $\dot{P}_2\setminus \{\dot{v}^2_j\}$ and returns the star graph $\dot{S}_{n_1-1}$ involving the considered nodes with node $\dot{v}^2_j$ being the center of the star.
    \end{IEEEproof}
\end{lem}

%--------------------------------------------------------------
\section{PROOF OF COROLLARY~\ref{cor:01}}
\label{app:03}
The proof follows by adopting similar reasoning as in Prop.~\ref{prop:02}. Specifically, we adopt Fig.~\ref{fig:05} labeling and we suppose, without any loss in generality, that the star graph state $\ket{\hat{S}_{n_i}}$ has to be shared among all the right-most QLAN clients. Accordingly, the final star graph state is $\ket{\hat{S}_{n_2}}$, corresponding to the star graph $\hat{S}_{n_2} = ( \{ \ddot{v}^2_i \}, \{\dot{v}^1_1 \} \cup \ddot{P}_2 \setminus \{ \ddot{v}^2_i \} ,\hat{E})$ with $\ddot{P}_2,$ defined in \eqref{app:1.1.bis} and with $\hat{E}=\ddot{v}^2_i \times 
\{\{\dot{v}^1_1 \} \cup \ddot{P}_2 \setminus \{ \ddot{v}^2_i \}\}$. Specifically, $\hat{S}_{n_2}$ corresponds to a star graph centered at one arbitrary client $\ddot{v}^2_i$ of second QLAN and connecting the super-node of first QLAN and the remaining clients of second QLAN. The thesis follows by performing: i) $(n_1-1)$-Pauli-$z$ measurements on the qubits stored at the clients of QLAN-1, and ii) a Pauli-$x$ measurement on the qubit stored at the super-node of the second QLAN, by setting $k_0$ equal to $\ddot{v}^2_i$. 
The effects of $(n_1-1)$-Pauli-$z$ measurements are the same as analyzed in App.~\ref{app:02}, by yielding so to the star graph $S'_{n_2+1}$ in \eqref{app:2.1}. Then, as proved in App.~\ref{app:00}, the Pauli-$x$ measurement on the qubit $\ddot{v}^1_1$ returns a star graph $\hat{S}_{n_2}$ shared between the set $P_1$ in \eqref{app:1.3} and super-node $\dot{v}^1_1$ with one client node $\ddot{v}^2_i$ in $P_1$ being the center of the star.

%--------------------------------------------------------------
\section{PROOF OF PROPOSITION~\ref{prop:03}}
\label{app:05}
Similarly to the previous proofs, we adopt Fig.~\ref{fig:05} labeling and we suppose, without any loss in generality, that the $n_i$-complete connected graph state $\ket{K_{n_i}}$ has to be shared between one client node of the first QLAN and all the $(n_2-1)$-clients of the right-most QLAN. Accordingly, $\ket{K_{n_2}}$ corresponds to the complete graph $K_{n_2} = \big( \{\dot{v}^2_j\} \cup \ddot{P}_2, (\{\dot{v}^2_j\} \cup \ddot{P}_2)^2 \big)$ with $\ddot{P}_2$ defined in \eqref{app:1.1.bis}. The proof follows by performing: i) $(n_1-2)$-Pauli-$z$ measurements on the qubits stored at the clients of the left-most QLAN, with the exception of the client $\dot{v}^2_j$, ii) a Pauli-$y$ measurement at the super-node of the first QLAN, and iii) a Pauli-$y$ measurement at the super-node of the second QLAN.
From \eqref{eq:11}, by performing ($n_1- 2$) Pauli-$z$ measurements on the clients of first QLAN is equivalent to remove all the clients in $\dot{P}_2$ in \eqref{app:1.4}, except for the client node $\dot{v}^2_j$. Thus the resulting graph is $S_{n_1,n_2} - \big(\dot{P}_2 \setminus \{\dot{v}^2_j \}\big)$ which is equal to:
\begin{align}
    \label{app:3.2}
     \quad \big( \underbrace{P_1 \cup \{\dot{v}^2_j, \ddot{v}^1_1\}}_{V'},\underbrace{ \{\ddot{v}^1_1\} \times P_1 \cup \{(\dot{v}^2_j, \dot{v}^1_1)\}}_{E'} \big) \eqdef G'
\end{align}

Then, a Pauli-$y$ measurement is performed on the qubit -- associated to the vertex $\dot{v}^1_1$ -- stored at super-node of the first QLAN. From \eqref{eq:11}, and by reasoning as in Appendix~\ref{app:02}, this yields to the graph :
\begin{align}
    \label{app:3.3}
    &\tau_{\dot{v}^1_1}(G')- \dot{v}^1_1=\big(V', E'\cup N_{\dot{v}^1_1}^2 \setminus E_{N_{\dot{v}^1_1}} \big) - \dot{v}^1_1 = \nonumber \\
    & \quad =\big(V', E'\cup \{\ddot{v}^1_1,\dot{v}^2_j\}^2 \setminus \emptyset \big)- \dot{v}^1_1= \nonumber \\
    & \quad =\big( \ddot{P}_2 \cup \{\ddot{v}^1_1,\dot{v}^2_j\}, \{\ddot{v}^1_1\}\times (\ddot{P}_2 \cup \{\dot{v}^2_j\}) \big)= S''_{n_2+1}
\end{align}
We observe that the graph $S''_{n_2+1}$ in \eqref{app:3.3} corresponds to a star graph connecting the set $\ddot{P}_2 \cup \{\ddot{v}^1_1,\dot{v}^2_j\}$ with super-node $\ddot{v}^1_1$ being the center of the star. Then, a Pauli-$y$ measurement is performed on the qubit $\ddot{v}^1_1$ stored at super-node of second QLAN. From \eqref{eq:11}, this is equivalent to perform the following sequence of graph operations $\tau_{\ddot{v}^1_1} \left( S''_{n_2+1} \right)-\ddot{v}^1_1$. Step-by-step, the local complementation yields to the graph:
\begin{align}
    \label{eq:app:cc}
   \tau_{\ddot{v}^1_1} \left( S''_{n_2+1} \right)
    =
    \big( \ddot{P}_2 \cup \{ \dot{v}^2_j, \ddot{v}^1_1 \}, (\ddot{P}_2 \cup \{ \dot{v}^2_j, \ddot{v}^1_1 \})^2 \big)
\end{align}
This equals to adding to the edge-set of the star graph $S''_{n_2+1}$ all the possible edges among the subset $\ddot{P}_2$ with node $\{ \dot{v}^2_j \}$. As a result, we obtain the complete graph connecting the set $\ddot{P}_2 \cup \{ \ddot{v}^1_1, \dot{v}^2_j \}$, namely, the client nodes of the second QLAN, one client node $\{\dot{v}^2_j\}$ and the super-node of the first QLAN.
Then, by removing the super-node $\ddot{v}^1_1$, the proof follows:
\begin{align}
    \label{app:2.3.bis}
    \tau_{\ddot{v}^1_1} \left( S''_{n_2+1} \right) - \ddot{v}^1_1= \big(\{ \dot{v}^2_j \}\cup \ddot{P}_2 , (\{ \dot{v}^2_j \}\cup \ddot{P}_2 )^2 \big)=K_{n_2}
\end{align}

%--------------------------------------------------------------
\section{PROOF OF COROLLARY~\ref{cor:03}}
\label{app:06}

Similarly to the previous proofs, we adopt Fig.~\ref{fig:05} labeling and suppose that the state to be obtained is shared among all the right-most QLAN clients.
Accordingly, the final state has to be $\ket{S_{n_2}}$ corresponding to star graph $S_{n_2} = (\{\ddot{v}^2_i\},\{\dot{v}^2_j\} \cup \ddot{P}_2 \setminus \{\ddot{v}^2_i\}, E)$, 
with $\ddot{P}_2,$ defined in \eqref{app:1.1.bis} and with $E=\ddot{v}^2_i \times 
\{\{\dot{v}^j_1 \} \cup \ddot{P}_2 \setminus \{ \ddot{v}^2_i \}\}$. 
Thus the star graph is shared between the client node $\dot{v}^2_j$ of the left-most QLAN and the clients of the right-most QLAN with one client node $\ddot{v}^2_i$ of right-most QLAN being the center of the star. The proof follows by performing: i) $(n_1-2)$-Pauli-$z$ measurements on the qubits stored at the clients of the left-most QLAN except at the client 
$\dot{v}^2_j$, ii) a Pauli-$y$ measurement of the super-node of the first QLAN, and iii) a Pauli-$x$ measurement of the super-node of the second QLAN with the arbitrary node $k_0$ set equal to $\ddot{v}^2_i$.
By reasoning as in the Appendix~\ref{app:05}, it is easy to recognize that the graph before the $x$-measurement at the super-node $\ddot{v}^1_1$, is the star graph $S''_{n_2+1}$ expressed in \eqref{app:3.3}. By accounting for this, for the result in Appendix~\ref{app:00}, and by performing a Pauli-$z$ measurement on $\ddot{v}^1_1$, the star graph $S_{n_2}$ with the client node $\ddot{v}^2_i \in \ddot{P}_2$ being the center of star is obtained by setting $k_0=\ddot{v}^2_i$.

%--------------------------------------------------------------
\section{PROOF OF PROPOSITION~\ref{prop:04}}
\label{app:08}
We adopt Fig.~\ref{fig:05} labeling again, and the proof follows by performing: i) a Pauli-$x$ measurement on the qubit associated to the vertex $\dot{v}^1_1$ of the super-node of the first QLAN, with $k_0$ set equal to $\ddot{v}^1_1$, and ii) a Pauli-$z$ measurement on the qubit associated to the vertex $\ddot{v}^1_1$ of the super-node of the second QLAN.
From \eqref{eq:11}, a Pauli-$x$ measurement on  $\dot{v}^1_1$ with $k_0=\ddot{v}^1_1$ is equivalent to perform the sequence of graph operations $\tau_{\ddot{v}^1_1} \left( \tau_{\dot{v}^1_1}\big(\tau_{\ddot{v}^1_1}(S_{n_1,n_2})\big)-\dot{v}^1_1 \right)$.
Step-by-step, the local complementation at super-node $\ddot{v}^1_1$ adds all the possible edges having both endpoints belonging to the set $P_1$, by yielding to the graph $\tau_{\ddot{v}^1_1}(S_{n_1,n_2})$:

\begin{align}
    \label{app:4.3}
    \big( P_1 \cup P_2, ( P_1 \times \{ \ddot{v}^1_1 \} ) \cup ( P_2 \times \{ \dot{v}^1_1 \} ) \cup P_1^2 \big).
\end{align}
We observe that the neighborhood $N_{\dot{v}^1_1}$ of $\dot{v}^1_1$ in $\tau_{\ddot{v}^1_1}(S_{n_1,n_2}) $ is $P_1\cup P_2 \setminus \{\dot{v}^1_1\}$. Hence, the local complementation at the super-node $\dot{v}^1_1$ yields to the graph:
\begin{align}
    \label{app:4.4}
    \tau_{\dot{v}^1_1}(\tau_{\ddot{v}^1_1}(S_{n_1,n_2}) ) 
    = \big( P_1 \cup P_2, ( \{ \dot{v}^1_1 \} \times P_2) \cup P_2^2 \cup ( \ddot{P}_2 \times \dot{P}_2 ) \big),
\end{align}
where $\ddot{P}_2, \dot{P}_2$ are defined in \eqref{app:1.3} and \eqref{app:1.4}.
Then, we proceed by removing the super-node $\dot{v}^1_1$, which yields to the graph:
\begin{align}
    \label{app:4.5}
    \tau_{\dot{v}^1_1}(\tau_{\ddot{v}^1_1}(S_{n_1,n_2}) ) - \dot{v}^1_1 
    = \big( \ddot{P}_2 \cup P_2, P_2^2 \cup \big( \ddot{P}_2 \times \dot{P}_2 \big).
\end{align}
We observe that the neighborhood $N_{\ddot{v}^1_1}$ of $\ddot{v}^1_1$ in \eqref{app:4.5} corresponds to the set $\dot{P}_2$. Hence, the local complementation at the super-node $\ddot{v}^1_1$ yields to the graph:
\begin{align}
    \label{app:4.6}
    & \tau_{\ddot{v}^1_1} \left( \tau_{\dot{v}^1_1}\big(\tau_{\ddot{v}^1_1}(S_{n_1,n_2})\big)-\dot{v}^1_1 \right)
    = \nonumber \\
    & \quad =\big( \ddot{P}_2 \cup P_2, ( \{ \ddot{v}^1_1 \} \times \dot{P}_2 ) \cup \big( \ddot{P}_2 \times \dot{P}_2 \big) 
\end{align}

The proof follows after a Pauli-$z$ measurement on 
$\ddot{v}^1_1$:
\begin{align}
    \label{app:4.7}
    & \tau_{\ddot{v}^1_1} \left( \tau_{\dot{v}^1_1}\big(\tau_{\ddot{v}^1_1}(S_{n_1,n_2})\big)-\dot{v}^1_1 \right)- \{ \ddot{v}^1_1 \}
    = \nonumber \\
    & \quad = \big( \ddot{P}_2 \cup \dot{P}_2, \ddot{P}_2 \times \dot{P}_2 \big)
    = K_{n_1-1,n_2-1}.
\end{align}

%--------------------------------------------------------------
\section{PROOF OF COROLLARY~\ref{cor:05}}
\label{app:09}
Similarly to the previous proofs, we adopt Fig.~\ref{fig:05} labeling. The proof follows by performing: i) a Pauli-$x$ measurement on the qubit associated to vertex $\dot{v}_1^1$ of the super-node of the first QLAN with the arbitrary node $k_0$ set equal to $\dot{v}^2_j$, and ii) a Pauli-$x$ measurement on the qubit associated to vertex $\ddot{v}_1^1$ of the super-node of the second QLAN with the arbitrary node $k_0$ set equal to $\ddot{v}^2_i$.
From \eqref{eq:11}, a Pauli-$x$ measurement on  $\dot{v}^1_1$ with $k_0=\dot{v}^2_j$ is equivalent to perform the sequence of graph operations $\tau_{\dot{v}^2_j} \left( \tau_{\dot{v}^1_1}\big(\tau_{\dot{v}^2_j}(S_{n_1,n_2})\big)-\dot{v}^1_1 \right)$. By reasoning as in the previous appendix, it can be easily verified that one obtains the following graph: 
\begin{align}
    \label{app:5.4}
    & \tau_{\dot{v}^2_j} \left( \tau_{\dot{v}^1_1}\big(\tau_{\dot{v}^2_j}(S_{n_1,n_2})\big)-\dot{v}^1_1 \right)
    =\\
    & \quad = \big( \underbrace{\ddot{P}_2 \cup P_2}_{\widetilde{V}}, \underbrace{( \{ \ddot{v}^1_1 \} \times \dot{P}_2 ) \cup \big( \{\dot{v}^2_j\} \times (P_2 \setminus \{\dot{v}^2_j\}) \big)}_{\widetilde{E}} \big) =\widetilde{G}. \nonumber 
\end{align}
We then perform a $x$-measurement on the super-node $\ddot{v}^1_1$ of second QLAN with the arbitrary node $k_0$ set equal to $\ddot{v}^2_i$. From \eqref{eq:11}, this is equivalent to perform the following operations:
\begin{equation}
    \label{app:5.5}
    \tau_{\ddot{v}^2_i} \left( \tau_{\ddot{v}^1_1}\big(\tau_{\ddot{v}^2_i}(\widetilde{G})\big)-\ddot{v}^1_1 \right).
\end{equation}
Step-by-step, the local complementation at client $\ddot{v}^2_i$ does not change the graph $\widetilde{G}$. 
Then the local complementation at super-node $\ddot{v}^1_1$ yields to the graph:
\begin{align}
    \label{app:5.6}
    \tau_{\ddot{v}^1_1}(\widetilde{G}) 
    = \big( \widetilde{V}, \underbrace{( \{\dot{v}^2_j\} \times \ddot{P}_2)  \cup  \ddot{P}_2^2}_{\widetilde{E}'} \cup \widetilde{E} \big) 
\end{align}
By removing $\ddot{v}^1_1$, the resulting graph is given by $\tau_{\ddot{v}^1_1}(\widetilde{G}) - \ddot{v}^1_1$:
\begin{align}
    \label{app:5.7}
    \big( \widetilde{V} \setminus \{\ddot{v}^1_1\}, \widetilde{E}' \cup \big( \{\dot{v}^2_j\} \times (\dot{P}_2 \setminus \{\dot{v}^2_j\}) \big) \big)
    =\widetilde{G}'
\end{align}
The proof follows by observing that the neighborhood $N_{\ddot{v}^2_i}$ of $\ddot{v}^2_i$ in $\widetilde{G}'$ corresponds to the set $\ddot{P}_2$. Thus, the local complementation at $\ddot{v}^2_i$ yields to the graph $\tau_{\ddot{v}^2_i} \big( \widetilde{G}'  \big)$:

\begin{align}
    \label{app:5.8}
    (\dot{P}_2, \ddot{P}_2, \big( \{\ddot{v}^2_i\} \times \dot{P}_2 \big) \cup \big( \{\dot{v}^2_j\} \times \ddot{P}_2 \big) = S_{n_1-1,n_2-1}.
\end{align}

\end{appendices}

\bibliographystyle{IEEEtran}
\bibliography{biblio.bib}

\begin{IEEEbiography}
[{\includegraphics[width=1in,height=1.25in,clip,keepaspectratio]{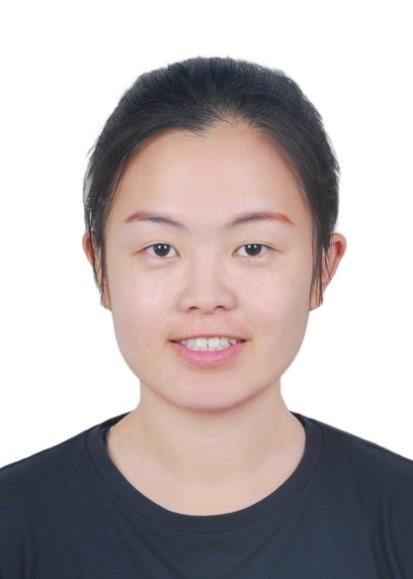}}]{Si-Yi Chen}\, received the Ph.D degree as 
an outstanding graduate in 2023 in School of Cyberspace Security from Beijing University of Posts and Telecommunications (China). Currently, she is a Postdoctoral fellow in University of Naples Federico II (Italy). She serves as TPC member in IEEE International Conference on Quantum Computing and Engineering 2024. Since 2022, she is a member of the Quantum Internet Research Group in University of Naples Federico II where she works on multiparty quantum networks, long-distance quantum communications and quantum entanglements.
\end{IEEEbiography}

\begin{IEEEbiography}[{\includegraphics[width=1in,height=1.25in,clip,keepaspectratio]{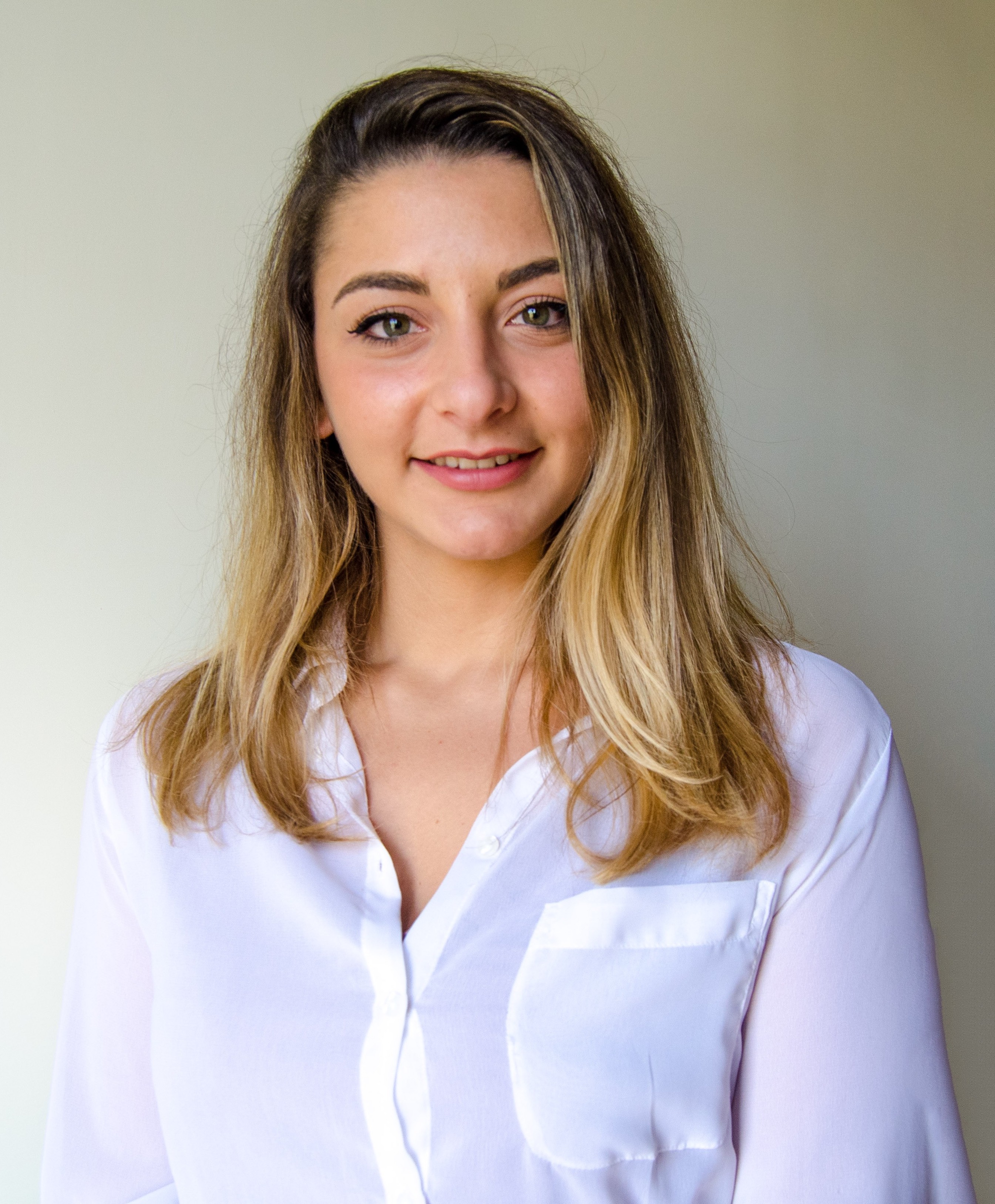}}]{Jessica Illiano}\, is an Assistant Professor at the University of Naples Federico II. She received the B.Sc degree in 2018 and then the M.Sc degree in 2020 both (summa cum laude) in Telecommunications Engineering from University of Naples Federico II (Italy). In 2020 she was winner of the scholarship "Quantum Communication Protocols for Quantum Security and Quantum Internet" fully funded by TIM S.p.A. and in 2024 she received her PhD degree in Information Technologies and Electrical Engineering at University of Naples Federico II. Since 2017, she is a member of the Quantum Internet Research Group, FLY: Future Communications Laboratory at the University of Naples Federico II. Currently, she serves as website co-chair of N2Women, student Associate Editor for IET Quantum Communication and Associate Editor for IEEE Communication Letters. Her research interests include quantum communications, quantum networks and quantum information processing.
\end{IEEEbiography}

\begin{IEEEbiography}
[{\includegraphics[width=1in,height=1.25in,clip,keepaspectratio]{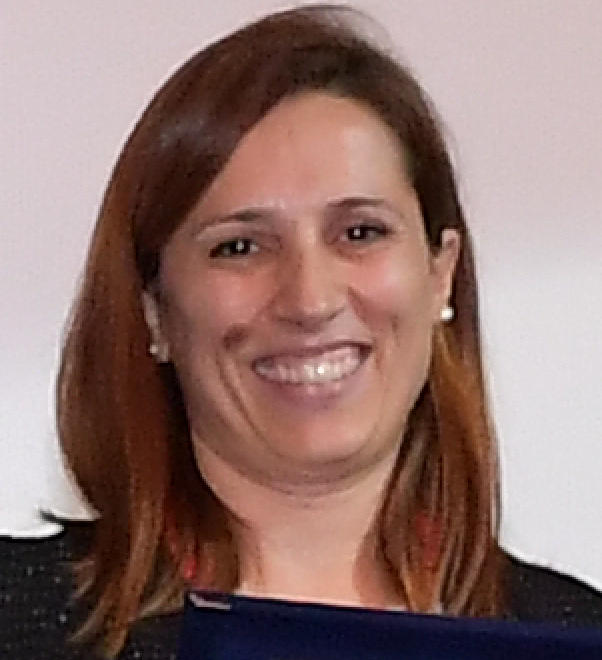}}]{Angela Sara Cacciapuoti} \, (Senior Member,
IEEE) is a Professor of Quantum Communications and Networks at the University of Naples Federico II (Italy). Her work has appeared in first tier IEEE journals and she received different awards, including the ``2024 IEEE ComSoc Award for Advances in Communication'', the ``2022 IEEE ComSoc Best Tutorial Paper Award'', the ``2022 WICE Outstanding Achievement Award'' for her contributions in the quantum communication and network fields, and ``2021 N2Women: Stars in Networking and Communications''. Lately, she also received the IEEE ComSoc Distinguished Service Award for EMEA 2023, assigned for the outstanding service to IEEE ComSoc in the EMEA Region. Currently, she is an IEEE ComSoc Distinguished Lecturer with lecture topics on the Quantum Internet design and Quantum Communications. And she serves also as Member of the TC on SPCOM within the IEEE Signal Processing Society. Moreover, she serves as Area Editor for IEEE Trans. on Communications and as Editor/Associate Editor for the journals: npj Quantum Information, IEEE Trans. on Quantum Engineering, IEEE Communications Surveys \& Tutorials. She served as Area Editor for IEEE Communications Letters(2019 - 2023), and she was the recipient of the 2017 Exemplary Editor Award of the IEEE Communications Letters. In 2023, she also served as Lead Guest Editor for IEEE JSAC special issue ''The Quantum Internet: Principles, Protocols, and Architectures''. From 2020 to 2021, Angela Sara was the Vice-Chair of the IEEE ComSoc Women in Communications Engineering. Previously, she has been appointed as Publicity Chair of WICE. From 2017 to 2020, she has been the Treasurer of the IEEE Women in Engineering (WIE) Affinity Group of the IEEE Italy Section. Her research interests are in Quantum Information Processing, Quantum Communications and Quantum Networks. She is also the PI of the ERC-CoG 2024 "QNattyNet".
\end{IEEEbiography}

\begin{IEEEbiography}
[{\includegraphics[width=1in,height=1.25in,clip,keepaspectratio]{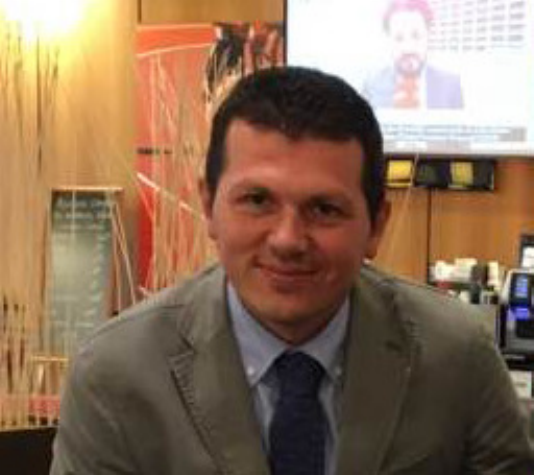}}]{Marcello Caleffi} \, (Senior Member,
IEEE) is currently Professor with the DIETI Department, University of Naples Federico II, where he co-lead the Quantum Internet Research Group. He is also with the National Laboratory of Multimedia Communications, National Inter- University Consortium for Telecommunications. From 2010 to 2011, he was with the Broadband Wireless Networking Laboratory with the Georgia Institute of Technology, as a Visiting Researcher. In 2011, he was also with the NaNoNetworking Center in Catalunya (N3Cat) with the Universitat Politecnica de Catalunya, as a Visiting Researcher. Since July 2018, he held the Italian National Habilitation as a Full Professor of Telecommunications Engineering. His work appeared in several premier IEEE Transactions and Journals, and he received multiple awards, including the ``2024 IEEE Communications Society Award for Advances in Communication and the ``2022 IEEE Communications Society Best Tutorial Paper Award''. He currently serves as an Editor/Associate Editor for IEEE TWC, IEEE TCOM, IEEE TQE, IEEE OJ-COMS and IEEE Internet Computing. He has served as the chair, the TPC chair, and a TPC member for several premier IEEE conferences. In 2017, he has been appointed as Distinguished Visitor Speaker from the IEEE Computer Society and he has been elected treasurer of the IEEE ComSoc/VT Italy Chapter. In 2019, he has been also appointed as a member of the IEEE New Initiatives Committee from the IEEE Board of Directors and, in 2023, he has been appointed as ComSoc Distinguished Lecturer.
\end{IEEEbiography}

\end{document}

%% file: Figures/Fig-01.tex
\tikzset{every picture/.style={line width=0.75pt}} %set default line width to 0.75pt        

\begin{tikzpicture}[x=0.75pt,y=0.75pt,yscale=-1,xscale=1]
%uncomment if require: \path (0,627); %set diagram left start at 0, and has height of 627

%Image [id:dp9201999854510197] 
\draw (630.74,199.15) node [rotate=-339.15,xscale=-1] {\includegraphics[width=245.45pt,height=5.07pt]{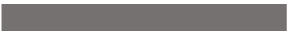}};
%Image [id:dp8473640390460608] 
\draw (724.44,375.74) node [rotate=-206.25,xscale=-1] {\includegraphics[width=413.78pt,height=5.51pt]{Figures/QchannelSS.png}};
%Image [id:dp8287034710911068] 
\draw (732.33,312.92) node [xslant=0.01] {\includegraphics[width=491.5pt,height=415.24pt]{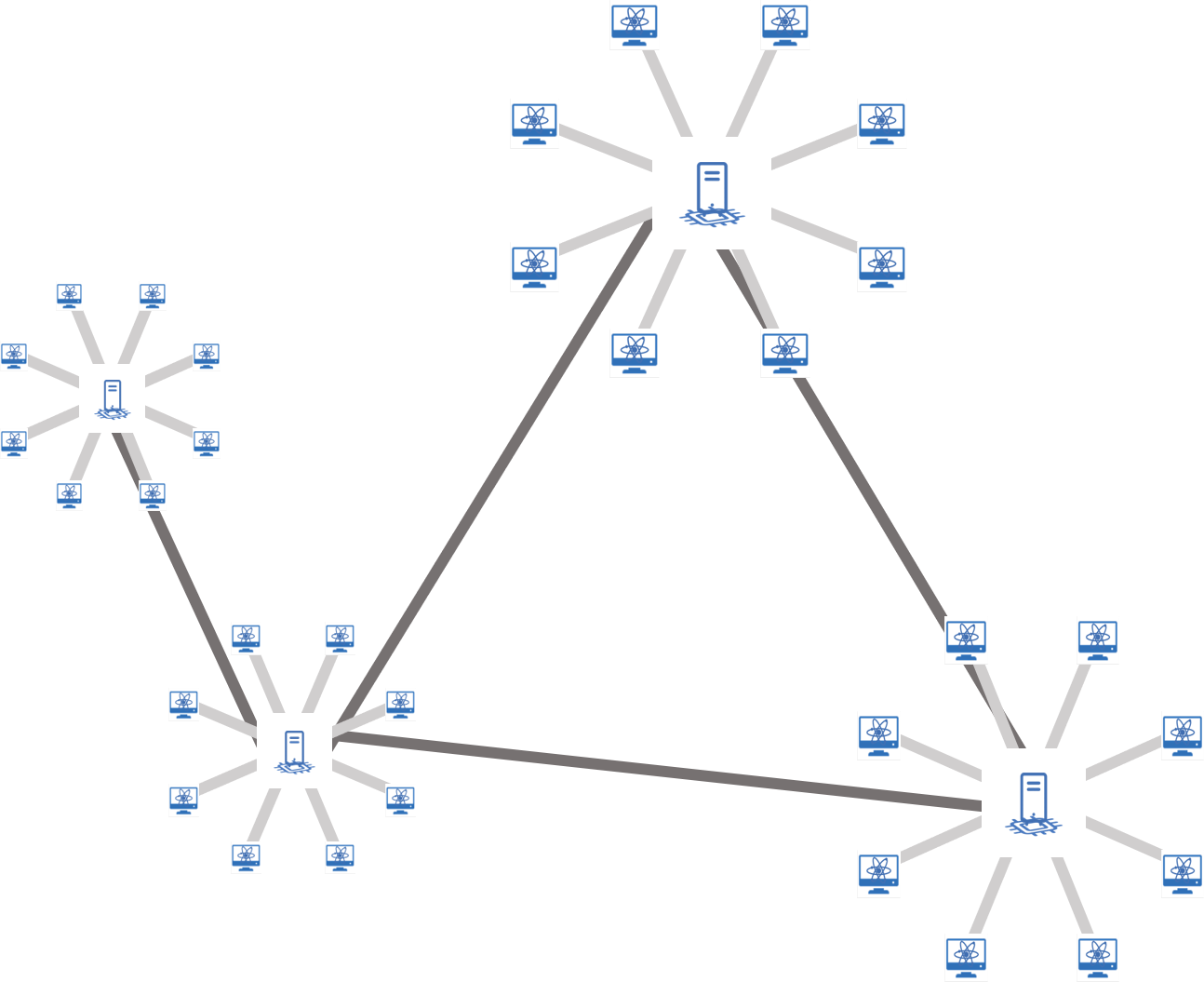}};

%Image [id:dp06531351316545742] 
\draw (984.96,53.26) node  {\includegraphics[width=37.56pt,height=41.2pt]{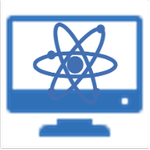}};
%Image [id:dp5548042400616655] 
\draw (1158.2,425) node  {\includegraphics[width=52pt,height=5.21pt]{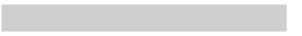}};
%Image [id:dp06672669768519612] 
\draw (1102.86,270.55) node  {\includegraphics[width=53pt,height=5.51pt]{Figures/QchannelSS.png}};
%Image [id:dp6599220746512912] 
\draw (1041.73,158.85) node  {\includegraphics[width=82.09pt,height=73.42pt]{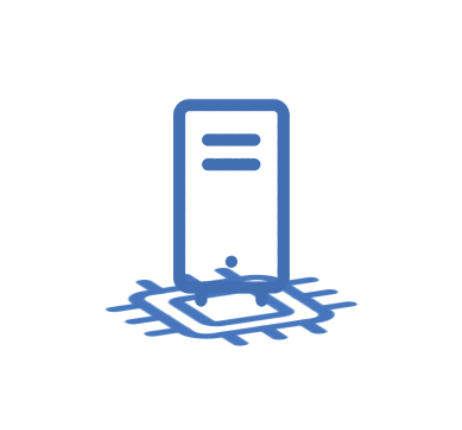}};
%Straight Lines [id:da9747753126043813] 
\draw [color={rgb, 255:red, 208; green, 2; blue, 2 }  ,draw opacity=1 ][line width=5.25]    (835,143.8) -- (991,143.8) ;
\draw [shift={(999,143.8)}, rotate = 180] [color={rgb, 255:red, 208; green, 2; blue, 2 }  ,draw opacity=1 ][line width=5.25]    (33.88,-10.2) .. controls (21.55,-4.33) and (10.25,-0.93) .. (0,0) .. controls (10.25,0.93) and (21.55,4.33) .. (33.88,10.2)   ;
%Straight Lines [id:da8884312685923723] 
\draw [color={rgb, 255:red, 208; green, 2; blue, 2 }  ,draw opacity=1 ][line width=5.25]    (875,49.8) -- (934,49.8) ;
\draw [shift={(942,49.8)}, rotate = 180] [color={rgb, 255:red, 208; green, 2; blue, 2 }  ,draw opacity=1 ][line width=5.25]    (33.88,-10.2) .. controls (21.55,-4.33) and (10.25,-0.93) .. (0,0) .. controls (10.25,0.93) and (21.55,4.33) .. (33.88,10.2)   ;
%Straight Lines [id:da018556595596147973] 
\draw [color={rgb, 255:red, 208; green, 2; blue, 2 }  ,draw opacity=1 ][line width=5.25]    (893.67,275.46) -- (1015,273.9) ;
\draw [shift={(1023,273.8)}, rotate = 179.26] [color={rgb, 255:red, 208; green, 2; blue, 2 }  ,draw opacity=1 ][line width=5.25]    (33.88,-10.2) .. controls (21.55,-4.33) and (10.25,-0.93) .. (0,0) .. controls (10.25,0.93) and (21.55,4.33) .. (33.88,10.2)   ;
%Rounded Rect [id:dp2272663585232838] 
\draw   (1005,132.2) .. controls (1005,122.59) and (1012.79,114.8) .. (1022.4,114.8) -- (1281.6,114.8) .. controls (1291.21,114.8) and (1299,122.59) .. (1299,132.2) -- (1299,184.4) .. controls (1299,194.01) and (1291.21,201.8) .. (1281.6,201.8) -- (1022.4,201.8) .. controls (1012.79,201.8) and (1005,194.01) .. (1005,184.4) -- cycle ;
%Rounded Rect [id:dp46780988294565473] 
\draw   (946,29.72) .. controls (946,22.03) and (952.23,15.8) .. (959.93,15.8) -- (1212.07,15.8) .. controls (1219.77,15.8) and (1226,22.03) .. (1226,29.72) -- (1226,71.5) .. controls (1226,79.19) and (1219.77,85.42) .. (1212.07,85.42) -- (959.93,85.42) .. controls (952.23,85.42) and (946,79.19) .. (946,71.5) -- cycle ;
%Rounded Rect [id:dp5531602798911208] 
\draw   (1113.01,402.37) .. controls (1113.01,390.71) and (1122.46,381.26) .. (1134.11,381.26) -- (1427.89,381.26) .. controls (1439.55,381.26) and (1449,390.71) .. (1449,402.37) -- (1449,465.69) .. controls (1449,477.35) and (1439.55,486.8) .. (1427.89,486.8) -- (1134.11,486.8) .. controls (1122.46,486.8) and (1113.01,477.35) .. (1113.01,465.69) -- cycle ;

%Rounded Rect [id:dp3841525260568438] 
\draw   (1049.01,244.37) .. controls (1049.01,232.71) and (1058.46,223.26) .. (1070.11,223.26) -- (1377.89,223.26) .. controls (1389.55,223.26) and (1399,232.71) .. (1399,244.37) -- (1399,307.69) .. controls (1399,319.35) and (1389.55,328.8) .. (1377.89,328.8) -- (1070.11,328.8) .. controls (1058.46,328.8) and (1049.01,319.35) .. (1049.01,307.69) -- cycle ;

%Straight Lines [id:da7635147553144701] 
\draw [color={rgb, 255:red, 208; green, 2; blue, 2 }  ,draw opacity=1 ][line width=5.25]    (1001,422.8) -- (1075,421.89) ;
\draw [shift={(1083,421.8)}, rotate = 179.3] [color={rgb, 255:red, 208; green, 2; blue, 2 }  ,draw opacity=1 ][line width=5.25]    (33.88,-10.2) .. controls (21.55,-4.33) and (10.25,-0.93) .. (0,0) .. controls (10.25,0.93) and (21.55,4.33) .. (33.88,10.2)   ;

% Text Node
\draw (1086.93,145) node [anchor=north west][inner sep=0.75pt]  [font=\huge] [align=left] {\textsc{Super-node}};
% Text Node
\draw (1030.7,35) node [anchor=north west][inner sep=0.75pt]  [font=\huge] [align=left] {\textsc{Client node}};
% Text Node
\draw (1200,386.08) node [anchor=north west][inner sep=0.75pt]  [font=\huge] [align=left] 
%{\textsc{Intra-QLAN physical channel}};
{\begin{minipage}[lt]{164.7pt}\setlength\topsep{0pt}
\begin{center}
\textsc{Intra-QLAN \\physical channel} 
\end{center}
\end{minipage}};

% Text Node
\draw (1142.5,229.41) node [anchor=north west][inner sep=0.75pt]  [font=\huge] [align=left] {\begin{minipage}[lt]{164.7pt}\setlength\topsep{0pt}
\begin{center}
\textsc{Inter-QLANs \\physical channel} 
\end{center}

\end{minipage}};

\end{tikzpicture}

%% file: Figures/Fig-G.tex
\tikzset{every picture/.style={line width=0.75pt}} %set default line width to 0.75pt        

\begin{tikzpicture}[x=0.75pt,y=0.75pt,yscale=-1,xscale=1]
%uncomment if require: \path (0,472); %set diagram left start at 0, and has height of 472
%Shape: Rectangle [id:dp6902335119690582] 
\draw   [fill={rgb, 126:red, 67; green, 146; blue, 226 }  ,fill opacity=0.4 ] [dash pattern={on 0.84pt off 2.51pt}] (90,164) -- (565,164) -- (565,350) -- (90,350) -- cycle ;
%Shape: Rectangle [id:dp6902335119690582] 
\draw   [fill={rgb, 255:red, 74; green, 144; blue, 226 }  ,fill opacity=0.4 ] [dash pattern={on 0.84pt off 2.51pt}] (90,75) -- (565,75) -- (565,164) -- (90,164) -- cycle ;
%Straight Lines [id:da6418980704527419] 
%\draw [line width=1.5]  [dash pattern={on 1.69pt off 2.76pt}]  (40,164) -- (600,164) ;
%Straight Lines [id:da9485468890118289] freccia g g'
\draw    (130,94) -- (510,94) ;
\draw [shift={(510,94)}, rotate = 180] [color={rgb, 255:red, 0; green, 0; blue, 0 }  ][line width=0.75]    (10.93,-3.29) .. controls (6.95,-1.4) and (3.31,-0.3) .. (0,0) .. controls (3.31,0.3) and (6.95,1.4) .. (10.93,3.29)   ;
%Straight Lines [id:da1547083050821495] freccia ket g ket gtilde
\draw    (145,306) -- (355,306) ;
\draw [shift={(355,306)}, rotate = 180] [color={rgb, 255:red, 0; green, 0; blue, 0 }  ][line width=0.75]    (10.93,-3.29) .. controls (6.95,-1.4) and (3.31,-0.3) .. (0,0) .. controls (3.31,0.3) and (6.95,1.4) .. (10.93,3.29)   ;
%Straight Lines [id:da7866041280858421] bi-freccia g ket g
\draw    (109,120) -- (109,280) ;
\draw [shift={(109,280)}, rotate = 270] [color={rgb, 255:red, 0; green, 0; blue, 0 }  ][line width=0.75]    (10.93,-3.29) .. controls (6.95,-1.4) and (3.31,-0.3) .. (0,0) .. controls (3.31,0.3) and (6.95,1.4) .. (10.93,3.29)   ;
\draw [shift={(109,120)}, rotate = 90] [color={rgb, 255:red, 0; green, 0; blue, 0 }  ][line width=0.75]    (10.93,-3.29) .. controls (6.95,-1.4) and (3.31,-0.3) .. (0,0) .. controls (3.31,0.3) and (6.95,1.4) .. (10.93,3.29)   ;
%Straight Lines [id:da32327494183866334] bi- freccia g' ket g'
\draw    (530,112.9) -- (530,204.9) ;
\draw [shift={(530,206.9)}, rotate = 270] [color={rgb, 255:red, 0; green, 0; blue, 0 }  ][line width=0.75]    (10.93,-3.29) .. controls (6.95,-1.4) and (3.31,-0.3) .. (0,0) .. controls (3.31,0.3) and (6.95,1.4) .. (10.93,3.29)   ;
\draw [shift={(530,112.9)}, rotate = 90] [color={rgb, 255:red, 0; green, 0; blue, 0 }  ][line width=0.75]    (10.93,-3.29) .. controls (6.95,-1.4) and (3.31,-0.3) .. (0,0) .. controls (3.31,0.3) and (6.95,1.4) .. (10.93,3.29)   ;
%Straight Lines [id:da31541739763511145] bi-freccia ket g' ket g tilde
\draw    (530,245) -- (530,284) ;
\draw [shift={(530,284)}, rotate = 270] [color={rgb, 255:red, 0; green, 0; blue, 0 }  ][line width=0.75]    (10.93,-3.29) .. controls (6.95,-1.4) and (3.31,-0.3) .. (0,0) .. controls (3.31,0.3) and (6.95,1.4) .. (10.93,3.29)   ;
\draw [shift={(530,245)}, rotate = 90] [color={rgb, 255:red, 0; green, 0; blue, 0 }  ][line width=0.75]    (10.93,-3.29) .. controls (6.95,-1.4) and (3.31,-0.3) .. (0,0) .. controls (3.31,0.3) and (6.95,1.4) .. (10.93,3.29)   ;

% Text Node
\draw (95,85) node [anchor=north west][inner sep=0.75pt] [font=\Huge]  [align=center] {$G$};
% Text Node
\draw (243,45) node [anchor=north west][inner sep=0.75pt] [font=\LARGE]  [align=left] {\textbf{\textsc{Graph Domain}}};
% Text Node
\draw (210,360) node [anchor=north west][inner sep=0.75pt] [font=\LARGE]  [align=center] {
\textbf{\textsc{Graph States Domain}}};
% Text Node
\draw (90,290) node [anchor=north west][inner sep=0.75pt] [font=\Huge]  [align=left] {$\ket{G}$};
% Text Node
\draw (150,280) node [anchor=north west][inner sep=0.75pt]  [font=\LARGE] [align=center] {$\xi$- Pauli measurement \\ on qubit $i$};
% Text Node
\draw (517,80) node [anchor=north west][inner sep=0.75pt] [font=\Huge]  [align=left] {$G'$};
% Text Node
\draw (378,285) node [anchor=north west][inner sep=0.75pt]  [font=\Huge] [align=left] {$\ket{\xi,\pm}^{i}\otimes \ket{\Tilde{G}}$};
% Text Node
\draw (506,210) node [anchor=north west][inner sep=0.75pt]  [font=\Huge] [align=left] {$\ket{G'}$};
% Text Node
\draw (370,253) node [anchor=north west][inner sep=0.75pt]  [font=\LARGE] [align=left] {LU equivalence};
% Text Node
\draw (243,100) node [anchor=north west][inner sep=0.75pt] [font=\LARGE]  [align=left] {Graph operation};

\end{tikzpicture}

%% file: Figures/Fig-02a.tex
\tikzset{every picture/.style={line width=0.75pt}} %set default line width to 0.75pt        

\begin{tikzpicture}[x=0.75pt,y=0.75pt,yscale=-1,xscale=1]
%uncomment if require: \path (0,610); %set diagram left start at 0, and has height of 610

%Image [id:dp2366835241837406] 
\draw (594,269.83) node  {\includegraphics[width=756pt,height=290.74pt]{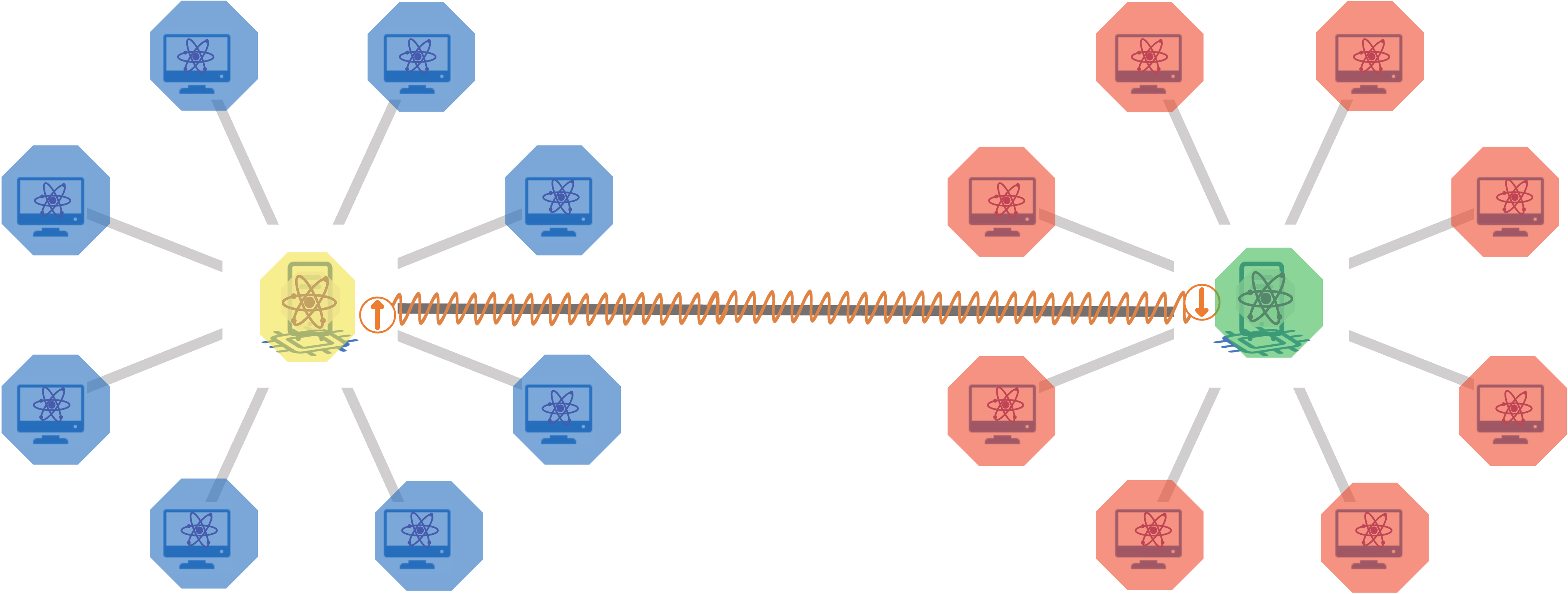}};
%Curve Lines [id:da5252548959449412] 
\draw [color={rgb, 255:red, 126; green, 67; blue, 146 }  ,draw opacity=1 ][line width=1.5]    (361.7,424.16) .. controls (321.54,396.66) and (287.63,317.79) .. (293.63,275.79) ;
%Curve Lines [id:da7547760928716795] 
\draw [color={rgb, 255:red, 126; green, 67; blue, 146 }  ,draw opacity=1 ][line width=1.5]    (459.83,340.76) .. controls (392.36,351.92) and (313.36,327.92) .. (294.43,275.48) ;
%Curve Lines [id:da6552344721188035] 
\draw [color={rgb, 255:red, 126; green, 67; blue, 146 }  ,draw opacity=1 ][line width=1.5]    (454.24,206.14) .. controls (431.67,259.05) and (355.36,299.92) .. (294.43,275.48) ;
%Curve Lines [id:da8567237921073346] 
\draw [color={rgb, 255:red, 126; green, 67; blue, 146 }  ,draw opacity=1 ][line width=1.5]    (363.24,113.14) .. controls (369.67,174.71) and (368.36,236.92) .. (294.43,275.48) ;
%Curve Lines [id:da37530083123862323] 
\draw [color={rgb, 255:red, 126; green, 67; blue, 146 }  ,draw opacity=1 ][line width=1.5]    (125.36,342.01) .. controls (144.74,297.24) and (226.74,255.24) .. (293.78,273.33) ;
%Curve Lines [id:da7207604811907508] 
\draw [color={rgb, 255:red, 126; green, 67; blue, 146 }  ,draw opacity=1 ][line width=1.5]    (220.83,423.25) .. controls (198.74,383.24) and (229.74,295.24) .. (293.78,278) ;
%Curve Lines [id:da7751922616758143] 
\draw [color={rgb, 255:red, 126; green, 67; blue, 146 }  ,draw opacity=1 ][line width=1.5]    (217.24,113.14) .. controls (273.05,148.44) and (314.74,202.24) .. (296.72,275.66) ;
%Curve Lines [id:da22210317930995538] 
\draw [color={rgb, 255:red, 126; green, 67; blue, 146 }  ,draw opacity=1 ][line width=1.5]    (124.93,208.94) .. controls (196.74,191.24) and (261.74,201.24) .. (296.72,275.66) ;
%Curve Lines [id:da2355727090571471] 
\draw [color={rgb, 255:red, 126; green, 67; blue, 146 }  ,draw opacity=1 ][line width=1.5]    (972.7,425.16) .. controls (932.54,397.66) and (898.63,318.79) .. (904.63,276.79) ;
%Curve Lines [id:da8505783258579247] 
\draw [color={rgb, 255:red, 126; green, 67; blue, 146 }  ,draw opacity=1 ][line width=1.5]    (1070.83,341.76) .. controls (1003.36,352.92) and (924.36,328.92) .. (905.43,276.48) ;
%Curve Lines [id:da169361805678641] 
\draw [color={rgb, 255:red, 126; green, 67; blue, 146 }  ,draw opacity=1 ][line width=1.5]    (1065.24,207.14) .. controls (1042.67,260.05) and (966.36,300.92) .. (905.43,276.48) ;
%Curve Lines [id:da7875791703531225] 
\draw [color={rgb, 255:red, 126; green, 67; blue, 146 }  ,draw opacity=1 ][line width=1.5]    (736.36,343.01) .. controls (755.74,298.24) and (837.74,256.24) .. (904.78,274.33) ;
%Curve Lines [id:da41589495787049247] 
\draw [color={rgb, 255:red, 126; green, 67; blue, 146 }  ,draw opacity=1 ][line width=1.5]    (831.83,424.25) .. controls (809.74,384.24) and (840.74,296.24) .. (904.78,279) ;
%Curve Lines [id:da047373231062882826] 
\draw [color={rgb, 255:red, 126; green, 67; blue, 146 }  ,draw opacity=1 ][line width=1.5]    (828.24,114.14) .. controls (884.05,149.44) and (925.74,203.24) .. (907.72,276.66) ;
%Curve Lines [id:da4709122194941807] 
\draw [color={rgb, 255:red, 126; green, 67; blue, 146 }  ,draw opacity=1 ][line width=1.5]    (735.93,209.94) .. controls (807.74,192.24) and (872.74,202.24) .. (907.72,276.66) ;
%Curve Lines [id:da2667549459738422] 
\draw [color={rgb, 255:red, 126; green, 67; blue, 146 }  ,draw opacity=1 ][line width=1.5]    (973.59,111.99) .. controls (980.02,173.55) and (978.7,235.77) .. (904.78,274.33) ;

% Text Node
\draw  [draw opacity=0][fill={rgb, 255:red, 255; green, 255; blue, 255 }  ,fill opacity=1 ]  (404,413.39) -- (474,413.39) -- (474,445.39) -- (404,445.39) -- cycle  ;
\draw (405,414.39) node [anchor=north west][inner sep=0.75pt]  [font=\Huge] [align=left] {$\dot{v}^2_{n_1-1}$};
% Text Node
\draw  [draw opacity=0][fill={rgb, 255:red, 255; green, 255; blue, 255 }  ,fill opacity=1 ]  (400.25,88) -- (464.25,88) -- (464.25,120) -- (400.25,120) -- cycle  ;
\draw (401.25,89) node [anchor=north west][inner sep=0.75pt]  [font=\Huge] [align=left] {$\dot{v}^2_i$};

% % Text Node
% \draw  [draw opacity=0][fill={rgb, 255:red, 255; green, 255; blue, 255 }  ,fill opacity=1 ]  (247,293.39) -- (315,293.39) -- (315,325.39) -- (247,325.39) -- cycle  ;
% \draw (248,294.39) node [anchor=north west][inner sep=0.75pt]  [font=\Huge] [align=left] {$\displaystyle Node\ 1$};

% Text Node
\draw  [draw opacity=0][fill={rgb, 255:red, 255; green, 255; blue, 255 }  ,fill opacity=1 ]  (265,207.43) -- (293,207.43) -- (293,239.43) -- (265,239.43) -- cycle  ;
\draw (266,208.43) node [anchor=north west][inner sep=0.75pt] [font=\Huge]  [align=left] {$\dot{v}^1_1$};

% Text Node
% \draw  [color={rgb, 255:red, 255; green, 187; blue, 0 }  ,draw opacity=1 ][fill={rgb, 255:red, 255; green, 255; blue, 255 }  ,fill opacity=1 ][line width=1.5]   (320.79,235.23) -- (346.79,235.23) -- (346.79,267.23) -- (320.79,267.23) -- cycle  ;
% \draw (321.79,236.23) node [anchor=north west][inner sep=0.75pt] [font=\Huge]  [align=left] {$\displaystyle e_{0}$};

% Text Node
\draw  [draw opacity=0][fill={rgb, 255:red, 255; green, 255; blue, 255 }  ,fill opacity=1 ]  (117,410.39) -- (185,410.39) -- (185,442.39) -- (117,442.39) -- cycle  ;
\draw (118,411.39) node [anchor=north west][inner sep=0.75pt]  [font=\Huge] [align=left] {$\dot{v}^2_1$};
% Text Node
\draw  [draw opacity=0][fill={rgb, 255:red, 255; green, 255; blue, 255 }  ,fill opacity=1 ]  (87,274.39) -- (155,274.39) -- (155,306.39) -- (87,306.39) -- cycle  ;
\draw (88,275.39) node [anchor=north west][inner sep=0.75pt]  [font=\Huge] [align=left] {$\dot{v}^2_2$};
% Text Node
\draw  [draw opacity=0][fill={rgb, 255:red, 255; green, 255; blue, 255 }  ,fill opacity=1 ]  (881,208.03) -- (931,208.03) -- (931,240.03) -- (881,240.03) -- cycle  ;
\draw (882,209.03) node [anchor=north west][inner sep=0.75pt] [font=\Huge]  [align=left] {$\ddot{v}^1_1$};

% Text Node
\draw  [draw opacity=0][fill={rgb, 255:red, 255; green, 255; blue, 255 }  ,fill opacity=1 ]  (1007,410) -- (1115,410) -- (1115,442) -- (1007,442) -- cycle  ;
\draw (1008,411) node [anchor=north west][inner sep=0.75pt]  [font=\Huge] [align=left] {$\ddot{v}^2_{n_2-1}$};

% Text Node
\draw  [draw opacity=0][fill={rgb, 255:red, 255; green, 255; blue, 255 }  ,fill opacity=1 ]  (990,140) -- (1088,140) -- (1088,172) -- (990,172) -- cycle  ;
\draw (991,141) node [anchor=north west][inner sep=0.75pt]  [font=\Huge] [align=left] {$\ddot{v}^2_j$};

% % Text Node
% \draw  [color={rgb, 255:red, 255; green, 187; blue, 0 }  ,draw opacity=1 ][line width=1.5]   (845.83,223.87) -- (871.83,223.87) -- (871.83,255.87) -- (845.83,255.87) -- cycle  ;
% \draw (846.83,224.87) node [anchor=north west][inner sep=0.75pt] [font=\Huge]  [align=left] {$\displaystyle e'_{0}$};

% % Text Node
% \draw  [draw opacity=0][fill={rgb, 255:red, 255; green, 255; blue, 255 }  ,fill opacity=1 ]  (856,312) -- (958,312) -- (958,344) -- (856,344) -- cycle  ;
% \draw (857,313) node [anchor=north west][inner sep=0.75pt]  [font=\Huge] [align=left] {$\displaystyle Node\ n+1$};

% Text Node
\draw  [draw opacity=0][fill={rgb, 255:red, 255; green, 255; blue, 255 }  ,fill opacity=1 ]  (691,407) -- (793,407) -- (793,439) -- (691,439) -- cycle  ;
\draw (692,408) node [anchor=north west][inner sep=0.75pt]  [font=\Huge] [align=left] {$\ddot{v}^2_1$};

%Image [id:dp5113430279365352] 
\draw (100,500) node  {\includegraphics[width=21.75pt,height=22.99pt]{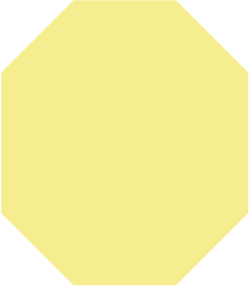}};
%Image [id:dp17299069323517247] 
\draw (100,550) node  {\includegraphics[width=21.75pt,height=25.24pt]{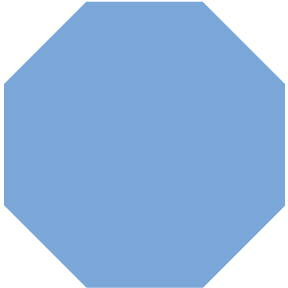}};

Image [id:dp5113430279365352] 
\draw (500,500) node  {\includegraphics[width=21.75pt,height=22.99pt]{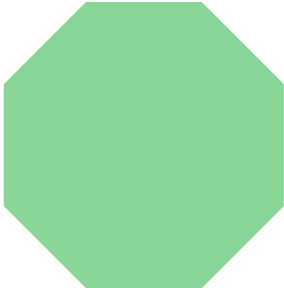}};
%Image [id:dp17299069323517247] 
\draw (500,550) node  {\includegraphics[width=21.75pt,height=25.24pt]{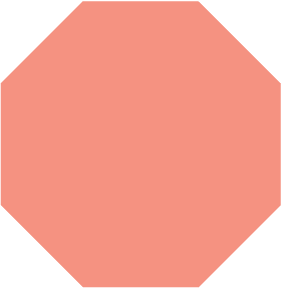}};

% Text Node LEGEND
\draw (120,480) node [anchor=north west][inner sep=0.75pt]  [font=\Huge] [align=left] {$\dot{P}_1=\{\dot{v}^1_1\}$};

\draw (120,530) node [anchor=north west][inner sep=0.75pt]  [font=\Huge] [align=left] {$\dot{P}_2=\{\dot{v}^2_1,\cdots,\dot{v}^2_{n_1-1}\}$};

% Text Node
\draw (520,480) node [anchor=north west][inner sep=0.75pt]  [font=\Huge] [align=left] {$\ddot{P}_1=\{\ddot{v}^1_1\}$};

\draw (520,530) node [anchor=north west][inner sep=0.75pt]  [font=\Huge] [align=left] {$\ddot{P}_2= \{\ddot{v}^2_1,\cdots,\ddot{v}^2_{n_2-1}\}$};

\end{tikzpicture}

%% file: Figures/Fig-02b.tex
\tikzset{every picture/.style={line width=0.75pt}} %set default line width to 0.75pt        

\begin{tikzpicture}[x=0.75pt,y=0.75pt,yscale=-1,xscale=1]
%uncomment if require: \path (0,494); %set diagram left start at 0, and has height of 494

%Image [id:dp29927173165777154] 
\draw (592,238.66) node  {\includegraphics[width=750pt,height=292.5pt]{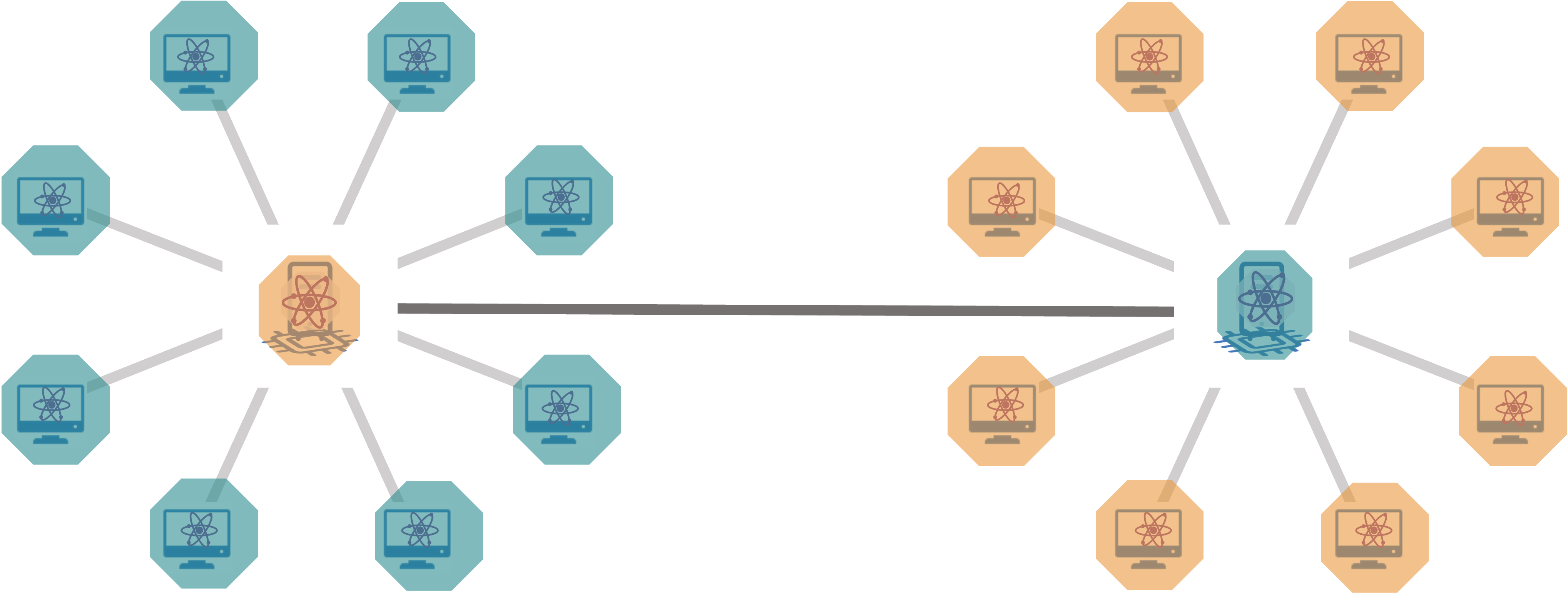}};
%Curve Lines [id:da3787551482280601] 
\draw [color={rgb, 255:red, 126; green, 67; blue, 146 }  ,draw opacity=1 ][line width=1.5]    (360.7,387.16) .. controls (320.54,359.66) and (286.63,280.79) .. (292.63,238.79) ;
%Curve Lines [id:da6125915453510902] 
\draw [color={rgb, 255:red, 126; green, 67; blue, 146 }  ,draw opacity=1 ][line width=1.5]    (458.83,303.76) .. controls (391.36,314.92) and (312.36,290.92) .. (293.43,238.48) ;
%Curve Lines [id:da702415746982079] 
\draw [color={rgb, 255:red, 126; green, 67; blue, 146 }  ,draw opacity=1 ][line width=1.5]    (453.24,169.14) .. controls (430.67,222.05) and (354.36,262.92) .. (293.43,238.48) ;
%Curve Lines [id:da12357547463764584] 
\draw [color={rgb, 255:red, 126; green, 67; blue, 146 }  ,draw opacity=1 ][line width=1.5]    (362.24,76.14) .. controls (368.67,137.71) and (367.36,199.92) .. (293.43,238.48) ;
%Curve Lines [id:da44300270180430723] 
\draw [color={rgb, 255:red, 126; green, 67; blue, 146 }  ,draw opacity=1 ][line width=1.5]    (124.36,305.01) .. controls (143.74,260.24) and (225.74,218.24) .. (292.78,236.33) ;
%Curve Lines [id:da4768945662810957] 
\draw [color={rgb, 255:red, 126; green, 67; blue, 146 }  ,draw opacity=1 ][line width=1.5]    (219.83,386.25) .. controls (197.74,346.24) and (228.74,258.24) .. (292.78,241) ;
%Curve Lines [id:da9908076256414639] 
\draw [color={rgb, 255:red, 126; green, 67; blue, 146 }  ,draw opacity=1 ][line width=1.5]    (216.24,76.14) .. controls (272.05,111.44) and (313.74,165.24) .. (295.72,238.66) ;
%Curve Lines [id:da09905963741432933] 
\draw [color={rgb, 255:red, 126; green, 67; blue, 146 }  ,draw opacity=1 ][line width=1.5]    (123.93,171.94) .. controls (195.74,154.24) and (260.74,164.24) .. (295.72,238.66) ;
%Curve Lines [id:da8165468750110914] 
\draw [color={rgb, 255:red, 126; green, 67; blue, 146 }  ,draw opacity=1 ][line width=1.5]    (971.7,388.16) .. controls (931.54,360.66) and (897.63,281.79) .. (903.63,239.79) ;
%Curve Lines [id:da07557141263597678] 
\draw [color={rgb, 255:red, 126; green, 67; blue, 146 }  ,draw opacity=1 ][line width=1.5]    (1069.83,304.76) .. controls (1002.36,315.92) and (923.36,291.92) .. (904.43,239.48) ;
%Curve Lines [id:da4236339445741386] 
\draw [color={rgb, 255:red, 126; green, 67; blue, 146 }  ,draw opacity=1 ][line width=1.5]    (1064.24,170.14) .. controls (1041.67,223.05) and (965.36,263.92) .. (904.43,239.48) ;
%Curve Lines [id:da49869689548490803] 
\draw [color={rgb, 255:red, 126; green, 67; blue, 146 }  ,draw opacity=1 ][line width=1.5]    (735.36,306.01) .. controls (754.74,261.24) and (836.74,219.24) .. (903.78,237.33) ;
%Curve Lines [id:da97726207240299] 
\draw [color={rgb, 255:red, 126; green, 67; blue, 146 }  ,draw opacity=1 ][line width=1.5]    (830.83,387.25) .. controls (808.74,347.24) and (839.74,259.24) .. (903.78,242) ;
%Curve Lines [id:da295408620413244] 
\draw [color={rgb, 255:red, 126; green, 67; blue, 146 }  ,draw opacity=1 ][line width=1.5]    (827.24,77.14) .. controls (883.05,112.44) and (924.74,166.24) .. (906.72,239.66) ;
%Curve Lines [id:da6694745051288383] 
\draw [color={rgb, 255:red, 126; green, 67; blue, 146 }  ,draw opacity=1 ][line width=1.5]    (734.93,172.94) .. controls (806.74,155.24) and (871.74,165.24) .. (906.72,239.66) ;
%Curve Lines [id:da5560403605823625] 
\draw [color={rgb, 255:red, 126; green, 67; blue, 146 }  ,draw opacity=1 ][line width=1.5]    (972.59,74.99) .. controls (979.02,136.55) and (977.7,198.77) .. (903.78,237.33) ;
%Curve Lines [id:da6645117840796705] 
\draw [color={rgb, 255:red, 126; green, 67; blue, 146 }  ,draw opacity=1 ][line width=1.5]    (584.47,239.63) .. controls (497.47,298.63) and (351.47,268.63) .. (292.78,241) ;
%Curve Lines [id:da15287658015333017] 
\draw [color={rgb, 255:red, 126; green, 67; blue, 146 }  ,draw opacity=1 ][line width=1.5]    (903.78,237.33) .. controls (784.47,206.63) and (664.47,190.63) .. (584.47,239.63) ;

% Text Node
\draw  [draw opacity=0][fill={rgb, 255:red, 255; green, 255; blue, 255 }  ,fill opacity=1 ]  (403,376.39) -- (473,376.39) -- (473,408.39) -- (403,408.39) -- cycle  ;
\draw (404,377.39) node [anchor=north west][inner sep=0.75pt]  [font=\Huge] [align=left] {$\dot{v}^2_{n_1-1}$};
% Text Node
\draw  [draw opacity=0][fill={rgb, 255:red, 255; green, 255; blue, 255 }  ,fill opacity=1 ]  (399.25,51) -- (463.25,51) -- (463.25,83) -- (399.25,83) -- cycle  ;
\draw (400.25,52) node [anchor=north west][inner sep=0.75pt]  [font=\Huge] [align=left] {$\dot{v}^2_i$};

% % Text Node
% \draw  [draw opacity=0][fill={rgb, 255:red, 255; green, 255; blue, 255 }  ,fill opacity=1 ]  (246,256.39) -- (314,256.39) -- (314,288.39) -- (246,288.39) -- cycle  ;
% \draw (247,257.39) node [anchor=north west][inner sep=0.75pt]  [font=\Huge] [align=left] {$\displaystyle Node\ 1$};

% Text Node
\draw  [draw opacity=0][fill={rgb, 255:red, 255; green, 255; blue, 255 }  ,fill opacity=1 ]  (264,170.43) -- (292,170.43) -- (292,202.43) -- (264,202.43) -- cycle  ;
\draw (265,171.43) node [anchor=north west][inner sep=0.75pt]  [font=\Huge] [align=left] {$\dot{v}^1_1$};

% Text Node
\draw  [draw opacity=0][fill={rgb, 255:red, 255; green, 255; blue, 255 }  ,fill opacity=1 ]  (116,373.39) -- (184,373.39) -- (184,405.39) -- (116,405.39) -- cycle  ;
\draw (150,374.39) node [anchor=north west][inner sep=0.75pt]  [font=\Huge] [align=left] {$\dot{v}^2_1$};
% Text Node
\draw  [draw opacity=0][fill={rgb, 255:red, 255; green, 255; blue, 255 }  ,fill opacity=1 ]  (86,237.39) -- (154,237.39) -- (154,269.39) -- (86,269.39) -- cycle  ;
\draw (87,238.39) node [anchor=north west][inner sep=0.75pt]  [font=\Huge] [align=left] {$\dot{v}^2_2$};
% Text Node
\draw  [draw opacity=0][fill={rgb, 255:red, 255; green, 255; blue, 255 }  ,fill opacity=1 ]  (880,171.03) -- (930,171.03) -- (930,203.03) -- (880,203.03) -- cycle  ;
\draw (881,172.03) node [anchor=north west][inner sep=0.75pt] [font=\Huge]  [align=left] {$\ddot{v}^1_1$};
% Text Node
\draw  [draw opacity=0][fill={rgb, 255:red, 255; green, 255; blue, 255 }  ,fill opacity=1 ]  (1006,373) -- (1114,373) -- (1114,405) -- (1006,405) -- cycle  ;
\draw (1007,374) node [anchor=north west][inner sep=0.75pt]  [font=\Huge] [align=left] {$\ddot{v}^2_{n_2-1}$};
% Text Node
\draw  [draw opacity=0][fill={rgb, 255:red, 255; green, 255; blue, 255 }  ,fill opacity=1 ]  (989,103) -- (1087,103) -- (1087,135) -- (989,135) -- cycle  ;
\draw (990,104) node [anchor=north west][inner sep=0.75pt]  [font=\Huge] [align=left] {$\ddot{v}^2_j$};

% % Text Node
% \draw  [draw opacity=0][fill={rgb, 255:red, 255; green, 255; blue, 255 }  ,fill opacity=1 ]  (855,275) -- (957,275) -- (957,307) -- (855,307) -- cycle  ;
% \draw (856,276) node [anchor=north west][inner sep=0.75pt]  [font=\Huge] [align=left] {$\displaystyle Node\ n+1$};

% Text Node
\draw  [draw opacity=0][fill={rgb, 255:red, 255; green, 255; blue, 255 }  ,fill opacity=1 ]  (690,370) -- (792,370) -- (792,402) -- (690,402) -- cycle  ;
\draw (760,371) node [anchor=north west][inner sep=0.75pt]  [font=\Huge] [align=left] {$\ddot{v}^2_1$};

Image [id:dp5113430279365352] 
\draw (100,470) node  {\includegraphics[width=25pt,height=26pt]{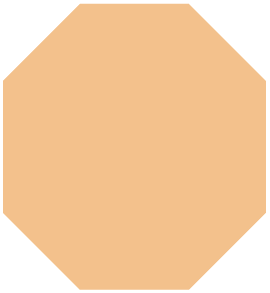}};
%Image [id:dp17299069323517247] 
\draw (100,530) node  {\includegraphics[width=25pt,height=27pt]{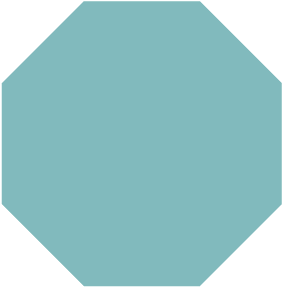}};

% Text Node LEGEND
\draw (120,450) node [anchor=north west][inner sep=0.75pt]  [font=\Huge] [align=left] {$P_1=\{\dot{v}^1_1,\ddot{v}^2_1,\cdots,\ddot{v}^2_{n_2-1}\}$};

\draw (120,510) node [anchor=north west][inner sep=0.75pt]  [font=\Huge] [align=left] {${P}_2=\{\ddot{v}^1_1,\dot{v}^2_1,\cdots,\dot{v}^2_{n_1-1}\}$};

\end{tikzpicture}

%% file: Figures/Fig-05-b.tex
\tikzset{every picture/.style={line width=0.75pt}} %set default line width to 0.75pt        

\begin{tikzpicture}[x=0.75pt,y=0.75pt,yscale=-1,xscale=1]
%uncomment if require: \path (0,494); %set diagram left start at 0, and has height of 494

%Image [id:dp39664821573719655] 
\draw (459,230.5) node  {\includegraphics[width=610.5pt,height=273.75pt]{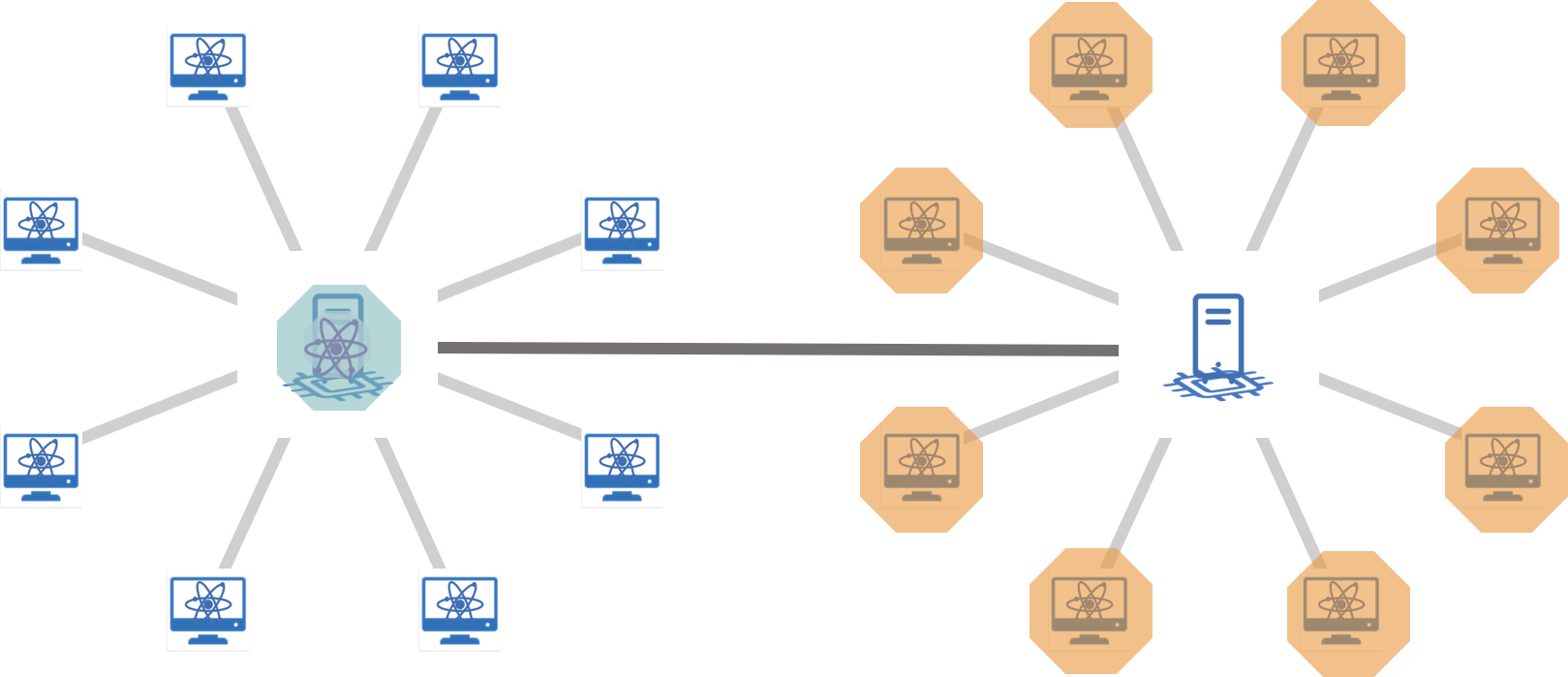}};

%Curve Lines [id:da3357606732957955] 
\draw [color={rgb, 255:red, 126; green, 67; blue, 146 }  ,draw opacity=1 ][line width=2.25]    (243.08,235.23) .. controls (373.08,134.23) and (580.08,76.23) .. (840.08,168.23) ;
%Curve Lines [id:da3565991329946939] 
\draw [color={rgb, 255:red, 126; green, 67; blue, 146 }  ,draw opacity=1 ][line width=2.25]    (235.71,238.86) .. controls (336.08,358.23) and (443.08,378.23) .. (628.08,376.23) ;
%Curve Lines [id:da43370471959377743] 
\draw [color={rgb, 255:red, 126; green, 67; blue, 146 }  ,draw opacity=1 ][line width=2.25]    (243.08,235.23) .. controls (358.08,124.23) and (400.08,89.23) .. (626.08,82.23) ;
%Curve Lines [id:da15925280594503888] 
\draw [color={rgb, 255:red, 126; green, 67; blue, 146 }  ,draw opacity=1 ][line width=2.25]    (243.08,235.23) .. controls (316.08,127.23) and (424.08,16.23) .. (757.08,81.23) ;
%Curve Lines [id:da8410402494198265] 
\draw [color={rgb, 255:red, 126; green, 67; blue, 146 }  ,draw opacity=1 ][line width=2.25]    (235.71,238.86) .. controls (369.08,149.23) and (480.08,157.23) .. (541.08,171.23) ;
%Curve Lines [id:da9866628497916163] 
\draw [color={rgb, 255:red, 126; green, 67; blue, 146 }  ,draw opacity=1 ][line width=2.25]    (540.08,300.23) .. controls (434.08,324.23) and (299.08,285.23) .. (235.71,238.86) ;
%Curve Lines [id:da88313220728278] 
\draw [color={rgb, 255:red, 126; green, 67; blue, 146 }  ,draw opacity=1 ][line width=2.25]    (235.71,238.86) .. controls (320.08,384.23) and (464.08,437.23) .. (758.08,378.23) ;
%Curve Lines [id:da7675811904293589] 
\draw [color={rgb, 255:red, 126; green, 67; blue, 146 }  ,draw opacity=1 ][line width=2.25]    (235.71,238.86) .. controls (377.08,360.23) and (597.08,374.23) .. (841.08,302.23) ;

%Shape: Polygon [id:dp8758498485740364] 
\draw  [color={rgb, 255:red, 126; green, 67; blue, 146 }  ,draw opacity=1 ][line width=2.25]  (837.79,292.4) -- (749.59,379.82) -- (625.56,379.27) -- (538.36,291.07) -- (539.07,166.89) -- (627.27,79.47) -- (751.3,80.02) -- (838.5,168.22) -- cycle ;
%Curve Lines [id:da31297034689336345] 
\draw [color={rgb, 255:red, 126; green, 67; blue, 146 }  ,draw opacity=1 ][line width=2.25]    (538.94,166.9) .. controls (607.38,181.52) and (690.89,155.7) .. (751.16,79.99) ;
%Curve Lines [id:da7659807161652639] 
\draw [color={rgb, 255:red, 126; green, 67; blue, 146 }  ,draw opacity=1 ][line width=2.25]    (627.11,79.47) .. controls (659.38,112.92) and (780.59,170.41) .. (838.41,168.16) ;
%Curve Lines [id:da9891950424386065] 
\draw [color={rgb, 255:red, 126; green, 67; blue, 146 }  ,draw opacity=1 ][line width=2.25]    (627.11,79.47) .. controls (630.49,147.11) and (764.27,262.8) .. (837.76,292.33) ;
%Curve Lines [id:da2733059375078791] 
\draw [color={rgb, 255:red, 126; green, 67; blue, 146 }  ,draw opacity=1 ][line width=2.25]    (627.11,79.47) .. controls (605.37,147.15) and (676.1,350.23) .. (749.59,379.76) ;
%Curve Lines [id:da7150990954023576] 
\draw [color={rgb, 255:red, 126; green, 67; blue, 146 }  ,draw opacity=1 ][line width=2.25]    (627.11,79.47) .. controls (586.34,148.42) and (589.33,301.84) .. (625.55,379.24) ;
%Curve Lines [id:da8165202438084969] 
\draw [color={rgb, 255:red, 126; green, 67; blue, 146 }  ,draw opacity=1 ][line width=2.25]    (627.11,79.47) .. controls (568.95,153.47) and (547.52,201.38) .. (538.29,291.07) ;
%Curve Lines [id:da027943686853034166] 
\draw [color={rgb, 255:red, 126; green, 67; blue, 146 }  ,draw opacity=1 ][line width=2.25]    (538.29,291.07) .. controls (649.04,258.55) and (694.13,229.84) .. (751.16,79.99) ;
%Curve Lines [id:da9653510515288823] 
\draw [color={rgb, 255:red, 126; green, 67; blue, 146 }  ,draw opacity=1 ][line width=2.25]    (625.55,379.24) .. controls (730.99,283.68) and (758.98,235.88) .. (751.16,79.99) ;
%Curve Lines [id:da49723167561043125] 
\draw [color={rgb, 255:red, 126; green, 67; blue, 146 }  ,draw opacity=1 ][line width=2.25]    (749.59,379.76) .. controls (772.23,325.69) and (779.74,140.95) .. (751.16,79.99) ;
%Curve Lines [id:da019055279849695794] 
\draw [color={rgb, 255:red, 126; green, 67; blue, 146 }  ,draw opacity=1 ][line width=2.25]    (837.76,292.33) .. controls (835.06,226.36) and (789.79,135.9) .. (751.16,79.99) ;
%Curve Lines [id:da9457787841330932] 
\draw [color={rgb, 255:red, 126; green, 67; blue, 146 }  ,draw opacity=1 ][line width=2.25]    (538.94,166.9) .. controls (571.21,200.35) and (796.91,214.76) .. (838.41,168.16) ;
%Curve Lines [id:da1446411231518192] 
\draw [color={rgb, 255:red, 126; green, 67; blue, 146 }  ,draw opacity=1 ][line width=2.25]    (538.29,291.07) .. controls (570.56,324.52) and (796.91,214.76) .. (838.41,168.16) ;
%Curve Lines [id:da6037442888235345] 
\draw [color={rgb, 255:red, 126; green, 67; blue, 146 }  ,draw opacity=1 ][line width=2.25]    (625.55,379.24) .. controls (674,372.23) and (817.38,235.26) .. (838.41,168.16) ;
%Curve Lines [id:da15329989209009465] 
\draw [color={rgb, 255:red, 126; green, 67; blue, 146 }  ,draw opacity=1 ][line width=2.25]    (749.59,379.76) .. controls (787.94,327.43) and (826.74,269.62) .. (838.41,168.16) ;
%Curve Lines [id:da19876031356038026] 
\draw [color={rgb, 255:red, 126; green, 67; blue, 146 }  ,draw opacity=1 ][line width=2.25]    (538.29,291.07) .. controls (607.02,358.67) and (651.42,369.37) .. (749.59,379.76) ;
%Curve Lines [id:da1686123923426217] 
\draw [color={rgb, 255:red, 126; green, 67; blue, 146 }  ,draw opacity=1 ][line width=2.25]    (538.29,291.07) .. controls (650.81,351.91) and (747.52,360.75) .. (837.76,292.33) ;
%Curve Lines [id:da11370135590700114] 
\draw [color={rgb, 255:red, 126; green, 67; blue, 146 }  ,draw opacity=1 ][line width=2.25]    (538.94,166.9) .. controls (607.85,271.69) and (702.51,307.45) .. (837.76,292.33) ;
%Curve Lines [id:da5647017448283775] 
\draw [color={rgb, 255:red, 126; green, 67; blue, 146 }  ,draw opacity=1 ][line width=2.25]    (538.94,166.9) .. controls (566.17,260.81) and (612.72,304.17) .. (749.59,379.76) ;

%-----------Labels nodes

% Text Node
\draw (185,180) node [anchor=north west][inner sep=0.75pt]  [font=\Huge] [align=left] {$ \dot{v}_{1}^{1}$};

% Text Node

\draw (157,420) node [anchor=north west][inner sep=0.75pt]  [font=\Huge] [align=left] {$ \dot{v}_{1}^{2}$};
% Text Node
\draw (70,340) node [anchor=north west][inner sep=0.75pt]  [font=\Huge] [align=left] {$\dot{v}_{2}^{2}$};
% Text Node

\draw (272,420) node [anchor=north west][inner sep=0.75pt]  [font=\Huge] [align=left] {$\dot{v}_{n_{1} -1}^{2}$};

% Text Node

\draw (725,180) node [anchor=north west][inner sep=0.75pt]  [font=\Huge] [align=left] {$\ddot{v}_{1}^{1}$};
% Text Node

\draw (620,400) node [anchor=north west][inner sep=0.75pt]  [font=\Huge] [align=left] {$\ddot{v}_{1}^{2}$};
% Text Node

\draw (810,350) node [anchor=north west][inner sep=0.75pt]  [font=\Huge] [align=left] {$\ddot{v}_{i}^{2}$};

\draw (750,400) node [anchor=north west][inner sep=0.75pt]  [font=\Huge] [align=left] {$\ddot{v}_{n_{2} -1}^{2}$};

\end{tikzpicture}

%% file: Figures/Fig-05-c.tex
\tikzset{every picture/.style={line width=0.75pt}} %set default line width to 0.75pt        

\begin{tikzpicture}[x=0.75pt,y=0.75pt,yscale=-1,xscale=1]
%uncomment if require: \path (0,494); %set diagram left start at 0, and has height of 494

%Image [id:dp39664821573719655] 
\draw (459,230.5) node  {\includegraphics[width=610.5pt,height=273.75pt]{Figures/Fig-4.a.png}};

%--------- Links

%Curve Lines [id:da5742208242934348] 
\draw [color={rgb, 255:red, 126; green, 67; blue, 146 }  ,draw opacity=1 ][line width=2.25]    (847.68,299.41) .. controls (676.97,397.27) and (404.92,410.73) .. (240.65,239.7) ;
%Curve Lines [id:da9997500153063955] 
\draw [color={rgb, 255:red, 126; green, 67; blue, 146 }  ,draw opacity=1 ][line width=2.25]    (847.68,299.41) .. controls (706.74,306.78) and (650.61,303.52) .. (543.61,173.21) ;
%Curve Lines [id:da45918724877979744] 
\draw [color={rgb, 255:red, 126; green, 67; blue, 146 }  ,draw opacity=1 ][line width=2.25]    (847.68,299.41) .. controls (831.74,343.78) and (796.61,356.21) .. (758.61,374.21) ;
%Curve Lines [id:da07214822810457133] 
\draw [color={rgb, 255:red, 126; green, 67; blue, 146 }  ,draw opacity=1 ][line width=2.25]    (847.68,299.41) .. controls (773.68,324.41) and (672.57,376.74) .. (543.61,301.21) ;
%Curve Lines [id:da6606780538711133] 
\draw [color={rgb, 255:red, 126; green, 67; blue, 146 }  ,draw opacity=1 ][line width=2.25]    (847.68,299.41) .. controls (736.74,274.78) and (672.67,227.58) .. (629.61,84.21) ;
%Curve Lines [id:da3584886728702822] 
\draw [color={rgb, 255:red, 126; green, 67; blue, 146 }  ,draw opacity=1 ][line width=2.25]    (847.68,299.41) .. controls (803.74,354.78) and (716.61,380.21) .. (628.61,378.21) ;
%Curve Lines [id:da330338745914573] 
\draw [color={rgb, 255:red, 126; green, 67; blue, 146 }  ,draw opacity=1 ][line width=2.25]    (847.68,299.41) .. controls (769.74,246.78) and (746.61,169.21) .. (757.61,81.21) ;
%Curve Lines [id:da9005140178988471] 
\draw [color={rgb, 255:red, 126; green, 67; blue, 146 }  ,draw opacity=1 ][line width=2.25]    (847.68,299.41) .. controls (819.74,232.78) and (804.61,227.21) .. (840.61,171.21) ;

%-----------Labels nodes

% Text Node
\draw (185,180) node [anchor=north west][inner sep=0.75pt]  [font=\Huge] [align=left] {$ \dot{v}_{1}^{1}$};

% Text Node

\draw (165,420) node [anchor=north west][inner sep=0.75pt]  [font=\Huge] [align=left] {$ \dot{v}_{1}^{2}$};
% Text Node
\draw (70,340) node [anchor=north west][inner sep=0.75pt]  [font=\Huge] [align=left] {$\dot{v}_{2}^{2}$};

% % Text Node
% \draw (70,200) node [anchor=north west][inner sep=0.75pt]  [font=\Huge] [align=left] {$\dot{v}_{3}^{2}$};

% Text Node

\draw (272,420) node [anchor=north west][inner sep=0.75pt]  [font=\Huge] [align=left] {$\dot{v}_{n_{1} -1}^{2}$};

% Text Node

\draw (810,350) node [anchor=north west][inner sep=0.75pt]  [font=\Huge] [align=left] {$\ddot{v}_{i}^{2}$};

% Text Node

\draw (725,180) node [anchor=north west][inner sep=0.75pt]  [font=\Huge] [align=left] {$\ddot{v}_{1}^{1}$};
% Text Node

\draw (620,400) node [anchor=north west][inner sep=0.75pt]  [font=\Huge] [align=left] {$\ddot{v}_{1}^{2}$};
% Text Node

% \draw (520,400) node [anchor=north west][inner sep=0.75pt]  [font=\Huge] [align=left] {$\ddot{v}_{2}^{2}$};
% % Text Node

\draw (750,410) node [anchor=north west][inner sep=0.75pt]  [font=\Huge] [align=left] {$\ddot{v}_{n_{2} -1}^{2}$};

\end{tikzpicture}

%% file: Figures/Fig-05.a.tex
\tikzset{every picture/.style={line width=0.75pt}} %set default line width to 0.75pt        

\begin{tikzpicture}[x=0.75pt,y=0.75pt,yscale=-1,xscale=1]
%uncomment if require: \path (0,494); %set diagram left start at 0, and has height of 494

%Image [id:dp39664821573719655] 
\draw (459,230.5) node  {\includegraphics[width=610.5pt,height=273.75pt]{Figures/Fig-4.a.png}};

%Curve Lines [id:da3357606732957955] 
\draw [color={rgb, 255:red, 126; green, 67; blue, 146 }  ,draw opacity=1 ][line width=2.25]    (243.08,235.23) .. controls (373.08,134.23) and (580.08,76.23) .. (840.08,168.23) ;
%Curve Lines [id:da3565991329946939] 
\draw [color={rgb, 255:red, 126; green, 67; blue, 146 }  ,draw opacity=1 ][line width=2.25]    (235.71,238.86) .. controls (336.08,358.23) and (443.08,378.23) .. (628.08,376.23) ;
%Curve Lines [id:da43370471959377743] 
\draw [color={rgb, 255:red, 126; green, 67; blue, 146 }  ,draw opacity=1 ][line width=2.25]    (243.08,235.23) .. controls (358.08,124.23) and (400.08,89.23) .. (626.08,82.23) ;
%Curve Lines [id:da15925280594503888] 
\draw [color={rgb, 255:red, 126; green, 67; blue, 146 }  ,draw opacity=1 ][line width=2.25]    (243.08,235.23) .. controls (316.08,127.23) and (424.08,16.23) .. (757.08,81.23) ;
%Curve Lines [id:da8410402494198265] 
\draw [color={rgb, 255:red, 126; green, 67; blue, 146 }  ,draw opacity=1 ][line width=2.25]    (235.71,238.86) .. controls (369.08,149.23) and (480.08,157.23) .. (541.08,171.23) ;
%Curve Lines [id:da9866628497916163] 
\draw [color={rgb, 255:red, 126; green, 67; blue, 146 }  ,draw opacity=1 ][line width=2.25]    (540.08,300.23) .. controls (434.08,324.23) and (299.08,285.23) .. (235.71,238.86) ;
%Curve Lines [id:da88313220728278] 
\draw [color={rgb, 255:red, 126; green, 67; blue, 146 }  ,draw opacity=1 ][line width=2.25]    (235.71,238.86) .. controls (320.08,384.23) and (464.08,437.23) .. (758.08,378.23) ;
%Curve Lines [id:da7675811904293589] 
\draw [color={rgb, 255:red, 126; green, 67; blue, 146 }  ,draw opacity=1 ][line width=2.25]    (235.71,238.86) .. controls (377.08,360.23) and (597.08,374.23) .. (841.08,302.23) ;

%-----------Labels nodes

% Text Node
\draw (185,180) node [anchor=north west][inner sep=0.75pt]  [font=\Huge] [align=left] {$ \dot{v}_{1}^{1}$};

% Text Node

\draw (157,420) node [anchor=north west][inner sep=0.75pt]  [font=\Huge] [align=left] {$ \dot{v}_{1}^{2}$};
% Text Node
\draw (70,340) node [anchor=north west][inner sep=0.75pt]  [font=\Huge] [align=left] {$\dot{v}_{2}^{2}$};
% Text Node

\draw (272,420) node [anchor=north west][inner sep=0.75pt]  [font=\Huge] [align=left] {$\dot{v}_{n_{1} -1}^{2}$};

% Text Node

\draw (725,180) node [anchor=north west][inner sep=0.75pt]  [font=\Huge] [align=left] {$\ddot{v}_{1}^{1}$};
% Text Node

\draw (620,400) node [anchor=north west][inner sep=0.75pt]  [font=\Huge] [align=left] {$\ddot{v}_{1}^{2}$};
% Text Node
% Text Node

\draw (810,350) node [anchor=north west][inner sep=0.75pt]  [font=\Huge] [align=left] {$\ddot{v}_{i}^{2}$};

% \draw (520,400) node [anchor=north west][inner sep=0.75pt]  [font=\Huge] [align=left] {$\ddot{v}_{2}^{2}$};
% % Text Node

\draw (750,400) node [anchor=north west][inner sep=0.75pt]  [font=\Huge] [align=left] {$\ddot{v}_{n_{2} -1}^{2}$};

\end{tikzpicture}

%% file: Figures/Fig-07.a.tex
\tikzset{every picture/.style={line width=0.75pt}} %set default line width to 0.75pt        

\begin{tikzpicture}[x=0.75pt,y=0.75pt,yscale=-1,xscale=1]
%uncomment if require: \path (0,1453); %set diagram left start at 0, and has height of 1453

%Image [id:dp07338920357128909] 
\draw (531.33,241.99) node  {\includegraphics[width=612pt,height=270.75pt]{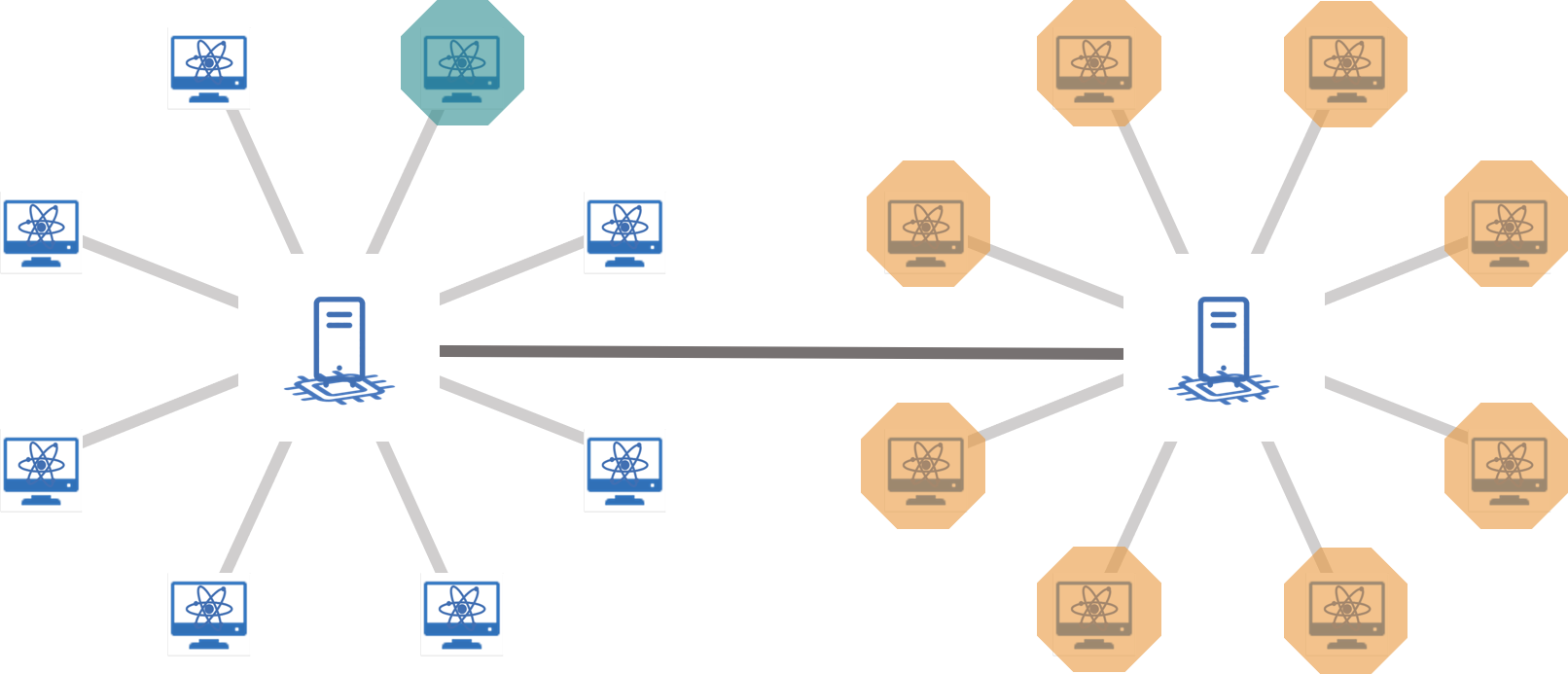}};
%Curve Lines [id:da8894377129064395] 
\draw [color={rgb, 255:red, 126; green, 67; blue, 146 }  ,draw opacity=1 ][line width=2.25]    (362.71,94.35) .. controls (566.33,5.49) and (739.58,47) .. (812.3,93.51) ;
%Curve Lines [id:da30327161280852066] 
\draw [color={rgb, 255:red, 126; green, 67; blue, 146 }  ,draw opacity=1 ][line width=2.25]    (362.71,94.35) .. controls (596.33,7.49) and (809.33,99.49) .. (901.62,181.38) ;
%Curve Lines [id:da13412374829339035] 
\draw [color={rgb, 255:red, 126; green, 67; blue, 146 }  ,draw opacity=1 ][line width=2.25]    (370.1,93.7) .. controls (597.33,36.49) and (868.79,235.89) .. (898.79,305.89) ;
%Curve Lines [id:da8249309942334604] 
\draw [color={rgb, 255:red, 126; green, 67; blue, 146 }  ,draw opacity=1 ][line width=2.25]    (362.71,94.35) .. controls (505.33,63.49) and (803.4,188.54) .. (818.4,390.68) ;
%Curve Lines [id:da7797227161369342] 
\draw [color={rgb, 255:red, 126; green, 67; blue, 146 }  ,draw opacity=1 ][line width=2.25]    (362.71,94.35) .. controls (573.33,138.49) and (623.33,255.49) .. (686.56,392.76) ;
%Curve Lines [id:da9265098541999388] 
\draw [color={rgb, 255:red, 126; green, 67; blue, 146 }  ,draw opacity=1 ][line width=2.25]    (362.71,94.35) .. controls (481.33,155.49) and (536.33,193.49) .. (599.29,304.56) ;
%Curve Lines [id:da8054802411104098] 
\draw [color={rgb, 255:red, 126; green, 67; blue, 146 }  ,draw opacity=1 ][line width=2.25]    (362.71,94.35) .. controls (524.33,59.49) and (612.33,71.49) .. (688.27,92.96) ;
%Shape: Polygon [id:dp30183190208130717] 
\draw  [color={rgb, 255:red, 126; green, 67; blue, 146 }  ,draw opacity=1 ][line width=2.25]  (898.79,305.89) -- (810.59,393.31) -- (686.56,392.76) -- (599.36,304.56) -- (600.07,180.38) -- (688.27,92.96) -- (812.3,93.51) -- (899.5,181.71) -- cycle ;
%Curve Lines [id:da9438708248125147] 
\draw [color={rgb, 255:red, 126; green, 67; blue, 146 }  ,draw opacity=1 ][line width=2.25]    (599.94,180.39) .. controls (668.38,195.01) and (751.89,169.19) .. (812.16,93.48) ;
%Curve Lines [id:da45720562801293174] 
\draw [color={rgb, 255:red, 126; green, 67; blue, 146 }  ,draw opacity=1 ][line width=2.25]    (688.11,92.96) .. controls (720.38,126.41) and (841.59,183.9) .. (899.41,181.65) ;
%Curve Lines [id:da6927795661334457] 
\draw [color={rgb, 255:red, 126; green, 67; blue, 146 }  ,draw opacity=1 ][line width=2.25]    (688.11,92.96) .. controls (691.49,160.6) and (825.27,276.29) .. (898.76,305.82) ;
%Curve Lines [id:da08450649024456669] 
\draw [color={rgb, 255:red, 126; green, 67; blue, 146 }  ,draw opacity=1 ][line width=2.25]    (688.11,92.96) .. controls (666.37,160.64) and (737.1,363.72) .. (810.59,393.25) ;
%Curve Lines [id:da5011766127817763] 
\draw [color={rgb, 255:red, 126; green, 67; blue, 146 }  ,draw opacity=1 ][line width=2.25]    (688.11,92.96) .. controls (647.34,161.91) and (650.33,315.33) .. (686.55,392.73) ;
%Curve Lines [id:da7250297276567421] 
\draw [color={rgb, 255:red, 126; green, 67; blue, 146 }  ,draw opacity=1 ][line width=2.25]    (688.11,92.96) .. controls (629.95,166.96) and (608.52,214.87) .. (599.29,304.56) ;
%Curve Lines [id:da29890995720745805] 
\draw [color={rgb, 255:red, 126; green, 67; blue, 146 }  ,draw opacity=1 ][line width=2.25]    (599.29,304.56) .. controls (710.04,272.04) and (755.13,243.33) .. (812.16,93.48) ;
%Curve Lines [id:da2736811279635083] 
\draw [color={rgb, 255:red, 126; green, 67; blue, 146 }  ,draw opacity=1 ][line width=2.25]    (686.55,392.73) .. controls (791.99,297.17) and (819.98,249.37) .. (812.16,93.48) ;
%Curve Lines [id:da6328036391455891] 
\draw [color={rgb, 255:red, 126; green, 67; blue, 146 }  ,draw opacity=1 ][line width=2.25]    (810.59,393.25) .. controls (833.23,339.18) and (840.74,154.44) .. (812.16,93.48) ;
%Curve Lines [id:da8498059802151203] 
\draw [color={rgb, 255:red, 126; green, 67; blue, 146 }  ,draw opacity=1 ][line width=2.25]    (898.76,305.82) .. controls (896.06,239.85) and (850.79,149.39) .. (812.16,93.48) ;
%Curve Lines [id:da4431138081060708] 
\draw [color={rgb, 255:red, 126; green, 67; blue, 146 }  ,draw opacity=1 ][line width=2.25]    (599.94,180.39) .. controls (632.21,213.84) and (857.91,228.25) .. (899.41,181.65) ;
%Curve Lines [id:da10553666385963378] 
\draw [color={rgb, 255:red, 126; green, 67; blue, 146 }  ,draw opacity=1 ][line width=2.25]    (599.29,304.56) .. controls (631.56,338.01) and (857.91,228.25) .. (899.41,181.65) ;
%Curve Lines [id:da5421219918119924] 
\draw [color={rgb, 255:red, 126; green, 67; blue, 146 }  ,draw opacity=1 ][line width=2.25]    (686.55,392.73) .. controls (735,385.72) and (878.38,248.75) .. (899.41,181.65) ;
%Curve Lines [id:da5064649658362033] 
\draw [color={rgb, 255:red, 126; green, 67; blue, 146 }  ,draw opacity=1 ][line width=2.25]    (810.59,393.25) .. controls (848.94,340.92) and (887.74,283.11) .. (899.41,181.65) ;
%Curve Lines [id:da11199916515907904] 
\draw [color={rgb, 255:red, 126; green, 67; blue, 146 }  ,draw opacity=1 ][line width=2.25]    (599.29,304.56) .. controls (668.02,372.16) and (712.42,382.86) .. (810.59,393.25) ;
%Curve Lines [id:da7324922264244318] 
\draw [color={rgb, 255:red, 126; green, 67; blue, 146 }  ,draw opacity=1 ][line width=2.25]    (599.29,304.56) .. controls (711.81,365.4) and (808.52,374.24) .. (898.76,305.82) ;
%Curve Lines [id:da7466625087153311] 
\draw [color={rgb, 255:red, 126; green, 67; blue, 146 }  ,draw opacity=1 ][line width=2.25]    (599.94,180.39) .. controls (668.85,285.18) and (763.51,320.94) .. (898.76,305.82) ;
%Curve Lines [id:da7144906158955804] 
\draw [color={rgb, 255:red, 126; green, 67; blue, 146 }  ,draw opacity=1 ][line width=2.25]    (599.94,180.39) .. controls (627.17,274.3) and (673.72,317.66) .. (810.59,393.25) ;

%Curve Lines [id:da33408832252332354] 
\draw [color={rgb, 255:red, 126; green, 67; blue, 146 }  ,draw opacity=1 ][line width=2.25]    (362.71,94.35) .. controls (435.33,98.49) and (539.33,113.49) .. (600.07,180.38) ;

\draw (300,100) node [anchor=north west][inner sep=0.75pt]  [font=\Huge] [align=left] {$ \dot{v}_{j}^{2}$};

% Text Node
\draw (250,200) node [anchor=north west][inner sep=0.75pt]  [font=\Huge] [align=left] {$ \dot{v}_{1}^{1}$};

% Text Node

\draw (215,430) node [anchor=north west][inner sep=0.75pt]  [font=\Huge] [align=left] {$ \dot{v}_{1}^{2}$};
% Text Node
\draw (90,350) node [anchor=north west][inner sep=0.75pt]  [font=\Huge] [align=left] {$\dot{v}_{2}^{2}$};
% Text Node

\draw (355,430) node [anchor=north west][inner sep=0.75pt]  [font=\Huge] [align=left] {$\dot{v}_{n_{1} -1}^{2}$};

% Text Node

\draw (740,200) node [anchor=north west][inner sep=0.75pt]  [font=\Huge] [align=left] {$\ddot{v}_{1}^{1}$};
% Text Node

\draw (650,430) node [anchor=north west][inner sep=0.75pt]  [font=\Huge] [align=left] {$\ddot{v}_{1}^{2}$};
% Text Node

\draw (800,430) node [anchor=north west][inner sep=0.75pt]  [font=\Huge] [align=left] {$\ddot{v}_{n_{2} -1}^{2}$};

\draw (900,350) node [anchor=north west][inner sep=0.75pt]  [font=\Huge] [align=left] {$\ddot{v}_{i}^{2}$};

\end{tikzpicture}

%% file: Figures/Fig-07.b.tex
\tikzset{every picture/.style={line width=0.75pt}} %set default line width to 0.75pt        

\begin{tikzpicture}[x=0.75pt,y=0.75pt,yscale=-1,xscale=1]
%uncomment if require: \path (0,504); %set diagram left start at 0, and has height of 504

%Image [id:dp5330508282073316] 
\draw (511,245.66) node  {\includegraphics[width=612pt,height=270.75pt]{Figures/Fig-4.b.png}};
%Curve Lines [id:da773779309718726] 
\draw [color={rgb, 255:red, 126; green, 67; blue, 146 }  ,draw opacity=1 ][line width=2.25]    (352.77,100.37) .. controls (657,106.16) and (781,240.16) .. (881.46,312.56) ;
%Curve Lines [id:da5420380554337849] 
\draw [color={rgb, 255:red, 126; green, 67; blue, 146 }  ,draw opacity=1 ][line width=2.25]    (670.78,99.62) .. controls (763,130.16) and (837,238.16) .. (881.43,312.48) ;
%Curve Lines [id:da38602723834616626] 
\draw [color={rgb, 255:red, 126; green, 67; blue, 146 }  ,draw opacity=1 ][line width=2.25]    (881.43,312.48) .. controls (878.73,246.51) and (846.63,160.07) .. (808,104.16) ;
%Curve Lines [id:da7134755098805916] 
\draw [color={rgb, 255:red, 126; green, 67; blue, 146 }  ,draw opacity=1 ][line width=2.25]    (590,316.16) .. controls (717.04,349.09) and (763,362.16) .. (881.43,312.48) ;
%Curve Lines [id:da03363083159018232] 
\draw [color={rgb, 255:red, 126; green, 67; blue, 146 }  ,draw opacity=1 ][line width=2.25]    (588,192.16) .. controls (666,256.16) and (738,316.16) .. (881.43,312.48) ;
%Curve Lines [id:da625374927248445] 
\draw [color={rgb, 255:red, 126; green, 67; blue, 146 }  ,draw opacity=1 ][line width=2.25]    (881.46,312.56) .. controls (895,254.16) and (893,258.16) .. (883,189.16) ;
%Curve Lines [id:da2134533664796039] 
\draw [color={rgb, 255:red, 126; green, 67; blue, 146 }  ,draw opacity=1 ][line width=2.25]    (881.46,312.56) .. controls (857,361.16) and (845,378.16) .. (802,396.16) ;
%Curve Lines [id:da6273468984981478] 
\draw [color={rgb, 255:red, 126; green, 67; blue, 146 }  ,draw opacity=1 ][line width=2.25]    (881.46,312.56) .. controls (847,354.16) and (752,386.16) .. (692,394.16) ;

\draw (280,100) node [anchor=north west][inner sep=0.75pt]  [font=\Huge] [align=left] {$ \dot{v}_{j}^{2}$};

% Text Node
\draw (230,200) node [anchor=north west][inner sep=0.75pt]  [font=\Huge] [align=left] {$ \dot{v}_{1}^{1}$};

% Text Node

\draw (195,430) node [anchor=north west][inner sep=0.75pt]  [font=\Huge] [align=left] {$ \dot{v}_{1}^{2}$};
% Text Node
\draw (70,350) node [anchor=north west][inner sep=0.75pt]  [font=\Huge] [align=left] {$\dot{v}_{2}^{2}$};
% Text Node

\draw (335,430) node [anchor=north west][inner sep=0.75pt]  [font=\Huge] [align=left] {$\dot{v}_{n_{1} -1}^{2}$};

% Text Node

\draw (700,200) node [anchor=north west][inner sep=0.75pt]  [font=\Huge] [align=left] {$\ddot{v}_{1}^{1}$};
% Text Node

\draw (630,430) node [anchor=north west][inner sep=0.75pt]  [font=\Huge] [align=left] {$\ddot{v}_{1}^{2}$};
% Text Node

\draw (780,430) node [anchor=north west][inner sep=0.75pt]  [font=\Huge] [align=left] {$\ddot{v}_{n_{2} -1}^{2}$};

\draw (880,350) node [anchor=north west][inner sep=0.75pt]  [font=\Huge] [align=left] {$\ddot{v}_{i}^{2}$};

\end{tikzpicture}

%% file: Figures/Fig-07.c.tex
\tikzset{every picture/.style={line width=0.75pt}} %set default line width to 0.75pt        

\begin{tikzpicture}[x=0.75pt,y=0.75pt,yscale=-1,xscale=1]
%uncomment if require: \path (0,1453); %set diagram left start at 0, and has height of 1453

%Image [id:dp365091129283001] 
\draw (525.33,689.32) node  {\includegraphics[width=612pt,height=270.75pt]{Figures/Fig-4.b.png}};
%Curve Lines [id:da1877919181564136] 
\draw [color={rgb, 255:red, 126; green, 67; blue, 146 }  ,draw opacity=1 ][line width=2.25]    (356.71,541.68) .. controls (560.33,452.82) and (733.58,494.33) .. (806.3,540.84) ;
%Curve Lines [id:da23153511720085773] 
\draw [color={rgb, 255:red, 126; green, 67; blue, 146 }  ,draw opacity=1 ][line width=2.25]    (356.71,541.68) .. controls (590.33,454.82) and (803.33,546.82) .. (895.62,628.72) ;
%Curve Lines [id:da9322719452895378] 
\draw [color={rgb, 255:red, 126; green, 67; blue, 146 }  ,draw opacity=1 ][line width=2.25]    (364.1,541.04) .. controls (591.33,483.82) and (862.79,683.22) .. (892.79,753.22) ;
%Curve Lines [id:da5226463452580747] 
\draw [color={rgb, 255:red, 126; green, 67; blue, 146 }  ,draw opacity=1 ][line width=2.25]    (356.71,541.68) .. controls (499.33,510.82) and (797.4,635.87) .. (812.4,838.01) ;
%Curve Lines [id:da3644808783061717] 
\draw [color={rgb, 255:red, 126; green, 67; blue, 146 }  ,draw opacity=1 ][line width=2.25]    (356.71,541.68) .. controls (567.33,585.82) and (617.33,702.82) .. (680.56,840.1) ;
%Curve Lines [id:da9751588260530102] 
\draw [color={rgb, 255:red, 126; green, 67; blue, 146 }  ,draw opacity=1 ][line width=2.25]    (356.71,541.68) .. controls (475.33,602.82) and (530.33,640.82) .. (593.29,751.89) ;
%Curve Lines [id:da2145435297982008] 
\draw [color={rgb, 255:red, 126; green, 67; blue, 146 }  ,draw opacity=1 ][line width=2.25]    (356.71,541.68) .. controls (518.33,506.82) and (606.33,518.82) .. (682.27,540.29) ;
%Curve Lines [id:da08092331359809979] 
\draw [color={rgb, 255:red, 126; green, 67; blue, 146 }  ,draw opacity=1 ][line width=2.25]    (356.71,541.68) .. controls (429.33,545.82) and (533.33,560.82) .. (594.07,627.71) ;

\draw (300,550) node [anchor=north west][inner sep=0.75pt]  [font=\Huge] [align=left] {$ \dot{v}_{j}^{2}$};

% Text Node
\draw (250,650) node [anchor=north west][inner sep=0.75pt]  [font=\Huge] [align=left] {$ \dot{v}_{1}^{1}$};

% Text Node

\draw (215,870) node [anchor=north west][inner sep=0.75pt]  [font=\Huge] [align=left] {$ \dot{v}_{1}^{2}$};
% Text Node
\draw (85,800) node [anchor=north west][inner sep=0.75pt]  [font=\Huge] [align=left] {$\dot{v}_{2}^{2}$};
% Text Node

\draw (355,870) node [anchor=north west][inner sep=0.75pt]  [font=\Huge] [align=left] {$\dot{v}_{n_{1} -1}^{2}$};

% Text Node

\draw (765,650) node [anchor=north west][inner sep=0.75pt]  [font=\Huge] [align=left] {$\ddot{v}_{1}^{1}$};
% Text Node

\draw (650,870) node [anchor=north west][inner sep=0.75pt]  [font=\Huge] [align=left] {$\ddot{v}_{1}^{2}$};
% Text Node

\draw (800,870) node [anchor=north west][inner sep=0.75pt]  [font=\Huge] [align=left] {$\ddot{v}_{n_{2} -1}^{2}$};

\draw (900,800) node [anchor=north west][inner sep=0.75pt]  [font=\Huge] [align=left] {$\ddot{v}_{i}^{2}$};

\end{tikzpicture}

%% file: Figures/Fig-10.tex
\tikzset{every picture/.style={line width=0.75pt}} %set default line width to 0.75pt        

\begin{tikzpicture}[x=0.75pt,y=0.75pt,yscale=-1,xscale=1]
%uncomment if require: \path (0,504); %set diagram left start at 0, and has height of 504
%Image [id:dp17497844902940407] 
\draw (485,250) node  {\includegraphics[width=616.5pt,height=274.5pt]{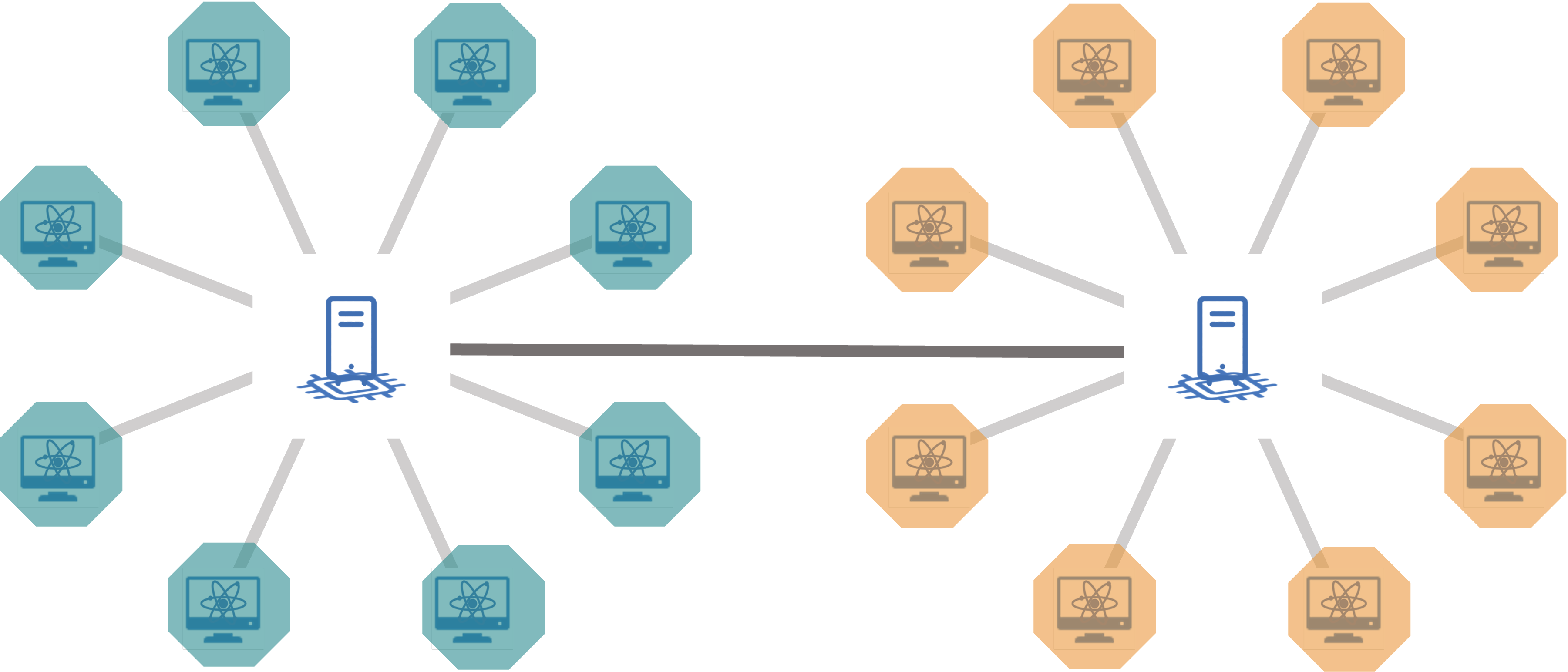}};
%Curve Lines [id:da8070495921675097] 
\draw [color={rgb, 255:red, 126; green, 67; blue, 146 }  ,draw opacity=1 ][line width=2.25]    (324.39,103.9) .. controls (459.22,19.85) and (640.22,15.85) .. (781.22,105.85) ;
%Curve Lines [id:da8708942707722517] 
\draw [color={rgb, 255:red, 126; green, 67; blue, 146 }  ,draw opacity=1 ][line width=2.25]    (324.52,106.52) .. controls (501.22,-4.15) and (718.22,71.85) .. (868.22,196.85) ;
%Curve Lines [id:da216771674757831] 
\draw [color={rgb, 255:red, 126; green, 67; blue, 146 }  ,draw opacity=1 ][line width=2.25]    (324.52,106.52) .. controls (463.22,85.85) and (526.22,133.85) .. (565.22,196.85) ;
%Curve Lines [id:da060267706916656394] 
\draw [color={rgb, 255:red, 126; green, 67; blue, 146 }  ,draw opacity=1 ][line width=2.25]    (324.52,106.52) .. controls (428.36,47.68) and (578.5,61.34) .. (651.22,107.85) ;
%Curve Lines [id:da9119313143381294] 
\draw [color={rgb, 255:red, 126; green, 67; blue, 146 }  ,draw opacity=1 ][line width=2.25]    (652.22,401.56) .. controls (517.39,485.6) and (336.39,489.6) .. (195.39,399.6) ;
%Curve Lines [id:da18904573752797393] 
\draw [color={rgb, 255:red, 126; green, 67; blue, 146 }  ,draw opacity=1 ][line width=2.25]    (652.08,398.93) .. controls (475.39,509.6) and (257.35,447) .. (107.35,322) ;
%Curve Lines [id:da9112457653555257] 
\draw [color={rgb, 255:red, 126; green, 67; blue, 146 }  ,draw opacity=1 ][line width=2.25]    (652.08,398.93) .. controls (513.39,419.6) and (445.71,379.86) .. (406.71,316.86) ;
%Curve Lines [id:da12541782433455473] 
\draw [color={rgb, 255:red, 126; green, 67; blue, 146 }  ,draw opacity=1 ][line width=2.25]    (652.08,398.93) .. controls (548.25,457.77) and (398.1,444.11) .. (325.39,397.6) ;
%Curve Lines [id:da4845430378361111] 
\draw [color={rgb, 255:red, 126; green, 67; blue, 146 }  ,draw opacity=1 ][line width=2.25]    (194.71,100.86) .. controls (352.35,39) and (520.35,-24) .. (781.22,105.85) ;
%Curve Lines [id:da8727306211212] 
\draw [color={rgb, 255:red, 126; green, 67; blue, 146 }  ,draw opacity=1 ][line width=2.25]    (194.71,100.86) .. controls (352.35,39) and (780.35,74) .. (861.08,322.23) ;
%Curve Lines [id:da060217973215935805] 
\draw [color={rgb, 255:red, 126; green, 67; blue, 146 }  ,draw opacity=1 ][line width=2.25]    (194.71,100.86) .. controls (352.35,39) and (781.35,125) .. (778.08,398.23) ;
%Curve Lines [id:da3009880860653137] 
\draw [color={rgb, 255:red, 126; green, 67; blue, 146 }  ,draw opacity=1 ][line width=2.25]    (194.71,100.86) .. controls (352.35,39) and (715.35,127) .. (652.08,398.93) ;
%Curve Lines [id:da8855999003755226] 
\draw [color={rgb, 255:red, 126; green, 67; blue, 146 }  ,draw opacity=1 ][line width=2.25]    (194.71,100.86) .. controls (349.35,73) and (455.35,133) .. (560.08,320.23) ;
%Curve Lines [id:da6876297329261024] 
\draw [color={rgb, 255:red, 126; green, 67; blue, 146 }  ,draw opacity=1 ][line width=2.25]    (107.71,187.86) .. controls (261.35,136) and (447.08,97.23) .. (560.08,320.23) ;
%Curve Lines [id:da2503325943630741] 
\draw [color={rgb, 255:red, 126; green, 67; blue, 146 }  ,draw opacity=1 ][line width=2.25]    (107.71,187.86) .. controls (261.35,136) and (319.35,86) .. (565.22,196.85) ;
%Curve Lines [id:da46293510715815744] 
\draw [color={rgb, 255:red, 126; green, 67; blue, 146 }  ,draw opacity=1 ][line width=2.25]    (107.71,187.86) .. controls (226.35,120) and (281.35,62) .. (644.35,107) ;
%Curve Lines [id:da477134322665303] 
\draw [color={rgb, 255:red, 126; green, 67; blue, 146 }  ,draw opacity=1 ][line width=2.25]    (107.71,187.86) .. controls (226.35,120) and (281.35,62) .. (644.35,107) ;
%Curve Lines [id:da9799471001194598] 
\draw [color={rgb, 255:red, 126; green, 67; blue, 146 }  ,draw opacity=1 ][line width=2.25]    (107.71,187.86) .. controls (176.35,91) and (511.35,-26) .. (776.35,103) ;
%Curve Lines [id:da14528614812400753] 
\draw [color={rgb, 255:red, 126; green, 67; blue, 146 }  ,draw opacity=1 ][line width=2.25]    (107.71,187.86) .. controls (346.35,64) and (603.22,67.85) .. (868.22,196.85) ;
%Curve Lines [id:da03513408512890415] 
\draw [color={rgb, 255:red, 126; green, 67; blue, 146 }  ,draw opacity=1 ][line width=2.25]    (107.71,187.86) .. controls (360.35,51) and (731.35,127) .. (858.35,318) ;
%Curve Lines [id:da1513878405589122] 
\draw [color={rgb, 255:red, 126; green, 67; blue, 146 }  ,draw opacity=1 ][line width=2.25]    (107.71,187.86) .. controls (360.35,51) and (651.08,207.23) .. (778.08,398.23) ;
%Curve Lines [id:da7469691117055364] 
\draw [color={rgb, 255:red, 126; green, 67; blue, 146 }  ,draw opacity=1 ][line width=2.25]    (107.71,187.86) .. controls (309.35,129) and (436.35,151) .. (560.08,320.23) ;
%Curve Lines [id:da01662311406096184] 
\draw [color={rgb, 255:red, 126; green, 67; blue, 146 }  ,draw opacity=1 ][line width=2.25]    (407.35,190) .. controls (471.35,217) and (496.35,222) .. (560.08,320.23) ;
%Curve Lines [id:da16931386144038918] 
\draw [color={rgb, 255:red, 126; green, 67; blue, 146 }  ,draw opacity=1 ][line width=2.25]    (560.08,320.23) .. controls (481.35,345) and (460.35,352) .. (406.71,316.86) ;
%Curve Lines [id:da8303195859619128] 
\draw [color={rgb, 255:red, 126; green, 67; blue, 146 }  ,draw opacity=1 ][line width=2.25]    (778.08,398.23) .. controls (579.35,396) and (460.35,352) .. (406.71,316.86) ;
%Curve Lines [id:da04244706666551756] 
\draw [color={rgb, 255:red, 126; green, 67; blue, 146 }  ,draw opacity=1 ][line width=2.25]    (853.35,319) .. controls (683.35,369) and (567.35,379) .. (411.35,320) ;
%Curve Lines [id:da6197679331575211] 
\draw [color={rgb, 255:red, 126; green, 67; blue, 146 }  ,draw opacity=1 ][line width=2.25]    (862.35,197) .. controls (761.35,304) and (562.71,375.86) .. (406.71,316.86) ;
%Curve Lines [id:da9236478167087268] 
\draw [color={rgb, 255:red, 126; green, 67; blue, 146 }  ,draw opacity=1 ][line width=2.25]    (781.22,105.85) .. controls (764.35,261) and (567.35,379) .. (411.35,320) ;
%Curve Lines [id:da9847194373715091] 
\draw [color={rgb, 255:red, 126; green, 67; blue, 146 }  ,draw opacity=1 ][line width=2.25]    (651.22,107.85) .. controls (634.35,263) and (579.35,308) .. (406.71,316.86) ;
%Curve Lines [id:da9667092491467683] 
\draw [color={rgb, 255:red, 126; green, 67; blue, 146 }  ,draw opacity=1 ][line width=2.25]    (565.22,196.85) .. controls (562.35,247) and (522.35,311) .. (411.35,320) ;
%Curve Lines [id:da6340434933093245] 
\draw [color={rgb, 255:red, 126; green, 67; blue, 146 }  ,draw opacity=1 ][line width=2.25]    (560.08,320.23) .. controls (457.35,391) and (420.35,390) .. (325.39,397.6) ;
%Curve Lines [id:da18306424189960468] 
\draw [color={rgb, 255:red, 126; green, 67; blue, 146 }  ,draw opacity=1 ][line width=2.25]    (778.08,398.23) .. controls (514.35,443) and (440.35,421) .. (325.39,397.6) ;
%Curve Lines [id:da2015755054688796] 
\draw [color={rgb, 255:red, 126; green, 67; blue, 146 }  ,draw opacity=1 ][line width=2.25]    (853.35,319) .. controls (600.35,444) and (440.35,421) .. (325.39,397.6) ;
%Curve Lines [id:da18149081635651676] 
\draw [color={rgb, 255:red, 126; green, 67; blue, 146 }  ,draw opacity=1 ][line width=2.25]    (868.22,196.85) .. controls (657.35,389) and (440.35,421) .. (325.39,397.6) ;
%Curve Lines [id:da46738050626510474] 
\draw [color={rgb, 255:red, 126; green, 67; blue, 146 }  ,draw opacity=1 ][line width=2.25]    (781.22,105.85) .. controls (684.35,322) and (564.35,411) .. (325.39,397.6) ;
%Curve Lines [id:da18707233939716905] 
\draw [color={rgb, 255:red, 126; green, 67; blue, 146 }  ,draw opacity=1 ][line width=2.25]    (781.22,105.85) .. controls (684.35,322) and (284.35,431) .. (107.35,322) ;
%Curve Lines [id:da32343107182037956] 
\draw [color={rgb, 255:red, 126; green, 67; blue, 146 }  ,draw opacity=1 ][line width=2.25]    (651.22,107.85) .. controls (554.35,324) and (284.35,431) .. (107.35,322) ;
%Curve Lines [id:da2530643245804639] 
\draw [color={rgb, 255:red, 126; green, 67; blue, 146 }  ,draw opacity=1 ][line width=2.25]    (565.22,196.85) .. controls (361.35,322) and (283.35,344) .. (107.35,322) ;
%Curve Lines [id:da36325608542450827] 
\draw [color={rgb, 255:red, 126; green, 67; blue, 146 }  ,draw opacity=1 ][line width=2.25]    (856.35,197) .. controls (652.49,322.15) and (280.35,374) .. (107.35,322) ;
%Curve Lines [id:da31347865279411946] 
\draw [color={rgb, 255:red, 126; green, 67; blue, 146 }  ,draw opacity=1 ][line width=2.25]    (858.35,318) .. controls (560.35,380) and (270.35,402) .. (101.35,318) ;
%Curve Lines [id:da012543996805587931] 
\draw [color={rgb, 255:red, 126; green, 67; blue, 146 }  ,draw opacity=1 ][line width=2.25]    (778.08,398.23) .. controls (643.25,482.28) and (336.39,489.6) .. (195.39,399.6) ;
%Curve Lines [id:da5372553888976734] 
\draw [color={rgb, 255:red, 126; green, 67; blue, 146 }  ,draw opacity=1 ][line width=2.25]    (858.35,318) .. controls (723.52,402.05) and (336.39,489.6) .. (195.39,399.6) ;
%Curve Lines [id:da4013761544934418] 
\draw [color={rgb, 255:red, 126; green, 67; blue, 146 }  ,draw opacity=1 ][line width=2.25]    (781.22,105.85) .. controls (638.35,299) and (464.35,376) .. (195.39,399.6) ;

%-----------Labels nodes

% Text Node
\draw (195,190) node [anchor=north west][inner sep=0.75pt]  [font=\Huge] [align=left] {$ \dot{v}_{1}^{1}$};

% Text Node

\draw (180,450) node [anchor=north west][inner sep=0.75pt]  [font=\Huge] [align=left] {$ \dot{v}_{1}^{2}$};
% Text Node
\draw (70,340) node [anchor=north west][inner sep=0.75pt]  [font=\Huge] [align=left] {$\dot{v}_{2}^{2}$};
% Text Node

\draw (300,450) node [anchor=north west][inner sep=0.75pt]  [font=\Huge] [align=left] {$\dot{v}_{n_{1} -1}^{2}$};

% Text Node

\draw (725,180) node [anchor=north west][inner sep=0.75pt]  [font=\Huge] [align=left] {$\ddot{v}_{1}^{1}$};
% Text Node

\draw (620,450) node [anchor=north west][inner sep=0.75pt]  [font=\Huge] [align=left] {$\ddot{v}_{1}^{2}$};
% Text Node

% \draw (520,400) node [anchor=north west][inner sep=0.75pt]  [font=\Huge] [align=left] {$\ddot{v}_{2}^{2}$};
% % Text Node
% Text Node

\draw (820,350) node [anchor=north west][inner sep=0.75pt]  [font=\Huge] [align=left] {$\ddot{v}_{n_{2} -2}^{2}$};

\draw (760,450) node [anchor=north west][inner sep=0.75pt]  [font=\Huge] [align=left] {$\ddot{v}_{n_{2} -1}^{2}$};

\end{tikzpicture}

%% file: Figures/Fig-10.b.tex
\tikzset{every picture/.style={line width=0.75pt}} %set default line width to 0.75pt        

\begin{tikzpicture}[x=0.75pt,y=0.75pt,yscale=-1,xscale=1]
%uncomment if require: \path (0,504); %set diagram left start at 0, and has height of 504

%Image [id:dp8782621407439193] 
\draw (506.33,279.67) node  {\includegraphics[width=616.5pt,height=274.5pt]{Figures/Fig-10.png}};
%Curve Lines [id:da45497779343632794] 
\draw [color={rgb, 255:red, 126; green, 67; blue, 146 }  ,draw opacity=1 ][line width=2.25]    (345.86,126.52) .. controls (449.69,67.68) and (599.83,81.34) .. (672.55,127.85) ;
%Curve Lines [id:da44863406273916784] 
\draw [color={rgb, 255:red, 126; green, 67; blue, 146 }  ,draw opacity=1 ][line width=2.25]    (673.55,421.56) .. controls (630.34,269.14) and (678.34,159.14) .. (672.55,127.85) ;
%Curve Lines [id:da2177637363684488] 
\draw [color={rgb, 255:red, 126; green, 67; blue, 146 }  ,draw opacity=1 ][line width=2.25]    (580.71,217.51) .. controls (607.34,175.14) and (642.34,146.14) .. (672.55,127.85) ;
%Curve Lines [id:da17287219867105752] 
\draw [color={rgb, 255:red, 126; green, 67; blue, 146 }  ,draw opacity=1 ][line width=2.25]    (806.34,426.14) .. controls (816.26,379.05) and (791.55,235.85) .. (672.55,127.85) ;
%Curve Lines [id:da3681007948576993] 
\draw [color={rgb, 255:red, 126; green, 67; blue, 146 }  ,draw opacity=1 ][line width=2.25]    (874.69,339) .. controls (852.34,299.14) and (822.34,220.14) .. (677.34,131.14) ;
%Curve Lines [id:da5262187541571024] 
\draw [color={rgb, 255:red, 126; green, 67; blue, 146 }  ,draw opacity=1 ][line width=2.25]    (802.55,125.85) .. controls (748.34,114.14) and (737.34,113.14) .. (672.55,127.85) ;
%Curve Lines [id:da052998726474708935] 
\draw [color={rgb, 255:red, 126; green, 67; blue, 146 }  ,draw opacity=1 ][line width=2.25]    (345.86,126.52) .. controls (396.34,255.14) and (423.34,274.14) .. (349.34,423.14) ;
%Curve Lines [id:da1820193108860797] 
\draw [color={rgb, 255:red, 126; green, 67; blue, 146 }  ,draw opacity=1 ][line width=2.25]    (345.86,126.52) .. controls (410.34,184.14) and (431.34,274.14) .. (428.04,336.86) ;
%Curve Lines [id:da2174624300715633] 
\draw [color={rgb, 255:red, 126; green, 67; blue, 146 }  ,draw opacity=1 ][line width=2.25]    (581.41,340.23) .. controls (601.34,306.14) and (655.34,140.14) .. (672.55,127.85) ;
%Curve Lines [id:da6803182469291718] 
\draw [color={rgb, 255:red, 126; green, 67; blue, 146 }  ,draw opacity=1 ][line width=2.25]    (889.55,216.85) .. controls (840.34,193.14) and (787.52,151.25) .. (672.55,127.85) ;
%Curve Lines [id:da27569687375278806] 
\draw [color={rgb, 255:red, 126; green, 67; blue, 146 }  ,draw opacity=1 ][line width=2.25]    (345.86,126.52) .. controls (338.91,265.2) and (320.24,350.54) .. (218.34,424.14) ;
%Curve Lines [id:da518310076644626] 
\draw [color={rgb, 255:red, 126; green, 67; blue, 146 }  ,draw opacity=1 ][line width=2.25]    (345.86,126.52) .. controls (404.34,123.14) and (444.34,212.14) .. (427.34,214.14) ;
%Curve Lines [id:da5912842809631949] 
\draw [color={rgb, 255:red, 126; green, 67; blue, 146 }  ,draw opacity=1 ][line width=2.25]    (345.86,126.52) .. controls (203.34,254.14) and (186.34,256.14) .. (127.34,344.14) ;
%Curve Lines [id:da9403096025192836] 
\draw [color={rgb, 255:red, 126; green, 67; blue, 146 }  ,draw opacity=1 ][line width=2.25]    (345.86,126.52) .. controls (199.34,171.14) and (224.34,174.14) .. (125.34,214.14) ;
%Curve Lines [id:da9216502239314532] 
\draw [color={rgb, 255:red, 126; green, 67; blue, 146 }  ,draw opacity=1 ][line width=2.25]    (345.86,126.52) .. controls (309.34,126.14) and (310.34,123.14) .. (214.34,125.14) ;

\draw (620,60) node [anchor=north west][inner sep=0.75pt]  [font=\Huge] [align=left] {$\ddot{v}_{i}^{2}$};
% Text Node

\draw (300,60) node [anchor=north west][inner sep=0.75pt]  [font=\Huge] [align=left] {$\dot{v}_{j}^{2}$};
% Text Node

\draw (225,220) node [anchor=north west][inner sep=0.75pt]  [font=\Huge] [align=left] {$ \dot{v}_{1}^{1}$};

% Text Node

\draw (200,470) node [anchor=north west][inner sep=0.75pt]  [font=\Huge] [align=left] {$ \dot{v}_{1}^{2}$};
% Text Node
\draw (110,380) node [anchor=north west][inner sep=0.75pt]  [font=\Huge] [align=left] {$\dot{v}_{2}^{2}$};
% Text Node

\draw (300,470) node [anchor=north west][inner sep=0.75pt]  [font=\Huge] [align=left] {$\dot{v}_{n_{1} -1}^{2}$};

% Text Node

\draw (725,220) node [anchor=north west][inner sep=0.75pt]  [font=\Huge] [align=left] {$\ddot{v}_{1}^{1}$};
% Text Node

\draw (620,450) node [anchor=north west][inner sep=0.75pt]  [font=\Huge] [align=left] {$\ddot{v}_{1}^{2}$};
% Text Node

% \draw (520,400) node [anchor=north west][inner sep=0.75pt]  [font=\Huge] [align=left] {$\ddot{v}_{2}^{2}$};
% % Text Node
% Text Node

\draw (850,380) node [anchor=north west][inner sep=0.75pt]  [font=\Huge] [align=left] {$\ddot{v}_{n_{2} -2}^{2}$};

\draw (750,450) node [anchor=north west][inner sep=0.75pt]  [font=\Huge] [align=left] {$\ddot{v}_{n_{2} -1}^{2}$};

\draw (760,20) node [anchor=north west][inner sep=0.75pt] [font=\tiny]  [align=left] { };

\end{tikzpicture}

%% file: Figures/FigAv1.tex
\tikzset{every picture/.style={line width=0.75pt}} %set default line width to 0.75pt        

\begin{tikzpicture}[x=0.75pt,y=0.75pt,yscale=-1,xscale=1]
%uncomment if require: \path (0,682); %set diagram left start at 0, and has height of 682

%Image [id:dp09158841082861091] 
\draw (390.5,329.66) node  {\includegraphics[width=528.75pt,height=253.5pt]{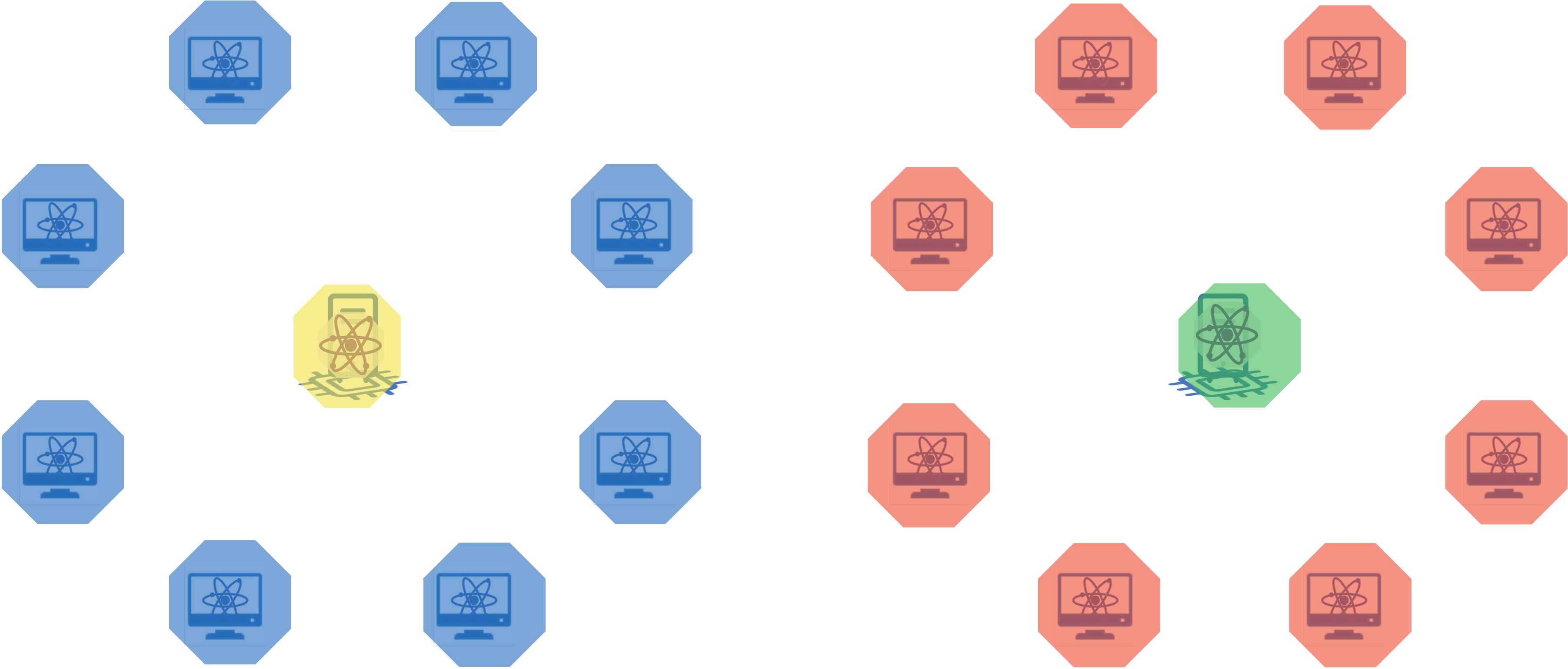}};
%Image [id:dp17653540978315285] 
\draw (1958.5,329.66) node  {\includegraphics[width=528.75pt,height=253.5pt]{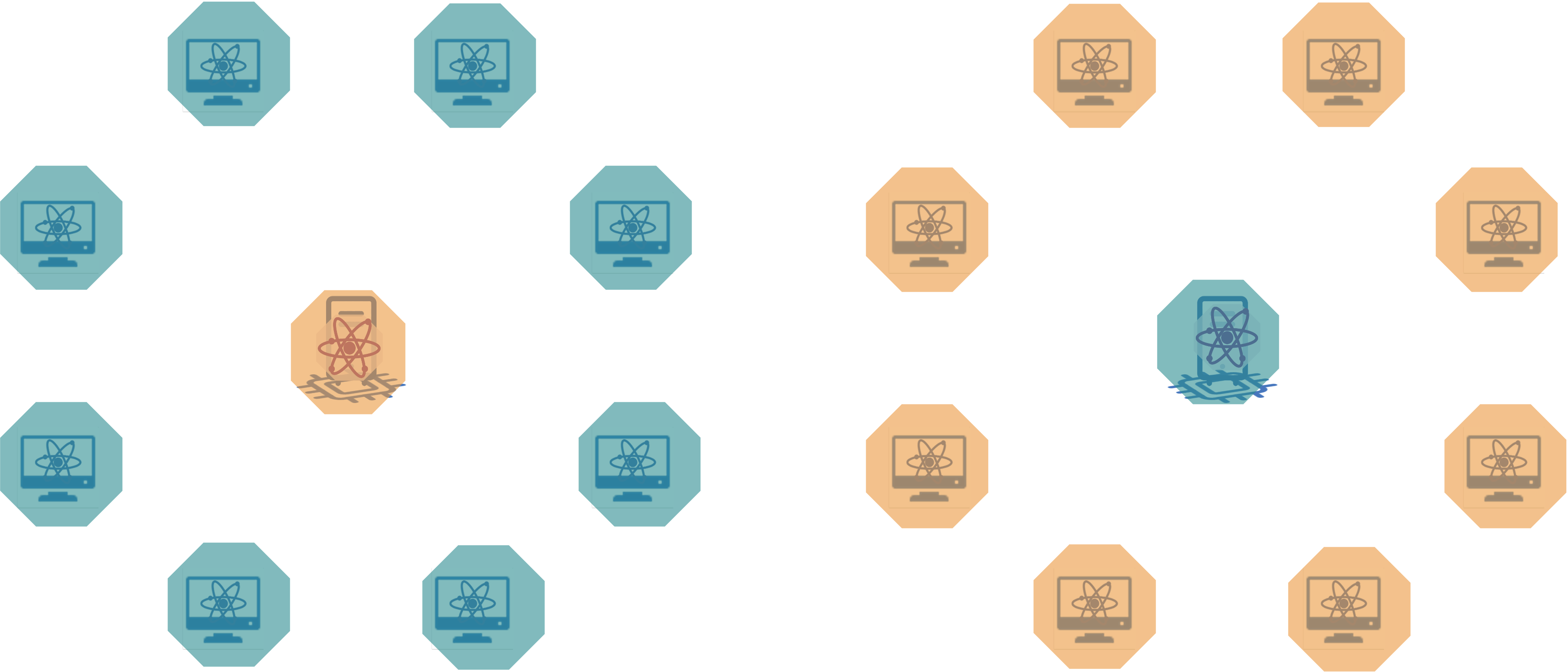}};
%Image [id:dp541735328745913] 
\draw (1163.5,329.66) node  {\includegraphics[width=528.75pt,height=253.5pt]{Figures/Fig-02.av2.png}};
%Straight Lines [id:da9894520501527102] freccia remote
\draw [line width=1.5]    (707.59,580) -- (930,580) ;
\draw [shift={(933,580)}, rotate = 180] [fill={rgb, 255:red, 0; green, 0; blue, 0 }  ][line width=0.08]  [draw opacity=0] (15.6,-3.9) -- (0,0) -- (15.6,3.9) -- cycle    ;
%Straight Lines [id:da4392244176921605] separatore bc
\draw  [dash pattern={on 0.84pt off 2.51pt}]  (1580,130) -- (1580,651.9);
%Straight Lines [id:da4392244176921605] separatore ab
\draw  [dash pattern={on 0.84pt off 2.51pt}]  (800,130) -- (800,651.9);

%Straight Lines [id:da6624249743866606]  freccia coloring
\draw [line width=1.5]    (1470,580) -- (1665,580) ;
\draw [shift={(1667,580)}, rotate = 180] [fill={rgb, 255:red, 0; green, 0; blue, 0 }  ][line width=0.08]  [draw opacity=0] (15.6,-3.9) -- (0,0) -- (15.6,3.9) -- cycle    ;
%Rounded Rect [id:dp797589735915788] 
\draw  [dash pattern={on 0.84pt off 2.51pt}] (25,200.6) .. controls (25,164.37) and (54.37,135) .. (90.6,135) -- (287.4,135) .. controls (323.63,135) and (353,164.37) .. (353,200.6) -- (353,453.06) .. controls (353,489.29) and (323.63,518.66) .. (287.4,518.66) -- (90.6,518.66) .. controls (54.37,518.66) and (25,489.29) .. (25,453.06) -- cycle ;
%Rounded Rect [id:dp6045629499129002] 
\draw [dash pattern={on 0.84pt off 2.51pt}]  (419,200.6) .. controls (419,164.37) and (448.37,135) .. (484.6,135) -- (681.4,135) .. controls (717.63,135) and (747,164.37) .. (747,200.6) -- (747,453.06) .. controls (747,489.29) and (717.63,518.66) .. (681.4,518.66) -- (484.6,518.66) .. controls (448.37,518.66) and (419,489.29) .. (419,453.06) -- cycle ;
%Rounded Rect [id:dp30926599771578067] 
\draw [dash pattern={on 0.84pt off 2.51pt}]  (1597,211.73) .. controls (1597,169.35) and (1631.35,135) .. (1673.73,135) -- (2254.27,135) .. controls (2296.65,135) and (2331,169.35) .. (2331,211.73) -- (2331,441.93) .. controls (2331,484.3) and (2296.65,518.66) .. (2254.27,518.66) -- (1673.73,518.66) .. controls (1631.35,518.66) and (1597,484.3) .. (1597,441.93) -- cycle ;
%Curve Lines [id:da11956542939508619] 
\draw [color={rgb, 255:red, 126; green, 67; blue, 146 }  ,draw opacity=1 ][line width=2.25]    (1366.94,329.24) .. controls (1280.32,306.97) and (1222.94,315.24) .. (1148.94,357.24) ;
%Curve Lines [id:da1572139009440252] 
\draw [color={rgb, 255:red, 126; green, 67; blue, 146 }  ,draw opacity=1 ][line width=2.25]    (1148.94,357.24) .. controls (1077.34,390.76) and (1026.94,370.24) .. (973.94,334.24) ;

% Text Node
\draw (175,145) node [anchor=north west][inner sep=0.75pt]  [font=\Huge] [align=left] {$\dot{S}_{n_1}$};
% Text Node

% Text Node
\draw (565,145) node [anchor=north west][inner sep=0.75pt]  [font=\Huge] [align=left] {$\ddot{S}_{n_2}$};
% Text Node
\draw (1945,145) node [anchor=north west][inner sep=0.75pt]  [font=\Huge] [align=left] {$S_{n_1,n_2}$};

% Text Node
\draw (734,550) node [anchor=north west][inner sep=0.75pt]  [font=\Huge] [align=left] {\textbf{Remote CZ}};

% Text Node
\draw (1497,550) node [anchor=north west][inner sep=0.75pt] [font=\Huge]  [align=center] {\textbf{Color} \\\textbf{Relabeling}};
% Text Node

\draw (150,282.43) node [anchor=north west][inner sep=0.75pt]  [font=\Huge] [align=left] {$ \dot{v}_{1}^{1}$};
% Text Node
\draw  [draw opacity=0][fill={rgb, 255:red, 255; green, 255; blue, 255 }  ,fill opacity=1 ]  (156,465.43) -- (190,465.43) -- (190,505.43) -- (156,505.43) -- cycle  ;
\draw (157,466.43) node [anchor=north west][inner sep=0.75pt]  [font=\Huge] [align=left] {$ \dot{v}_{1}^{2}$};
% Text Node
\draw (47,424.43) node [anchor=north west][inner sep=0.75pt]  [font=\Huge] [align=left] {$\dot{v}_{2}^{2}$};
% Text Node

\draw (272,446.43) node [anchor=north west][inner sep=0.75pt]  [font=\Huge] [align=left] {$\dot{v}_{n_{1} -1}^{2}$};
% Text Node

\draw (614,290.43) node [anchor=north west][inner sep=0.75pt]  [font=\Huge] [align=left] {$\ddot{v}_{1}^{1}$};
% Text Node

\draw (562,450.43) node [anchor=north west][inner sep=0.75pt]  [font=\Huge] [align=left] {$\ddot{v}_{1}^{2}$};
% Text Node

\draw (484,388.43) node [anchor=north west][inner sep=0.75pt]  [font=\Huge] [align=left] {$\ddot{v}_{2}^{2}$};
% Text Node

\draw (671,444.43) node [anchor=north west][inner sep=0.75pt]  [font=\Huge] [align=left] {$\ddot{v}_{n_{2} -1}^{2}$};

% Text Node
\draw (924,283.43) node [anchor=north west][inner sep=0.75pt]  [font=\Huge] [align=left] {$\dot{v}_{1}^{1}$};
% Text Node

\draw (929,467.43) node [anchor=north west][inner sep=0.75pt]  [font=\Huge] [align=left] {$\dot{v}_{1}^{2}$};
% Text Node

\draw (819,425.43) node [anchor=north west][inner sep=0.75pt]  [font=\Huge] [align=left] {$\dot{v}_{2}^{2}$};
% Text Node

\draw (1044,447.43) node [anchor=north west][inner sep=0.75pt]  [font=\Huge] [align=left] {$\dot{v}_{n_{1} -1}^{2}$};
% Text Node

\draw (1386,291.43) node [anchor=north west][inner sep=0.75pt]  [font=\Huge] [align=left] {$\ddot{v}_{1}^{1}$};
% Text Node

\draw (1334,451.43) node [anchor=north west][inner sep=0.75pt]  [font=\Huge] [align=left] {$\ddot{v}_{1}^{2}$};
% Text Node

\draw (1256,389.43) node [anchor=north west][inner sep=0.75pt]  [font=\Huge] [align=left] {$\ddot{v}_{2}^{2}$};
% Text Node

\draw (1443,445.43) node [anchor=north west][inner sep=0.75pt]  [font=\Huge] [align=left] {$\ddot{v}_{n_{2} -1}^{2}$};
% Text Node

\draw (1719,276.43) node [anchor=north west][inner sep=0.75pt]  [font=\Huge] [align=left] {$ \dot{v}_{1}^{1}$};
% Text Node

\draw (1724,460.43) node [anchor=north west][inner sep=0.75pt]  [font=\Huge] [align=left] {$ \dot{v}_{1}^{2}$};
% Text Node

\draw (1614,418.43) node [anchor=north west][inner sep=0.75pt]  [font=\Huge] [align=left] {$\dot{v}_{2}^{2}$};
% Text Node

\draw (1839,440.43) node [anchor=north west][inner sep=0.75pt]  [font=\Huge] [align=left] {$\dot{v}_{n_{1} -1}^{2}$};
% Text Node
\draw (2181,284.43) node [anchor=north west][inner sep=0.75pt]  [font=\Huge] [align=left] {$\ddot{v}_{1}^{1}$};
% Text Node
\draw (2129,444.43) node [anchor=north west][inner sep=0.75pt]  [font=\Huge] [align=left] {$\ddot{v}_{1}^{2}$};
% Text Node

\draw (2051,382.43) node [anchor=north west][inner sep=0.75pt]  [font=\Huge] [align=left] {$\ddot{v}_{2}^{2}$};
% Text Node

\draw (2238,438.43) node [anchor=north west][inner sep=0.75pt]  [font=\Huge] [align=left] {$\ddot{v}_{n_{2} -1}^{2}$};

%-----------------Entangled links ------------------------

%Curve Lines [id:da7458251252512123] 
\draw [color={rgb, 255:red, 126; green, 67; blue, 146 }  ,draw opacity=1 ][line width=2.25]    (254.67,461.99) .. controls (226.87,439.6) and (194.61,369.29) .. (198.76,335.09) ;
%Curve Lines [id:da12900210469614393] 
\draw [color={rgb, 255:red, 126; green, 67; blue, 146 }  ,draw opacity=1 ][line width=2.25]    (329.67,391.99) .. controls (282.97,401.08) and (212.42,377.54) .. (199.31,334.84) ;
%Curve Lines [id:da8699689765198585] 
\draw [color={rgb, 255:red, 126; green, 67; blue, 146 }  ,draw opacity=1 ][line width=2.25]    (329.67,273.99) .. controls (314.05,317.07) and (241.48,354.74) .. (199.31,334.84) ;
%Curve Lines [id:da5870475509902352] 
\draw [color={rgb, 255:red, 126; green, 67; blue, 146 }  ,draw opacity=1 ][line width=2.25]    (255.67,187.99) .. controls (260.12,238.12) and (250.48,303.44) .. (199.31,334.84) ;
%Curve Lines [id:da4975903035757633] 
\draw [color={rgb, 255:red, 126; green, 67; blue, 146 }  ,draw opacity=1 ][line width=2.25]    (69.67,391.99) .. controls (83.08,355.54) and (152.47,318.36) .. (198.86,333.09) ;
%Curve Lines [id:da6962141028067751] 
\draw [color={rgb, 255:red, 126; green, 67; blue, 146 }  ,draw opacity=1 ][line width=2.25]    (142.67,459.99) .. controls (127.38,427.42) and (154.54,350.93) .. (198.86,336.89) ;
%Curve Lines [id:da022605611221807487] 
\draw [color={rgb, 255:red, 126; green, 67; blue, 146 }  ,draw opacity=1 ][line width=2.25]    (143.67,191.99) .. controls (182.3,220.73) and (213.37,275.21) .. (200.9,334.99) ;
%Curve Lines [id:da1564894432852041] 
\draw [color={rgb, 255:red, 126; green, 67; blue, 146 }  ,draw opacity=1 ][line width=2.25]    (68.67,271.99) .. controls (118.37,257.58) and (176.69,274.39) .. (200.9,334.99) ;

%Curve Lines [id:da08438053260402867] 
\draw [color={rgb, 255:red, 126; green, 67; blue, 146 }  ,draw opacity=1 ][line width=2.25]    (644.67,461.99) .. controls (616.87,439.6) and (584.61,369.29) .. (588.76,335.09) ;
%Curve Lines [id:da7329212847453385] 
\draw [color={rgb, 255:red, 126; green, 67; blue, 146 }  ,draw opacity=1 ][line width=2.25]    (719.67,391.99) .. controls (672.97,401.08) and (602.42,377.54) .. (589.31,334.84) ;
%Curve Lines [id:da8777810413094981] 
\draw [color={rgb, 255:red, 126; green, 67; blue, 146 }  ,draw opacity=1 ][line width=2.25]    (719.67,273.99) .. controls (704.05,317.07) and (631.48,354.74) .. (589.31,334.84) ;
%Curve Lines [id:da8607138745767958] 
\draw [color={rgb, 255:red, 126; green, 67; blue, 146 }  ,draw opacity=1 ][line width=2.25]    (645.67,187.99) .. controls (650.12,238.12) and (640.48,303.44) .. (589.31,334.84) ;
%Curve Lines [id:da796825892356693] 
\draw [color={rgb, 255:red, 126; green, 67; blue, 146 }  ,draw opacity=1 ][line width=2.25]    (459.67,391.99) .. controls (473.08,355.54) and (542.47,318.36) .. (588.86,333.09) ;
%Curve Lines [id:da575623517675816] 
\draw [color={rgb, 255:red, 126; green, 67; blue, 146 }  ,draw opacity=1 ][line width=2.25]    (532.67,459.99) .. controls (517.38,427.42) and (544.54,350.93) .. (588.86,336.89) ;
%Curve Lines [id:da0931381630260314] 
\draw [color={rgb, 255:red, 126; green, 67; blue, 146 }  ,draw opacity=1 ][line width=2.25]    (533.67,191.99) .. controls (572.3,220.73) and (603.37,275.21) .. (590.9,334.99) ;
%Curve Lines [id:da8171680565718034] 
\draw [color={rgb, 255:red, 126; green, 67; blue, 146 }  ,draw opacity=1 ][line width=2.25]    (458.67,271.99) .. controls (508.37,257.58) and (566.69,274.39) .. (590.9,334.99) ;

%Curve Lines [id:da12048832057983694] 
\draw [color={rgb, 255:red, 126; green, 67; blue, 146 }  ,draw opacity=1 ][line width=2.25]    (1024.67,462.99) .. controls (996.87,440.6) and (964.61,370.29) .. (968.76,336.09) ;
%Curve Lines [id:da05510215279454844] 
\draw [color={rgb, 255:red, 126; green, 67; blue, 146 }  ,draw opacity=1 ][line width=2.25]    (1099.67,392.99) .. controls (1052.97,402.08) and (982.42,378.54) .. (969.31,335.84) ;
%Curve Lines [id:da63160605951362] 
\draw [color={rgb, 255:red, 126; green, 67; blue, 146 }  ,draw opacity=1 ][line width=2.25]    (1099.67,274.99) .. controls (1084.05,318.07) and (1011.48,355.74) .. (969.31,335.84) ;
%Curve Lines [id:da1870710968726318] 
\draw [color={rgb, 255:red, 126; green, 67; blue, 146 }  ,draw opacity=1 ][line width=2.25]    (1025.67,188.99) .. controls (1030.12,239.12) and (1020.48,304.44) .. (969.31,335.84) ;
%Curve Lines [id:da47531181545021695] 
\draw [color={rgb, 255:red, 126; green, 67; blue, 146 }  ,draw opacity=1 ][line width=2.25]    (839.67,392.99) .. controls (853.08,356.54) and (922.47,319.36) .. (968.86,334.09) ;
%Curve Lines [id:da2452046632678272] 
\draw [color={rgb, 255:red, 126; green, 67; blue, 146 }  ,draw opacity=1 ][line width=2.25]    (912.67,460.99) .. controls (897.38,428.42) and (924.54,351.93) .. (968.86,337.89) ;
%Curve Lines [id:da18469471743524046] 
\draw [color={rgb, 255:red, 126; green, 67; blue, 146 }  ,draw opacity=1 ][line width=2.25]    (913.67,192.99) .. controls (952.3,221.73) and (983.37,276.21) .. (970.9,335.99) ;
%Curve Lines [id:da2893022548057481] 
\draw [color={rgb, 255:red, 126; green, 67; blue, 146 }  ,draw opacity=1 ][line width=2.25]    (838.67,272.99) .. controls (888.37,258.58) and (946.69,275.39) .. (970.9,335.99) ;

%Curve Lines [id:da14917128844493444] 
\draw [color={rgb, 255:red, 126; green, 67; blue, 146 }  ,draw opacity=1 ][line width=2.25]    (1419.67,461.99) .. controls (1391.87,439.6) and (1359.61,369.29) .. (1363.76,335.09) ;
%Curve Lines [id:da1172799532498241] 
\draw [color={rgb, 255:red, 126; green, 67; blue, 146 }  ,draw opacity=1 ][line width=2.25]    (1494.67,391.99) .. controls (1447.97,401.08) and (1377.42,377.54) .. (1364.31,334.84) ;
%Curve Lines [id:da15437348269721451] 
\draw [color={rgb, 255:red, 126; green, 67; blue, 146 }  ,draw opacity=1 ][line width=2.25]    (1494.67,273.99) .. controls (1479.05,317.07) and (1406.48,354.74) .. (1364.31,334.84) ;
%Curve Lines [id:da473521624852015] 
\draw [color={rgb, 255:red, 126; green, 67; blue, 146 }  ,draw opacity=1 ][line width=2.25]    (1420.67,187.99) .. controls (1425.12,238.12) and (1415.48,303.44) .. (1364.31,334.84) ;
%Curve Lines [id:da9489909588426053] 
\draw [color={rgb, 255:red, 126; green, 67; blue, 146 }  ,draw opacity=1 ][line width=2.25]    (1234.67,391.99) .. controls (1248.08,355.54) and (1317.47,318.36) .. (1363.86,333.09) ;
%Curve Lines [id:da8728824258303441] 
\draw [color={rgb, 255:red, 126; green, 67; blue, 146 }  ,draw opacity=1 ][line width=2.25]    (1307.67,459.99) .. controls (1292.38,427.42) and (1319.54,350.93) .. (1363.86,336.89) ;
%Curve Lines [id:da05397100197277682] 
\draw [color={rgb, 255:red, 126; green, 67; blue, 146 }  ,draw opacity=1 ][line width=2.25]    (1308.67,191.99) .. controls (1347.3,220.73) and (1378.37,275.21) .. (1365.9,334.99) ;
%Curve Lines [id:da6792822755051234] 
\draw [color={rgb, 255:red, 126; green, 67; blue, 146 }  ,draw opacity=1 ][line width=2.25]    (1233.67,271.99) .. controls (1283.37,257.58) and (1341.69,274.39) .. (1365.9,334.99) ;

%Curve Lines [id:da2121563932755164] 
\draw [color={rgb, 255:red, 126; green, 67; blue, 146 }  ,draw opacity=1 ][line width=2.25]    (1819.67,460.99) .. controls (1791.87,438.6) and (1759.61,368.29) .. (1763.76,334.09) ;
%Curve Lines [id:da0025196835042017307] 
\draw [color={rgb, 255:red, 126; green, 67; blue, 146 }  ,draw opacity=1 ][line width=2.25]    (1894.67,390.99) .. controls (1847.97,400.08) and (1777.42,376.54) .. (1764.31,333.84) ;
%Curve Lines [id:da6401044661694427] 
\draw [color={rgb, 255:red, 126; green, 67; blue, 146 }  ,draw opacity=1 ][line width=2.25]    (1894.67,272.99) .. controls (1879.05,316.07) and (1806.48,353.74) .. (1764.31,333.84) ;
%Curve Lines [id:da9925736687778418] 
\draw [color={rgb, 255:red, 126; green, 67; blue, 146 }  ,draw opacity=1 ][line width=2.25]    (1820.67,186.99) .. controls (1825.12,237.12) and (1815.48,302.44) .. (1764.31,333.84) ;
%Curve Lines [id:da5105835943714491] 
\draw [color={rgb, 255:red, 126; green, 67; blue, 146 }  ,draw opacity=1 ][line width=2.25]    (1634.67,390.99) .. controls (1648.08,354.54) and (1717.47,317.36) .. (1763.86,332.09) ;
%Curve Lines [id:da6320189253450761] 
\draw [color={rgb, 255:red, 126; green, 67; blue, 146 }  ,draw opacity=1 ][line width=2.25]    (1707.67,458.99) .. controls (1692.38,426.42) and (1719.54,349.93) .. (1763.86,335.89) ;
%Curve Lines [id:da50017499581131] 
\draw [color={rgb, 255:red, 126; green, 67; blue, 146 }  ,draw opacity=1 ][line width=2.25]    (1708.67,190.99) .. controls (1747.3,219.73) and (1778.37,274.21) .. (1765.9,333.99) ;
%Curve Lines [id:da8764087009183164] 
\draw [color={rgb, 255:red, 126; green, 67; blue, 146 }  ,draw opacity=1 ][line width=2.25]    (1633.67,270.99) .. controls (1683.37,256.58) and (1741.69,273.39) .. (1765.9,333.99) ;

%Curve Lines [id:da4478094476238734] 
\draw [color={rgb, 255:red, 126; green, 67; blue, 146 }  ,draw opacity=1 ][line width=2.25]    (2215.67,462.99) .. controls (2187.87,440.6) and (2155.61,370.29) .. (2159.76,336.09) ;
%Curve Lines [id:da15557184295061444] 
\draw [color={rgb, 255:red, 126; green, 67; blue, 146 }  ,draw opacity=1 ][line width=2.25]    (2290.67,392.99) .. controls (2243.97,402.08) and (2173.42,378.54) .. (2160.31,335.84) ;
%Curve Lines [id:da1112657227749928] 
\draw [color={rgb, 255:red, 126; green, 67; blue, 146 }  ,draw opacity=1 ][line width=2.25]    (2290.67,274.99) .. controls (2275.05,318.07) and (2202.48,355.74) .. (2160.31,335.84) ;
%Curve Lines [id:da2788508896133094] 
\draw [color={rgb, 255:red, 126; green, 67; blue, 146 }  ,draw opacity=1 ][line width=2.25]    (2216.67,188.99) .. controls (2221.12,239.12) and (2211.48,304.44) .. (2160.31,335.84) ;
%Curve Lines [id:da5690319389210059] 
\draw [color={rgb, 255:red, 126; green, 67; blue, 146 }  ,draw opacity=1 ][line width=2.25]    (2030.67,392.99) .. controls (2044.08,356.54) and (2113.47,319.36) .. (2159.86,334.09) ;
%Curve Lines [id:da924081249480234] 
\draw [color={rgb, 255:red, 126; green, 67; blue, 146 }  ,draw opacity=1 ][line width=2.25]    (2103.67,460.99) .. controls (2088.38,428.42) and (2115.54,351.93) .. (2159.86,337.89) ;
%Curve Lines [id:da23324915802656498] 
\draw [color={rgb, 255:red, 126; green, 67; blue, 146 }  ,draw opacity=1 ][line width=2.25]    (2104.67,192.99) .. controls (2143.3,221.73) and (2174.37,276.21) .. (2161.9,335.99) ;
%Curve Lines [id:da004546390447689341] 
\draw [color={rgb, 255:red, 126; green, 67; blue, 146 }  ,draw opacity=1 ][line width=2.25]    (2029.67,272.99) .. controls (2079.37,258.58) and (2137.69,275.39) .. (2161.9,335.99) ;

%Curve Lines [id:da014903102825976844] 
\draw [color={rgb, 255:red, 126; green, 67; blue, 146 }  ,draw opacity=1 ][line width=2.25]    (2160.94,331.24) .. controls (2074.32,308.97) and (2016.94,317.24) .. (1942.94,359.24) ;
%Curve Lines [id:da12654818327898487] 
\draw [color={rgb, 255:red, 126; green, 67; blue, 146 }  ,draw opacity=1 ][line width=2.25]    (1942.94,359.24) .. controls (1871.34,392.76) and (1820.94,372.24) .. (1767.94,336.24) ;

%------------Legenda--------

%Image [id:dp5113430279365352] 
\draw (50,627.33) node  {\includegraphics[width=21.75pt,height=22.99pt]{Figures/blu.png}};
%Image [id:dp17299069323517247] 
\draw (50,567.83) node  {\includegraphics[width=21.75pt,height=25.24pt]{Figures/giallo.png}};
%Image [id:dp11621001860668345] 
\draw (420,561.33) node  {\includegraphics[width=21.75pt,height=22.99pt]{Figures/verde.png}};
%Image [id:dp6683370430932896] 
\draw (420,624.66) node  {\includegraphics[width=22pt,height=23.49pt]{Figures/rosso.png}};
%Image [id:dp7602197634871326] 
\draw (1760,554.73) node  {\includegraphics[width=21.75pt,height=22.99pt]{Figures/arancione.png}};
%Image [id:dp9860480691383868] 
\draw (1760,616.56) node  {\includegraphics[width=21.75pt,height=22.99pt]{Figures/azzurro.png}};

% Text Node
\draw (1800,540) node [anchor=north west][inner sep=0.75pt]  [font=\Huge] [align=left] {$P_1=\{\dot{v}^1_1,\ddot{v}^2_1,\cdots,\ddot{v}^2_{n_1-1}\}$};

% Text Node
\draw (1800,600) node [anchor=north west][inner sep=0.75pt]  [font=\Huge] [align=left] {$P_2=\{\ddot{v}^1_1,\dot{v}^2_1,\cdots,\dot{v}^2_{n_2-1}\}$};

% Text Node
\draw (80,540) node [anchor=north west][inner sep=0.75pt]  [font=\Huge] [align=left] {$\dot{P}_1=\{\dot{v}^1_1\}$};

\draw (80,600) node [anchor=north west][inner sep=0.75pt]  [font=\Huge] [align=left] {$\dot{P}_2=\{\dot{v}^2_1,\cdots,\dot{v}^2_{n_1-1}\}$};

% Text Node
\draw (450,540) node [anchor=north west][inner sep=0.75pt]  [font=\Huge] [align=left] {$\ddot{P}_1=\{\ddot{v}^1_1\}$};

\draw (450,600) node [anchor=north west][inner sep=0.75pt]  [font=\Huge] [align=left] {$\ddot{P}_2= \{\ddot{v}^2_1,\cdots,\ddot{v}^2_{n_2-1}\}$};

%------------------------------------------------

\draw (370,680) node [anchor=north west][inner sep=0.75pt]  [font=\Huge] [align=left] {(a)};

\draw (1150,680) node [anchor=north west][inner sep=0.75pt]  [font=\Huge] [align=left] {(b)};

\draw (1930,680) node [anchor=north west][inner sep=0.75pt]  [font=\Huge] [align=left] {(c)};

\end{tikzpicture}